\documentstyle[aps,prd,epsf,floats]{revtex}

\renewcommand{\Re}{{\mathrm{Re}}}
\renewcommand{\Im}{{\mathrm{Im}}}

\makeatletter 
\def\preprint#1{
   \def\@preprint{\noindent\hfill\hbox{#1}\vskip 10pt}
}
\makeatother

\preprint{
   \begin{tabular}{l}
	SLAC--PUB--8406 \\
	CPHT--S001--0100 \\ 
	hep-ph/0003233
   \end{tabular}
}

\title{Exclusive production of pion pairs in $\gamma^* \gamma $
collisions at large $Q^2$ \thanks{Work supported by Department of
Energy contract DE--AC03--76SF00515 and by TMR contracts
FMRX--CT96--0008 and FMRX--CT98--0194}}

\author{M. Diehl \thanks{Supported by the Feodor Lynen Program of the
Alexander von Humboldt Foundation}}

\address{Stanford Linear Accelerator Center, Stanford University,
Stanford, CA 94309, U.S.A.}

\author{T. Gousset}

\address{SUBATECH, B.P.~20722, 44307 Nantes, France \thanks{Unit\'e
mixte 6457 de l'Universit\'e de Nantes, de l'\'Ecole des Mines de
Nantes et de l'IN2P3/CNRS}}

\author{B. Pire}

\address{CPhT, \'Ecole Polytechnique, 91128 Palaiseau, France
\thanks{Unit\'e mixte C7644 du CNRS}}

\date{March 2000}

\begin{document}

\maketitle

\begin{abstract}
We perform a QCD analysis of the exclusive production of two mesons in
$\gamma^* \gamma$ collisions in the kinematical domain of large photon
virtuality $Q$ and small hadronic invariant mass $W$. This reaction is
dominated by a scale invariant mechanism which factorizes into a
perturbative subprocess, $\gamma^* \gamma \to q \bar q$ or $\gamma^*
\gamma \to g g$, and a generalized two-meson distribution
amplitude. We develop in detail the phenomenology of this process at
$e^+e^-$ colliders. Using a simple model for the two-pion distribution
amplitude, based on its general properties, we estimate the cross
section for the kinematics accessible at BABAR, BELLE, CLEO and LEP.
\end{abstract}




\section{Introduction}

Exclusive hadron production in two-photon collisions provides a tool
to study a variety of fundamental aspects of QCD and has long been a
subject of great interest (cf., e.g., \cite{Terazawa,Bud,photon-conf}
and references therein). Recently a new facet of this has been pointed
out, namely the physics of the process $\gamma^* \gamma \to \pi \pi$
in the region where $Q^2$ is large but $W^2$ small~\cite{DGPT}. This
process factorizes~\cite{MRG,Freund} into a perturbatively calculable,
short-distance dominated scattering $\gamma^* \gamma \to q \bar q$ or
$\gamma^* \gamma \to g g$, and non-perturbative matrix elements
measuring the transitions $q \bar q \to \pi \pi$ and $g g \to \pi
\pi$. We have called these matrix elements generalized distribution
amplitudes (GDAs) to emphasize their close connection to the
distribution amplitudes introduced many years ago in the QCD
description of exclusive hard processes~\cite{LepageBrodsky}.

Indeed it is instructive to consider $\gamma^* \gamma \to \pi \pi$ as
a generalization of the process $\gamma^* \gamma \to \pi^0$, where the
distribution amplitude of a single pion appears. The $\gamma$--$\pi$
transition form factor has been the subject of detailed theoretical
studies~\cite{TFF}. The experimental data~\cite{CLEO} are well
reproduced by a description based on QCD factorization and provide one
of the best constraints so far on the form of the single-pion
distribution amplitude.

{}From a different point of view $\gamma^* \gamma \to \pi \pi$ is the
crossed channel of virtual Compton scattering on a pion. The
kinematical region we consider here is closely related to deeply
virtual Compton scattering (DVCS), which has attracted considerable
attention in the context of skewed parton distributions~\cite{DVCS}.

Our reaction can also be seen as the exclusive limit of a
hadronization process. The hadronization of a $q \bar q$-pair
originating from a hard, short-distance process such as a $\gamma^*
\gamma $ collision is usually formulated in terms of fragmentation
functions which describe in a universal way semi-inclusive reactions,
specifically the transition from a quark or antiquark to a final-state
hadron when one integrates over all final states containing this
hadron. We specialize here to the case where the final state consists
of two mesons with specified four-momenta, and nothing else.

Like other hadronic matrix elements the GDAs are process
independent. It has recently been pointed out~\cite{rho-pi-pi} that
they occur in the hard exclusive process $\gamma^* p\to \pi\pi\, p$,
where the pion pair is or is not the decay product of a $\rho$ meson,
and that the analysis of that reaction would benefit from the
measurement of the two-pion GDA in $\gamma^* \gamma \to \pi \pi$.

All these aspects lead us to consider GDAs as a promising new tool for
hadronic physics, which may be used to unveil some of the mysteries of
hadronization and the confining regime of QCD. The process $\gamma^*
\gamma \to \pi \pi$ is well suited to access these quantities
experimentally. In the present paper, we develop in detail the
phenomenology of this reaction and emphasize the feasibility of its
investigation at existing $e^+ e^-$ colliders.

In Sect.~\ref{process} we discuss the kinematics of our process,
recall its main properties in the factorization regime we are
interested in, and sketch the crossing relation between
$\gamma^*\gamma\to \pi\pi$ and deep virtual Compton scattering. In
Sect.~\ref{general} we list the general properties of generalized
distribution amplitudes and in particular review their
QCD evolution equations. These properties lead us to construct a
simple model of the two-pion GDA, which is described in
Sect.~\ref{simple-model}. Section~\ref{one-pion} gives a comparison
between one-pion and two-pion production in $\gamma^* \gamma$
collisions. Relations with the inclusive production of hadrons,
commonly described by the photon structure function, are discussed in
Sect.~\ref{structure-function}. The phenomenology of our process in $e
\gamma$ collisions is described in detail in
Sect.~\ref{phenomenology}, with special emphasis on the information
contained in angular distributions and in the interference with the
bremsstrahlung mechanism. In Sect.~\ref{cross-sections} we give
estimates for the cross section for various experimental setups at
existing $e^+e^-$ colliders. Section~\ref{summary} contains our
conclusions. In Appendix~\ref{pion-states} we specify our sign
conventions for pion states, and in Appendix~\ref{beam-polar} we
discuss what additional information can be obtained with polarized
beams.

\section{The process $\gamma^* \gamma \to \pi \pi$}
\label{process}

\subsection{Kinematics in the $\gamma^* \gamma$ center of mass.}
\label{kinematics}

The reaction we are interested in is
\begin{equation}
e(k) + \gamma (q') \to e(k')  +\pi^i(p) + \pi^j(p') ,
\label{egamma}
\end{equation}
where four-momenta are indicated in parentheses. We further use
\begin{equation}
q = k - k', \hspace{2em}
Q^2 = -q^2, \hspace{2em}
P = p+p', \hspace{2em}
W^2 = P^2 .
\end{equation}
The pions may be charged ($i=+$, $j=-$) or neutral ($i=j=0$), and the
lepton $e$ may be an electron or a positron. Scattered with large
momentum transfer this lepton radiates a virtual photon $\gamma^*
(q)$, and for the $\gamma^* \gamma$ subprocess we introduce the
Bjorken variable
\begin{equation}
x = \frac{Q^2}{2 q\cdot q'} = \frac{Q^2}{Q^2+W^2}.
\end{equation}

In $e^+e^-$ collisions the photon $\gamma (q')$ can be obtained by
bremsstrahlung from the other beam lepton, so that the overall process
is
\begin{equation}
e(k) + e(l) \to e(k') + e(l') +\pi^i(p) + \pi^j(p') 
\label{ee}
\end{equation}
with $q' = l - l'$. In the spirit of the equivalent photon
approximation we approximate $q'^2$ as zero and the momenta $q'$ and
$l$ as collinear. We write $E_1 = k^0$, $E_2 = l^0$ and $q'^0 = x_2\,
l^0$ for the energies in the laboratory frame.\footnote{ We neglect
the small finite crossing angle between the beams at BELLE, so that in
our parlance the lepton beams are collinear in the ``laboratory
frame''.} For the c.m.\ energies of the $ee$ and $e\gamma$ collisions
we have
\begin{equation}
s_{ee} = (k+l)^2 , \hspace{2em}
s_{e\gamma} = (k+q')^2 = x_2\, s_{ee} .
\end{equation}

\begin{figure}
   \begin{center}
        \leavevmode
        \epsfxsize=0.65\hsize
        \epsfbox{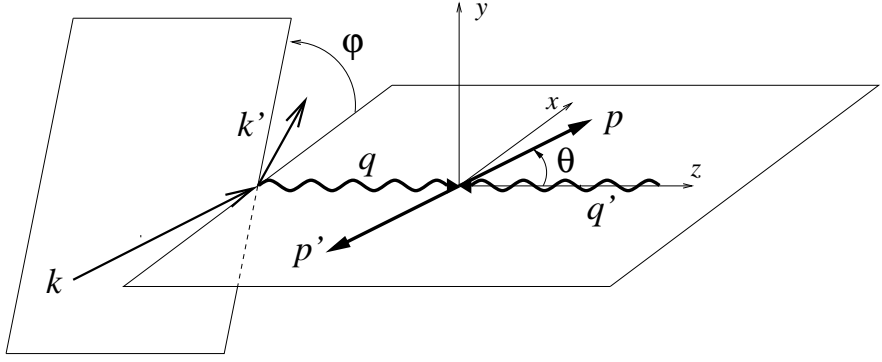}
   \end{center}
\caption{\label{fig-kinematics} The kinematics of $e(k) + \gamma (q')
\to e(k') +\pi^i(p) + \pi^j(p')$ in the center of mass of the pion
pair.}
\end{figure}

Let us now discuss the kinematics in the $\gamma^* \gamma$ center of
mass frame. We use a coordinate system with the $z$ axis along
$\mathbf q$, and with $x$ and $y$ axes such that $\mathbf p$ lies in
the $x$-$z$ plane and has a positive $x$ component, i.e.,
\begin{equation}
q = (q^0, 0, 0, |{\mathbf q}|) , \hspace{3em}
p = (p^0, |{\mathbf p}| \sin\theta, 0, |{\mathbf p}| \cos\theta) ,
\end{equation}
where we have introduced the polar angle $\theta$ of $\mathbf
p$. Another natural variable for our process in this frame is the
azimuth $\varphi$ of ${\mathbf k}'$, which is the angle between the
leptonic and the hadronic planes, cf.\ Fig.~\ref{fig-kinematics}. In
terms of Lorentz invariants these angles can be obtained from
\begin{eqnarray}
\cos\theta &=& \frac{2 q\cdot (p'-p)}{\beta\, (Q^2+W^2)} ,
 \nonumber \\
\cos\varphi &=& \frac{2 k\cdot (p'-p)\, (Q^2+W^2) +
             \beta \cos\theta\, 
             [ Q^2 (s_{e\gamma}-Q^2-W^2) - s_{e\gamma} W^2 ]}{2 \beta
             \sin\theta
             \sqrt{s_{e\gamma}\, Q^2 W^2 (s_{e\gamma}-Q^2-W^2)
                   \rule{0em}{1.8ex} }} ,
 \nonumber \\
\sin\varphi &=& \frac{4 \epsilon_{\alpha\beta\gamma\delta} \,
             (p+p')^\alpha\, p^\beta\, k^\gamma\, q^\delta }{\beta
             \sin\theta
             \sqrt{s_{e\gamma}\, Q^2 W^2 (s_{e\gamma}-Q^2-W^2)
                   \rule{0em}{1.8ex} }}
\end{eqnarray}
with $\epsilon_{0123} = +1$ and the velocity
\begin{equation}
\beta = \sqrt{1-\frac{4m_\pi^2}{W^2}}
\end{equation}
of the pions. A further quantity we will use is the usual $y$-variable
for the $e\gamma$ collision,
\begin{equation}
y = \frac{q\cdot q'}{k\cdot q'} = \frac{Q^2+W^2}{s_{e\gamma}} ,
  \label{y-def}
\end{equation}
which can be traded for
\begin{equation}
\epsilon = \frac{1-y}{1-y+y^2/2} ,
\end{equation}
the ratio of longitudinal to transverse polarization of the
virtual photon $\gamma^* (q)$.

We finally define light cone components $a^{\pm} = (a^0 \pm a^3)
/\sqrt{2}$ for any four-vector $a$ and introduce the fraction
\begin{equation}
\zeta=\frac{p^+}{P^+} = \frac{1+\beta\cos\theta}{2}
\end{equation}
of light cone momentum carried by $\pi^i(p)$ with respect to the
pion pair.

\subsection{Factorization at large $Q^2$ and small $W^2$}
\label{factorization}

\begin{figure}
   \begin{center}
        \leavevmode
        \hspace{0.05\hsize} \epsfxsize=0.45\hsize
        \epsfbox{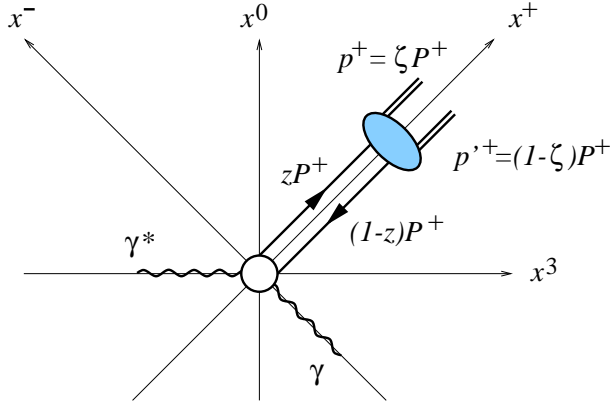}
   \end{center}
\caption{\label{fig-spacetime} Spacetime diagram of $\gamma^*\gamma
  \to \pi\pi$ in the Breit frame.}
\end{figure}

Let us briefly review how $\gamma^*\gamma\to \pi\pi$ factorizes in the
kinematical regime we are interested in. Firstly, we require $Q^2$ to
be large compared with the scale $\Lambda^2 \sim 1$~GeV$^2$ of soft
interactions, thus providing a hard scale for the process. Secondly,
we ask $W^2$ to be small compared with this large scale $Q^2$. In this
regime the dynamics of the process is conveniently represented in the
Breit frame, obtained by boosting from the $\gamma^*\gamma$ center of
mass along the $z$ axis. The spacetime cartoon of the process one can
derive
from power counting and factorization arguments is shown in
Fig.~\ref{fig-spacetime}.

In the Breit frame the real photon moves fast in the negative $z$
direction and is scattered into an energetic hadronic system moving in
the positive $z$ direction. The
hard part of this process takes place at the level of elementary
constituents, and the minimal number of quarks and gluons compatible
with conservation laws (color etc.) are produced. At Born level one
simply has $\gamma^*\gamma\to q\bar{q}$, but through a quark box the
photons can also couple to two gluons. Each quark or gluon carries a
fraction $z$ or $1-z$ of the large light-cone momentum component
$P^+$. Subsequently the soft part of the reaction, i.e., hadronization
into a pion pair, takes place.

\begin{figure}
   \begin{center}
        \leavevmode
        \epsfxsize=0.75\hsize
        \epsfbox{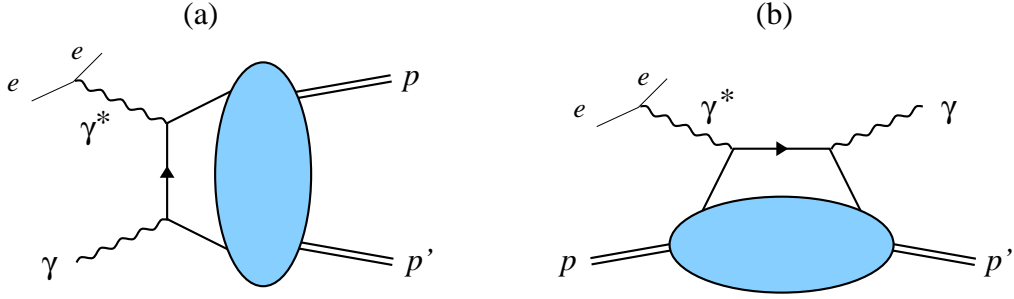}
   \end{center}
\caption{\label{fig-diagrams} (a) Factorization of the process
$\gamma^*\gamma \to \pi\pi$ in the region $Q^2 \gg W^2,
\Lambda^2$. The hard scattering is shown at Born level, with a second
diagram being obtained by interchanging the two photon vertices. The
blob denotes the two-pion GDA. (b) Crossing relates this process to
deep virtual Compton scattering, $\gamma^*\pi \to \gamma\pi$. The blob
now denotes the (skewed) quark distribution in the pion.}
\end{figure}

At leading order in $\alpha_S$ the amplitude is given by the diagram
of Fig.~\ref{fig-diagrams} (a) and the one where the two photon
vertices are interchanged. One calculates for the hadronic
tensor~\cite{DGPT}
\begin{equation}
  \label{tensor-two-pi}
T^{\mu\nu} = i\int\! d^4 x\, e^{-i q\cdot x} \,
    \langle \pi(p) \pi(p') |\,
    T J_{\mathrm{em}}^\mu(x) J_{\mathrm{em}}^\nu(0) \,| 0 \rangle
    \nonumber \\
 = -g^{\mu\nu}_T
\sum_q \frac{e_q^2}{2}\int_0^1\! dz\,{2z-1\over z(1-z)}\,
\Phi^{\pi\pi}_q(z,\zeta,W^2) ,
\end{equation}
where $g^{\mu\nu}_T$ denotes the metric tensor in transverse space
($g^{11}=-1$). The sum on the r.h.s.\ runs over all quarks flavors,
$e_q$ is the charge of quark $q$ in units of the positron charge $e$,
and $e J_{\mathrm{em}}^\mu(x)$ is the electromagnetic current. While
the expression of the hard subprocess $\gamma^*\gamma\to q\bar{q}$ is
explicit in Eq.~(\ref{tensor-two-pi}), the soft part of
$\gamma^*\gamma \to \pi\pi$ is parameterized by the generalized
distribution amplitude
\begin{equation}
\Phi^{\pi\pi}_q(z,\zeta,W^2)
=\int\frac{dx^-}{2\pi}e^{-iz(P^+ x^-)}\,
\langle\pi(p) \pi(p') 
|\,\bar{q}(x^-)\gamma^+ q(0)\,|0\rangle
 \label{gda-def}
\end{equation}
for each quark flavor $q$. We work in light cone gauge $A^+ = 0$,
otherwise the usual path ordered exponential of gluon potentials
appears between the quark fields. $\Phi_q$ depends on the light-cone
fraction $z$ of the quark with respect to the pion pair, on the
kinematical variables $\zeta$ and $W^2$ of the pions, and on a
factorization scale. The latter dependence, not displayed in
Eq.~(\ref{gda-def}), will be discussed in Sect.~\ref{evolution}.

In Eq.~(\ref{tensor-two-pi}) a scaling behavior for our process is
manifest: at fixed $\zeta$ and $W^2$ the $\gamma^\ast \gamma$
amplitude is independent of $Q^2$, up to logarithmic scaling
violations from radiative corrections to the hard scattering and from
the evolution of the two-pion distribution amplitude. This scaling
property is central to all processes where a factorization theorem
holds, and it is the basic signature one looks for when testing
whether the asymptotic analysis developed here applies to an
experimental situation at finite $Q^2$. There will of course be power
corrections in $\Lambda /Q$ and $W /Q$ to this leading
mechanism. Examples are the hadronic component of the real photon, and
the effect in the hard scattering of the transverse momentum of the
produced parton pair. We note that the crossed channel, i.e., virtual
Compton scattering has been analyzed in detail within the operator
product expansion~\cite{MRG,Wat,Leipzig}, which may provide a framework
for a systematic study of higher twist effects.

Contracting the hadronic tensor (\ref{tensor-two-pi}) with the photon
polarization vectors we see that in order to give a nonzero
$\gamma^\ast \gamma \to \pi \pi$ amplitude the virtual photon must
have the same helicity as the real one. As in the case of deep virtual
Compton scattering this is a direct consequence of chiral invariance
in the collinear hard-scattering process~\cite{DGPR,CD} and is valid
at all orders in $\alpha_S$. In the case of the $\gamma^\ast \gamma
\to g g$ subprocess the photon helicities can also be opposite
\cite{Kivel}. In any case the virtual photon must be transverse. As a
consequence nonleading twist effects can be studied in the amplitude
for longitudinal $\gamma^*$ polarization, without any ``background''
from leading twist pieces. We will develop in
Sect.~\ref{phenomenology} how the different $\gamma^\ast \gamma$
helicity amplitudes are experimentally accessible.

As we already mentioned, there is a close analogy of two-pion
production in the region $Q^2 \gg W^2, \Lambda^2$ with the one-pion
channel, commonly described in terms of the $\gamma$--$\pi$ transition
form factor. There again a factorization theorem holds, which allows
the hadronic tensor $T^{\mu\nu}$ to be expressed in terms of the
single-pion distribution amplitude $\phi^{\pi}$ as
\begin{equation}
  \label{tensor-one-pi}
T^{\mu\nu} = i\int\! d^4 x\, e^{-i q\cdot x} \,
    \langle \pi^0 |\,
    T J_{\mathrm{em}}^\mu(x) J_{\mathrm{em}}^\nu(0) \,| 0 \rangle
    \nonumber \\
 =  \epsilon^{\mu\nu}_T
\sum_q \frac{e_q^2}{2}\int_0^1\! dz\,{1\over z(1-z)}\,
\phi^{\pi}_q(z)
\end{equation}
to leading order in $\alpha_S$, where $\epsilon^{\mu\nu}_T$ is the
antisymmetric tensor in transverse space ($\epsilon^{12}_T = 1$) and
\begin{equation}
\phi^{\pi}_q(z) = i \int\frac{dx^-}{2\pi}e^{-iz(P^+ x^-)}\,
\langle\pi^0(P)|\,\bar{q}(x^-)\gamma^+ \gamma_5\, q(0)\, |0\rangle .
 \label{da-def}
\end{equation}
Notice the different Dirac structures in the matrix elements
(\ref{gda-def}) and (\ref{da-def}), due to the different parity
transformation properties of one- and two-pion states~\cite{DGPT}.

The theoretical analysis of this process has been highly developed
\cite{TFF}. Its generalization to the production of $\eta$ and $\eta'$
is also important, in particular with respect to the $SU(3)$ flavor
structure of the QCD evolution equations and the mixing of the quark
singlet and gluon channels~\cite{BG-rev}. In Sect.~\ref{one-pion} we
will
further compare the production of a single pion with that of a pion
pair.

\subsection{Relation with deep virtual Compton scattering and
parton distributions in the pion}
\label{Compton}

The process $\gamma^*\gamma\to\pi\pi$ at large $Q^2$ and $s\ll Q^2$ is
related by $s$--$t$ crossing to deep virtual Compton scattering on a
pion, i.e., to $\gamma^* \pi \to \gamma \pi$ at large $Q^2$ and $-t
\ll Q^2$. It turns out that factorization works in completely
analogous ways for both cases, as is shown in
Fig.~\ref{fig-diagrams}. The non-perturbative matrix elements
occurring in the Compton process are skewed parton distributions
\cite{DVCS}, defined in the pion case as~\cite{Poly-W}
\begin{equation}
H_q(x,\xi,t) = \frac{1}{2} \int \frac{d z^-}{2\pi}\, e^{ix (P^+z^-)}\,
  \langle \pi(p') | \bar{q}(-z^- /2) \gamma^+ q(z^- /2) | \pi(p)
  \rangle
\end{equation}
with $P = (p+p')/2$. They have been recognized as objects of
considerable interest and have triggered intensive theoretical and
experimental work. The processes $\gamma^* \gamma \to \pi \pi$ and
$\gamma^* \pi \to \gamma \pi$ share many common features, from their
scaling behavior and the details of their helicity selection rules to
the possibilities of phenomenological analysis, which we will develop
in Sect.~\ref{phenomenology}.

The imaginary part of the forward virtual Compton amplitude, $\gamma^*
\pi \to \gamma^* \pi$, obtained from Fig.~\ref{fig-diagrams} (b) by
replacing the $\gamma$ with a second $\gamma^*$, gives the cross
section for inclusive deep inelastic scattering, $\gamma^*\pi \to X$,
where the ordinary parton distributions in a pion occur.

As observed in~\cite{Poly-W} it is useful to implement crossing at the
level of moments in momentum fractions ($z$ and $\zeta$ for GDAs, $x$
and $\xi$ for SPDs), which depend only on a factorization scale and a
Lorentz invariant ($s$ for GDAs, $t$ for SPDs). The moments of GDAs
and of SPDs are connected by analytic continuation in that
invariant. In particular, analytic continuation to the point $t=0$
leads to moments of the ordinary parton distributions in the pion,
which we will use as an input for our model of GDAs in
Sect.~\ref{simple-model}.

\section{General properties of GDAs}
\label{general}

\subsection{Charge conjugation and isospin properties}

Let us start by compiling some symmetry properties which will be
useful in the following. For the quark GDAs (\ref{gda-def}) the
invariance of strong interactions under charge conjugation $C$ implies
\begin{equation}
 \label{C_invariance}
\Phi_q^{\pi\pi}(z,\zeta,W^2)=-\Phi_q^{\pi\pi}(1-z,1-\zeta,W^2).
\end{equation}
It is useful to project GDAs for charged pions on eigenstates of $C$
parity,
\begin{equation}
\Phi^{\pm}_q(z,\zeta,W^2)={1\over 2}\left(
\Phi^{\pi^+\pi^-}_{q}(z,\zeta,W^2)\pm
\Phi^{\pi^+\pi^-}_{q}(z,1-\zeta,W^2)\right) ,
\end{equation}
so that
\begin{equation}
\Phi^{\pi^+\pi^-}_q=\Phi^{+}_q(z,\zeta,W^2)+
\Phi^{-}_q(z,\zeta,W^2) .
\end{equation}
In the $C$ even sector Eq.~(\ref{C_invariance}) reduces to
\begin{equation}
 \label{C-parity}
\Phi^{+}_q(z,\zeta,W^2)=-\Phi^{+}_q(1-z,\zeta,W^2) .
\end{equation}
Our process is only sensitive to the $C$ even part of
$\Phi^{\pi^+\pi^-}_q$ since the initial state two-photon state has
positive $C$ parity. Of course a $\pi^0\pi^0$ pair has positive $C$
parity as well, so that $\Phi^{\pi^0\pi^0}_q$ has no $C$-odd part at
all.

Let us now turn to isospin symmetry. The $C$ odd component of a
two-pion state has total isospin $I=1$, whereas its $C$ even component
contains both $I=0$ and $I=2$ pieces. The quark operator in
$\Phi_q^{\pi\pi}$ has only components with isospin $I=0$ or
$I=1$. Hence it is a consequence of the leading twist production
mechanism and of isospin invariance that in our process the pion pair
is in a state of zero isospin, i.e., that no component with $I=2$ is
produced. Another consequence of isospin invariance is that
\begin{equation}
\Phi^{\pi^0\pi^0}_q(z,\zeta,W^2) = \Phi^{+}_q(z,\zeta,W^2) ,
 \label{isospin}
\end{equation}
so that the production amplitudes for neutral and charged pion pairs
are equal. Deviations from isospin symmetry in the present reaction
would be interesting, but since one can expect them to be small we
will assume isospin invariance to hold throughout the rest of our
study. Isospin invariance also implies that
\begin{equation}
\Phi^{+}_u = \Phi^{+}_d , \hspace{4em}
\Phi^{-}_u = - \Phi^{-}_d ,
 \label{more-isospin}
\end{equation}
so that in the $C$ even sector we only need to know the $SU(2)$ flavor
singlet combination $\Phi^{+}_u + \Phi^{+}_d$.

The connection between the notation $\Phi_{||}^{I=0,1}$ of
Polyakov~\cite{Pol} and ours is
\begin{equation}
 \label{compare} \Phi_{||}^{I=0} = \Phi^{+}_u , \hspace{4em}
\Phi_{||}^{I=1} = \Phi^{-}_u .
\end{equation}
We remark that the second term in Eq.~(2.6) of Ref.~\cite{Pol} should
come with a minus sign \cite{private}. Our relation
$\smash{\Phi_{||}^{I=1} = \Phi^{-}_u}$ takes this correction into
account.

Notice that the signs in Eqs.~(\ref{isospin}) and (\ref{compare})
depend on the choice of relative phases in the definition of charged
pion states. We specify our convention in Appendix~\ref{pion-states}.

\subsection{Evolution equation}
\label{evolution}

In the process of factorization generalized distribution amplitudes
acquire a scale dependence in the same way as usual distributions
do. This scale dependence can be computed within perturbative QCD, and
there is nothing special with multiparticle states since the scale
dependence is a property of the nonlocal product of fields under
consideration, rather than one of a particular hadronic matrix element
(see~\cite{BB} for an approach exploiting this feature). The scale
dependence of GDAs can be cast in the form of an ERBL evolution
equation~\cite{ERBL}, and the only complication in the
channel we are concerned with here is the mixing of quark and gluon
distribution amplitudes. The leading-logarithmic form of the evolution
equations has been studied in detail for the parity-odd
sector~\cite{BG-rev}, where the relevant quark operator is $\bar{q}
\gamma^+\gamma_5\, q$. Our application to pion pairs leads us to
consider the parity-even sector, where the quark operator is $\bar{q}
\gamma^+ q$ instead, see our remark after Eq.~(\ref{da-def}). For
completeness we give here the basic steps for deriving and solving the
evolution equation in this channel, following the procedure outlined
in~\cite{BG-odd}. Taking into account the different normalization
conventions we find agreement with the results of Baier and
Grozin~\cite{BG-even}, who reported a sign discrepancy with
Chase~\cite{Chase} for the gluon evolution kernel.

We are then concerned with the generalized quark and gluon
distribution amplitudes in $A^+=0$ gauge:
\begin{eqnarray}
  \label{quark-gluon}
\Phi_q(z,\zeta,W^2)&=&\int\!\frac{dx^-}{2\pi}\,e^{-iz (P^+ x^-)}
\langle\pi(p)\pi(p')|\bar q(x^-)\gamma^+ q(0)|0\rangle, \nonumber \\ 
\Phi_g(z,\zeta,W^2)&=&\frac{1}{P^+}\int\!\frac{dx^-}{2\pi}\,
e^{-iz (P^+ x^-)}
\langle\pi(p)\pi(p')|F^{+\mu}(x^-){F_\mu}^+(0)|0\rangle, \nonumber \\
&=&z(1-z) P^+\int\!\frac{dx^-}{2\pi}\,e^{-iz (P^+ x^-)}
\langle\pi(p)\pi(p')|A^{\mu}(x^-)A_{\mu}(0)|0\rangle.
\end{eqnarray}
Our gluon distribution amplitude $\Phi_g(z,\zeta,W^2)$ coincides with
$\Phi^G(z,\zeta,W^2)$ introduced in \cite{Kivel}. From the
definition~(\ref{quark-gluon}) one readily obtains
\begin{equation}
\Phi_g(z,\zeta,W^2)=\Phi_g(1-z,\zeta,W^2) ,
  \label{gluon-symmetry}
\end{equation}
and from $C$ invariance one has
\begin{equation}
\Phi_g(z,\zeta,W^2)=\Phi_g(1-z,1-\zeta,W^2) .
  \label{C_invariance_gluon}
\end{equation}
Here we have given definitions for a two-pion state, but as stated
above the evolution equation for DAs and GDAs is not specific to the
details of the hadronic system. The considerations of this and the
following subsection thus apply to any state in the parity even sector
which has four-momentum $P$ and total angular momentum $J_z=0$ along
the axis defining the light cone variables.

We now study the evolution of the distributions for gluons and of
quarks in the singlet combination of $n_f$ flavors. For convenience we
introduce
\begin{eqnarray}
z\bar z\,f_Q(z)&=&\sum_{q=1}^{n_f} \Phi_q(z),\\
z^2\bar z^2\,f_G(z)&=&\Phi_g(z),
\end{eqnarray}
where we use the notation $\bar{z} = 1-z$. In the end we will return
to the amplitudes $\Phi_q$ and $\Phi_g$.

The scale dependence is controlled by the parameter
\begin{equation}
\xi(\mu^2,\mu_0^2)=
\frac{2}{\beta_1}\ln
\left( \frac{\alpha_S(\mu_0^2)}{\alpha_S(\mu^2)} \right) ,
 \label{xi-def}
\end{equation}
where $\alpha_S$ is the one-loop running coupling and
$\beta_1=11-2n_f/3$. This parameter describes how the distribution
amplitude evolves when one changes the factorization point from
$\mu_0$ to $\mu$. The evolution equation takes the form
\begin{equation}
\label{evolution_eq}
\frac{\partial}{\partial\xi}f(z,\xi)=V*f= \int_0^1 du\,
V(z,u)\,f(u,\xi).
\end{equation}
where $f$ is a two-component vector
\begin{equation}
f=\left(
\begin{array}{c}
f_Q\\ f_G
\end{array}
\right),
\end{equation}
and $V$ is the $2\times 2$ matrix kernel
\begin{equation}
V=\left(
\begin{array}{cc}
V_{QQ}&V_{QG}\\ V_{GQ}&V_{GG}
\end{array}
\right).
\end{equation}

\begin{figure}
   \begin{center}
        \leavevmode
        \epsfxsize=.43\hsize
        \epsfbox{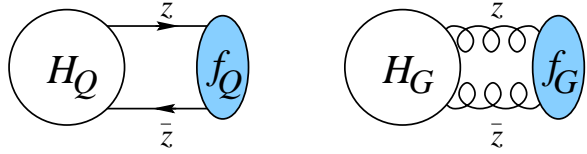}
   \end{center}
\caption{\label{amplitude_fig} The scattering amplitude $H*f$ with $f$
denoting the soft matrix elements and $H$ the hard scattering
kernels.}
\end{figure}

\begin{figure}
   \begin{center}
        \leavevmode
        \epsfxsize=.68\hsize
        \epsfbox{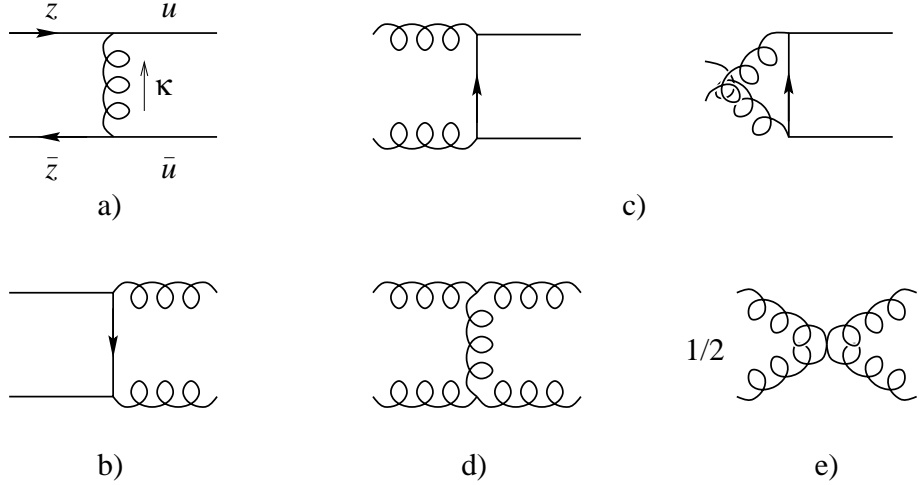}
   \end{center}
\caption{\label{evolution_fig} One-loop insertions, to be supplemented
by self-energy insertions on every line appearing in
Fig.~\protect\ref{amplitude_fig}. The sum of all insertions gives the
evolution kernel $\xi V$. We remark that the one-loop graph (e) must
be multiplied by $1/2$ to avoid double counting. $u$ and $z$ denote
light cone plus momentum fractions, and $\kappa$ the loop
four-momentum.}
\end{figure}

To obtain the leading logarithmic evolution equation it is sufficient
to consider one-loop corrections to the scattering amplitude. The
latter is depicted in Fig.~\ref{amplitude_fig} and has the form $H*f$,
where $H=(H_Q,H_G)$ denotes the hard-scattering kernels. It turns out
that in light cone gauge $A^+=0$ the relevant one-loop diagrams
consist of an insertion between $H$ and $f$ of the graphs shown in
Fig.~\ref{evolution_fig} (a) to (e), supplemented by (renormalized)
self-energy insertions on each line connecting $H$ to $f$ in
Fig.~\ref{amplitude_fig}. Calling the sum of these insertions $\xi V$
the one-loop diagrams have the structure $H*\xi V*f$.

The evolution from zeroth to first order of the generalized
distribution amplitude may thus be written as
\begin{equation}
\label{one-loop}
f^{(1)}(z)=f^{(0)}(z)+\xi\int_0^1 du\,V(z,u)\,f^{(0)}(u).
\end{equation}
In the computation of the diagrams, the $\kappa^-$ integral is
performed by the Cauchy method of contour integration in the complex
plane, and $\xi$ is the result of the integral over transverse
momentum from $\kappa_T=\mu_0$ to $\kappa_T=\mu$:
\begin{equation}
\xi(\mu^2,\mu_0^2)=\int_{\mu_0^2}^{\mu^2}\frac{d\kappa_T^2}{\kappa_T^2}
\frac{\alpha_S(\kappa_T^2)}{2\pi}.
 \label{xi}
\end{equation}
Despite the presence of $\alpha_S$ in Eq.~(\ref{xi}), $\xi$ is not
small if $\mu^2\gg\mu_0^2$, and this signals the necessity of an
all-order analysis. This analysis leads to the evolution equation,
with the feature that $V$ is the same matrix in
Eqs.~(\ref{evolution_eq}) and~(\ref{one-loop}). We refer the reader to
the literature for a general discussion~\cite{CZ}.

The integration over $\kappa^+$ may be reexpressed as an integral over
the incoming light cone fraction $u$. The evolution kernels contain
the remaining part of the dynamics, in particular they describe the
change of light cone fractions from $u$ to $z$. We get
\begin{eqnarray}
V_{QQ}(z,u)&=&C_F\left[
\theta(z-u)\frac{u}{z}\left(1+\frac{1}{z-u}\right)+
\{u\leftrightarrow\bar u,\,z\leftrightarrow\bar z\} \right]_+ ,
\nonumber \\
V_{QG}(z,u)&=&2 n_f T_F\left[\theta(z-u)\frac{u}{z}(2z-u)
-\{u\leftrightarrow\bar u,\,z\leftrightarrow\bar z\}\right] ,
\nonumber \\
V_{GQ}(z,u)&=&\frac{C_F}{z\bar z}\left[\theta(z-u)\frac{u}{z} (\bar
z-2\bar u) -\{u\leftrightarrow\bar u,\,z\leftrightarrow\bar
z\}\right] , \nonumber \\ 
V_{GG}(z,u)&=&\frac{C_A}{z\bar z}\left[\theta(z-u)
\left(\frac{u\bar u}{z-u}-u\bar u
-\frac{u}{2z}[(2z-1)^2+(2u-1)^2]\right) 
+ \{u\leftrightarrow\bar u,\,z\leftrightarrow\bar z\}\right]_+
-\frac{2}{3}\,n_f T_F\,\delta(u-z) ,
  \label{kernels}
\end{eqnarray}
where the color factors are $C_F=4/3$, $T_F=1/2$ and $C_A=3$. The
subscript $+$ stands for the $+$ distributions, whose action on a
function $f$ may be expressed symbolically as
\begin{equation}
[\cdots]_+\,f(u)=[\cdots]\,(f(u)-f(z)).
\end{equation}
The kernels (\ref{kernels}) give the finite parts that remain after
the cancellation of infrared divergences between graph (a), resp.\
(d), and quark self-energy, resp.\ gluon self-energy insertions. A
simple way to obtain self-energy corrections is to notice their
relation to parton splitting~\cite{BG-odd}, that is
\begin{eqnarray}
f_Q^{(1)}(z)\Big|_{SE}&=&\left[1-\xi\int dx\,P_{QQ}(x)\right]
f_Q^{(0)}(z)=\left[1-\xi\int dx\,P_{GQ}(x)\right] f_Q^{(0)}(z)
\nonumber \\ 
f_G^{(1)}(z)\Big|_{SE}&=&\left[1-\xi\int dx \left(
\frac{1}{2}P_{GG}(x)+n_f P_{QG}(x)\right)\right]f_G^{(0)}(z),
  \label{self-energies}
\end{eqnarray}
with the unregularized DGLAP splitting functions
\begin{eqnarray}
P_{QQ}(x)&=&C_F\frac{1+x^2}{1-x},\nonumber \\
P_{QG}(x)&=&T_F\left[x^2+(1-x)^2\right], \phantom{\frac{1}{2}}
\nonumber \\
P_{GQ}(x)&=&C_F\frac{1+(1-x)^2}{x},\nonumber \\
P_{GG}(x)&=&2\,C_A\left[\frac{x}{1-x}+\frac{1-x}{x}+x(1-x)\right].
\end{eqnarray}
The integrals (\ref{self-energies}) are not defined in the limit $x\to
0,1$, which is a manifestation of the infrared divergence of
self-energy graphs.

\subsection{Solution}

We will now solve the evolution equation (\ref{evolution_eq}). Given
our application we restrict ourselves to the $C$ even parts $\Phi^+_q$
of the quark distributions, the gluon distribution being of course
even under $C$ from the start.

We look for solutions of the form
\begin{equation}
f(z,\xi)=f(z)\, e^{-\gamma\xi} .
\end{equation}
To this end it is convenient to change variables, introducing $y =
2u-1$ and $x = 2z-1$, and to study the convolution of the matrix
kernel $V$ with
\begin{equation}
\left(\begin{array}{c} x^n\\ 0\end{array}\right),\quad\quad
\left(\begin{array}{c}0\\ x^{n-1}\end{array}\right),
\end{equation}
where $n$ is an odd integer to accommodate the symmetry properties
(\ref{C-parity}) and (\ref{gluon-symmetry}). One finds
\begin{eqnarray}
&&V_{QQ}*y^n=-\gamma_{QQ}(n)\,x^n+O(x^{n-2}),\phantom{^{n-1}}\quad
V_{QG}*y^{n-1}=-\gamma_{QG}(n)\,x^n+O(x^{n-2}),
\nonumber \\
&&V_{GQ}*y^n=-\gamma_{GQ}(n)\,x^{n-1}+O(x^{n-3}),\phantom{^n}\quad
V_{GG}*y^{n-1}=-\gamma_{GG}(n)\,x^{n-1}+O(x^{n-3}),
\end{eqnarray}
with anomalous dimensions
\begin{eqnarray}
 \label{anomalous}
\gamma_{QQ}(n)&=&C_F\left(\frac{1}{2}-\frac{1}{(n+1)(n+2)}
+2\sum_{k=2}^{n+1}\frac{1}{k}\right), \nonumber \\ 
\gamma_{QG}(n)&=&-n_f T_F\frac{n^2+3n+4}{n(n+1)(n+2)}, \nonumber \\
\gamma_{GQ}(n)&=&-2C_F\frac{n^2+3n+4}{(n+1)(n+2)(n+3)}, \nonumber \\
\gamma_{GG}(n)&=&C_A\left(\frac{1}{6}
-\frac{2}{n(n+1)}-\frac{2}{(n+2)(n+3)}
+2\sum_{k=2}^{n+1}\frac{1}{k}\right)+\frac{2}{3} n_f T_F.
\end{eqnarray}
Since for a given $n_0$ the space of solutions with $n\le n_0$ is
stable under the application of the kernel one can find polynomials
$p_n(x)$ and $q_{n-1}(x)$ satisfying
\begin{eqnarray}
&&V_{QQ}*p_n=-\gamma_{QQ}(n)\,p_n,\phantom{_{n-1}}\quad
V_{QG}*q_{n-1}=-\gamma_{QG}(n)\,p_n,\nonumber\\
\label{polynomials}
&&V_{GQ}*p_n=-\gamma_{GQ}(n)\,q_{n-1},\phantom{_n}\quad
V_{GG}*q_{n-1}=-\gamma_{GG}(n)\,q_{n-1}.
\end{eqnarray}
The symmetry properties of the kernels
\begin{eqnarray}
(1-x^2)\, V_{QQ}(x,y)&=&(1-y^2)\, V_{QQ}(y,x), \nonumber \\
2 C_F(1-x^2)\, V_{QG}(x,y)&=& n_f T_F(1-y^2)^2\,
V_{GQ}(y,x), \nonumber \\ 
(1-x^2)^2\, V_{GG}(x,y)&=&(1-y^2)^2\,
V_{GG}(y,x),
\end{eqnarray}
then imply that the $(p_n)$ are orthogonal polynomials on the interval
$[-1,\,1]$ with weight $1-x^2$, i.e., they are proportional to the
Gegenbauer polynomials $C^{(3/2)}_n(x)$, whereas the $(q_{n-1})$ are
orthogonal on $[-1,\,1]$ with weight $(1-x^2)^2$, that is,
proportional to the Gegenbauer polynomials $C^{(5/2)}_{n-1}(x)$. To
complete the identification it is necessary to take into account the
standard normalization of Gegenbauer polynomials. One finds that
$p_n=C^{(3/2)}_n$ and $q_{n-1}=C^{(5/2)}_{n-1}$ fulfill
Eq.~(\ref{polynomials}), provided one makes the replacements
\begin{equation}
\gamma^{\phantom{'}}_{QG}(n) \to 
\gamma'_{QG}(n) = \frac{n}{3}\, \gamma_{QG}^{\phantom{'}}(n) , 
\hspace{3em}
\gamma^{\phantom{'}}_{GQ}(n) \to
\gamma'_{GQ}(n) = \frac{3}{n}\, \gamma_{GQ}^{\phantom{'}}(n) .
\end{equation}

The final step is to diagonalize the $2\times 2$ anomalous dimension
matrices for each value of $n$. The eigenvalues are
\begin{equation}
\Gamma^{(\pm)}_n=\frac{1}{2}\left[ \, 
          \gamma^{\phantom{'}}_{QQ}(n)+\gamma^{\phantom{'}}_{GG}(n)
\pm\sqrt{[\gamma^{\phantom{'}}_{QQ}(n)-\gamma^{\phantom{'}}_{GG}(n)]^2
+4\gamma'_{QG}(n)\gamma'_{GQ}(n)} \; \right],
\end{equation}
and the eigenvectors of the kernel matrix read
\begin{equation}
v^{(\pm)}_n(x)=\left(
\begin{array}{c}
C^{(3/2)}_n(x)\\
g_n^{(\pm)}\, C^{(5/2)}_{n-1}(x)
\end{array}
\right) ,
\end{equation}
where
\begin{equation}
g_n^{(\pm)} = \frac{\Gamma^{(\pm)}_n-\gamma_{QQ}(n)}{\gamma'_{QG}(n)} .
\end{equation}
The general $C$ even solution of Eq.~(\ref{evolution_eq}) may then be
written as
\begin{equation}
f(x,\xi)=\sum_{{\mathrm{odd}}\ n}\left\{A^{(+)}_n
v^{(+)}_n(x)\,e^{-\Gamma^{(+)}_n\xi}+A^{(-)}_n v^{(-)}_n(x)
\,e^{-\Gamma^{(-)}_n\xi}\right\}
\end{equation}
with integration constants $A^{(\pm)}_n$.

We now return to the amplitudes $\Phi_q$, $\Phi_g$ and explicitly
express $\xi$ in terms of $\mu$ and $\mu_0$. The key result of this
section then reads
\begin{eqnarray}
\sum_{q=1}^{n_f} \Phi_q^+(z,\mu^2)&=&z(1-z)\sum_{{\mathrm{odd}}\ n}
A_n(\mu^2) \, C^{(3/2)}_n(2z-1), \nonumber \\
\Phi_g(z,\mu^2)&=&z^2(1-z)^2\sum_{{\mathrm{odd}}\ n}
A'_n(\mu^2) \, C^{(5/2)}_{n-1}(2z-1) ,
\label{expansion-poly}
\end{eqnarray}
with
\begin{eqnarray}
A_n(\mu^2) &=& A^{(+)}_n 
\left(\frac{\alpha_S(\mu^2)}{\alpha_S(\mu_0^2)}\right)^{K^{(+)}_n}
+ A^{(-)}_n
\left(\frac{\alpha_S(\mu^2)}{\alpha_S(\mu_0^2)}\right)^{K^{(-)}_n} ,
\nonumber \\
A'_n(\mu^2) &=& g_n^{(+)} A^{(+)}_n 
\left(\frac{\alpha_S(\mu^2)}{\alpha_S(\mu_0^2)}\right)^{K^{(+)}_n}
+ g_n^{(-)} A^{(-)}_n
\left(\frac{\alpha_S(\mu^2)}{\alpha_S(\mu_0^2)}\right)^{K^{(-)}_n} ,
 \label{z-coefficients}
\end{eqnarray}
and exponents $K^{(\pm)}_n = 2 \Gamma^{(\pm)}_n/\beta_1$, which are
positive except for $K^{(-)}_1 = 0$. For $n_f=2, 3, 4$, one explicitly
finds
\begin{equation}
K^{(+)}_1 = \frac{32+6n_f}{99-6n_f}=0.51,\ 0.62,\ 0.75 , \hspace{3em}
K^{(-)}_3 = 0.71,\ 0.76,\ 0.82 , \hspace{3em}
K^{(+)}_3 = 1.45,\ 1.64,\ 1.85 .
\end{equation}
>From Eq.~(\ref{z-coefficients}) we easily see that the integration
constants $A^{(\pm)}_n$ depend on the starting scale $\mu_0$ of the
evolution through a factor $\alpha_S(\mu_0^2)^{K^{(\pm)}_n}$.

\subsection{Expansion in $\zeta$}
\label{zeta-expansion}

For a two-meson state, the coefficients $A^{\phantom{'}}_n$ and $A'_n$
are functions of the factorization scale $\mu^2$ and of the remaining
kinematical variables $\zeta$ and $W^2$. From the definition of GDAs
in term of fields given in Eq.~(\ref{quark-gluon}) one obtains moments
\begin{eqnarray}
\int_0^1 dz\,z^n\,\Phi_q(z)&=&\frac{1}{(P^+)^{n+1}}\Bigl[
(-i \partial^+)^n
\langle\pi(p)\pi(p')|\, 
       \bar q(x)\gamma^+ q(0) \,|0\rangle\Bigr]_{x=0} ,
\nonumber \\
\int_0^1 dz\,z^{n-1}\,\Phi_g(z)&=&\frac{1}{(P^+)^{n+1}}\Bigl[
(-i\partial^+)^{n-1}
\langle\pi(p)\pi(p')|\, 
       F^{+\mu}(x){F_\mu}^+(0) \,|0\rangle\Bigr]_{x=0} .
 \label{local-operators}
\end{eqnarray}
These local matrix elements are the plus-components of tensors that
can be decomposed on a basis built up with the metric $g^{\mu\nu}$ and
the vectors $(p+p')^\mu$ and $(p-p')^\mu$. Since $(p+p')^+=P^+$ and
$(p-p')^+=(2\zeta-1)P^+$ the moments (\ref{local-operators}) are then
polynomials in $2\zeta-1$ with degree at most $n+1$. The
$A^{\phantom{'}}_n$ and $A'_n$ are Gegenbauer moments of
$\sum_q\Phi_q$ and $\Phi_g$, respectively, and therefore have the same
polynomiality properties in $\zeta$. Following~\cite{Pol} we expand
them on the Legendre polynomials, writing
\begin{equation}
A_n(\zeta,W^2)= 6 n_f  \sum_{{\mathrm{even}}\ l}^{n+1}
B_{nl}(W^2)\, P_l(2\zeta-1) 
 \label{Legendre}
\end{equation}
and the analogous expression for $A'_n$ with coefficients $B'_{nl}$.
The $C$ invariance properties (\ref{C_invariance}) and
(\ref{C_invariance_gluon}) restrict $l$ to even integers in the $C$
even sector. The expansion coefficients $B_{nl}$ are linear
combinations of the local operator matrix elements in
Eq.~(\ref{local-operators}) and are therefore analytic functions in
$W^2$. As we mentioned in Sect.~\ref{Compton} their continuation to
zero or spacelike $W^2$ leads to the moments of parton distributions
in the pion.

{}From Eq.~(\ref{z-coefficients}) the factorization scale dependence
of the $B_{nl}$ may be written as
\begin{equation}
B_{nl}(W^2, \mu^2) = 
B^{(+)}_{nl}(W^2) 
\left(\frac{\alpha_S(\mu^2)}{\alpha_S(\mu_0^2)}\right)^{K^{(+)}_n}
+ B^{(-)}_{nl}(W^2)
\left(\frac{\alpha_S(\mu^2)}{\alpha_S(\mu_0^2)}\right)^{K^{(-)}_n} ,
  \label{zeta-coefficients}
\end{equation}
with an analogous equation for $B'_{nl}$ involving the factors
$g_n^{(\pm)}$.

In the limit $\mu\to\infty$ only the terms with the smallest exponent
$K^{(-)}_1=0$ in the coefficients (\ref{z-coefficients}) survive. The
asymptotic form of the distribution amplitudes thus has only $n=1$ in
the Gegenbauer expansion (\ref{expansion-poly}) and reads
\begin{eqnarray}
\sum_{q=1}^{n_f} \Phi_q^+(z,\zeta,W^2) &=& 18 n_f z(1-z) (2z-1) \,
\left[ B_{10}^{(-)}(W^2)+B_{12}^{(-)}(W^2)\, P_2(2\zeta-1) \right] , 
\nonumber \\
\Phi_g(z,\zeta,W)&=& 48 z^2(1-z)^2 \,
\left[ B_{10}^{(-)}(W^2)+B_{12}^{(-)}(W^2)\, P_2(2\zeta-1) \right] ,
 \label{asy-solution}
\end{eqnarray}
where $P_2(2\zeta-1)=1-6\zeta(1-\zeta)$. Note that $B_{10}^{(-)}$ and
$B_{12}^{(-)}$ do not depend on a starting scale $\mu_0$ because
$K^{(-)}_1=0$. For reasons that will become clear we will also keep
the terms with the first nonzero exponent $K^{(+)}_1$ in our model for
the GDAs to be developed in Sect.~\ref{simple-model}. For the quark
distribution amplitudes this simply amounts to replacing
$B_{10}^{(-)}$ and $B_{12}^{(-)}$ in the first line of
Eq.~(\ref{asy-solution}) with the $\mu$-dependent coefficients
$B_{10}$ and $B_{12}$.

Let us finally remark that, as discussed in~\cite{Kivel}, there is
another generalized gluon distribution amplitude, with an operator
different from the one in Eq.~(\ref{quark-gluon}). It corresponds to
pion pairs with angular momentum $J_z= \pm 2$ and gives the
leading-twist part of the amplitudes $\gamma^*\gamma \to \pi\pi$ where
the photon helicities are opposite. The evolution of this helicity-two
distribution amplitude does not mix with any quark distribution. Its
smallest anomalous dimension is positive, so that this distribution
amplitude tends logarithmically to zero as $\mu\to\infty$. The study
of this distribution would be very interesting. Nothing is, however,
known about its size at present, and in our phenomenological analysis
we will neglect its contribution.

\subsection{Partial wave expansion}
\label{partial-waves}

The decomposition of generalized distribution amplitudes on Legendre
polynomials performed in the previous section translates into a
partial waves decomposition~\cite{Pol} if one transforms from
polynomials $P_l(2\zeta-1)$ to $P_l(\cos\theta)$ using that $2\zeta-1
= \beta \cos\theta$. The rearranged series reads
\begin{equation}
\sum_{q=1}^{n_f} \Phi_q^+ = 6 n_f\, z(1-z)
\sum_{\scriptstyle n=1 \atop \scriptstyle {\rm odd}}^{\infty}
\sum_{\scriptstyle l=0 \atop \scriptstyle {\rm even}}^{n+1}
\tilde{B}_{nl}(W^2)\, C_n^{(3/2)}(2 z-1)\, P_l(\cos\theta)
\end{equation}
for quarks, where the coefficients $\tilde{B}_{nl}(W^2)$ are linear
combinations of the form
\begin{equation}
\tilde{B}_{nl} = \beta^l \, \left[ B_{nl} + c_{l,\,l+2}\, B_{n,\,l+2} 
    + \ldots + c_{l,\,n+1}\, B_{n,\,n+1} \right]
\end{equation}
with polynomials $c_{l,\,l'}$ in $\beta^2$. Keeping only $n=1$ in the
Gegenbauer expansion one is restricted to an $S$- and a $D$-wave:
\begin{eqnarray}
\sum_{q=1}^{n_f} \Phi_q^+ 
&=& 18 n_f\, z(1-z) (2z-1) \left[ B_{10}(W^2) + B_{12}(W^2)\,
P_2(2\zeta-1) \right] \nonumber \\
&=& 18 n_f\, z(1-z) (2z-1) \left[
\tilde{B}_{10}(W^2) + \tilde{B}_{12}(W^2)\, P_2(\cos\theta) \right]
  \label{expansion-asy}
\end{eqnarray}
with
\begin{eqnarray}
\tilde{B}_{10}(W^2) &=& B_{10}(W^2) - \frac{1-\beta^2}{2} B_{12}(W^2) ,
        \nonumber \\
\tilde{B}_{12}(W^2) &=& \beta^2 B_{12}(W^2) .
  \label{expansion-transform}
\end{eqnarray}

It is a remarkable consequence of the condition $l \le n+1$ that the
presence of high partial waves implies a departure of the two-pion
distribution amplitude from its asymptotic form. The
$\theta$-distribution of the produced pion pair thus contains
information about the dependence of the GDAs on $z$, which as a loop
variable is integrated over in the amplitude of the process, cf.\
Eq.~(\ref{tensor-two-pi}).

One-meson distribution amplitudes are real valued functions due to
time reversal invariance. This is not true for generalized
distribution amplitudes: the two-pion ``out'' state in the definition
(\ref{gda-def}) of $\Phi^{\pi\pi}$ is transformed into an ``in'' state
under time reversal, and these states are different because hadrons
interact with each other. Below the inelastic threshold, however,
two-pion ``in'' and ``out'' states with definite angular momentum are
related in a simple way via the phase shifts of elastic
$\pi\pi$-scattering. With the aid of Watson's theorem one then obtains
the relation $\tilde{B}_{nl}^* = \tilde{B}_{nl}^{\phantom{*}}
\exp(-2i\delta_l)$~\cite{Pol}. This fixes the phase of the expansion
coefficient $\tilde{B}_{nl}$ up to its overall sign:
\begin{equation}
\tilde{B}_{nl} = \eta_{nl} |\tilde{B}_{nl}|\, \exp(i\delta_l) ,
\hspace{3em} \eta_{nl} = \pm 1 ,
  \label{Watson}
\end{equation}
where $\delta_l$ is the $\pi\pi$ phase shift for the $l$-th partial
wave in the $I=0$ channel.

\subsection{Momentum sum rule}

Of particular interest are the moments~\cite{DGPT,Poly-W,Pol}
\begin{eqnarray}
  \label{quark-sum-rule}
\int_0^1 dz\, (2z-1) \Phi^{+}_q(z,\zeta,W^2) &=&
        \frac{2}{(P^+)^2}\,
        \langle \pi^+(p)\pi^-(p')|\, 
        T_q^{++}(0) \,| 0\rangle , \\
  \label{gluon-sum-rule}
\int_0^1 dz\, \Phi_g(z,\zeta,W^2) &=&
        \frac{1}{(P^+)^2}\,
        \langle \pi^+(p)\pi^-(p')|\, 
        T_g^{++}(0) \,| 0\rangle ,
\end{eqnarray}
where $T^{\mu\nu}_q(x)$ and $T^{\mu\nu}_g(x)$ respectively denote the
Belinfante improved energy-momentum tensors for quarks of flavor $q$
and for gluons. After summing (\ref{quark-sum-rule}) over all flavors
these moments project out the coefficients $B_{10}(W^2)$,
$B_{12}(W^2)$ and $B'_{10}(W^2)$, $B'_{12}(W^2)$.

To proceed one decomposes $\langle \pi^+(p)\pi^-(p')|\,
T_q^{\mu\nu}(0) \,| 0\rangle$ on form factors. Their analytical
continuation to zero or negative $W^2$ leads to the form factors of
the matrix elements $\langle\pi^+(p) |\, T_q^{\mu\nu}(0) \,| \pi^+(p')
\rangle$ between one-pion states, with $W^2=0$ corresponding to
$p=p'$. At that point we get from Eq.~(\ref{quark-sum-rule})
\begin{equation}
B_{12}(0) = \frac{10}{9n_f}\, R_{\pi} ,
  \label{constraint}
\end{equation}
where $R_\pi$ is the fraction of light-cone momentum carried by quarks
and antiquarks in the pion. No constraint on $B_{10}(0)$ is obtained
this way, since the corresponding form factor in the decomposition of
$\langle\pi^+(p) |\, T_q^{\mu\nu}(0) \,| \pi^+(p') \rangle$ is
multiplied by a tensor that vanishes for $p=p'$. In an analogous
fashion one obtains an expression for $B'_{12}(0)$ from the sum rule
(\ref{gluon-sum-rule}).

We emphasize that both sides of Eq.~(\ref{constraint}) depend on the
renormalization scale $\mu$. Only the total energy-momentum tensor,
i.e., the sum $T^{\mu\nu} = \sum_q T_q^{\mu\nu} + T_g^{\mu\nu}$ over
quarks and gluons is conserved, so that its matrix elements are
renormalization scale independent. The appropriate sum of the moments
(\ref{quark-sum-rule}) and (\ref{gluon-sum-rule}) leads to a linear
combination of $B_{12}$ and $B'_{12}$ where the scale dependent term
with $B^{(+)}_{12}$ indeed drops out and only $B^{(-)}_{12}$ is
left. The normalization of $\langle\pi^+(p) |\, T^{\mu\nu}(0) \,|
\pi^+(p) \rangle$ thus fixes the expansion coefficient
\begin{equation}
B^{(-)}_{12}(0) = \frac{10}{9 n_f + 48} ,
\end{equation}
which through the relation (\ref{constraint}) gives the asymptotic
value
\begin{equation}
R_\pi \stackrel{\mu\to\infty}{\to} \frac{3 n_f}{3 n_f + 16} ,
 \label{asy-ratio}
\end{equation}
in agreement with the well-known result from the evolution of singlet
parton distributions~\cite{Alt}.

\section{A simple model of the GDA}
\label{simple-model}

So far no experimental information exists on the two-pion GDA. In the
numerical studies to follow we will therefore use a simple ansatz for
$\Phi^+_q(z,\zeta,W^2)$, which is based on the general properties we
have discussed in the previous section.

We only consider the contributions from $u$- and $d$-quarks, i.e., we
take $n_f=2$. As already mentioned we will use the isospin relations
(\ref{isospin}) and (\ref{more-isospin}), and take the asymptotic form
of the $z$ dependence given in Eq.~(\ref{expansion-asy}). It thus
remains to make an ansatz for the coefficients $B_{10}(W^2)$ and
$B_{12}(W^2)$, or equivalently for $\tilde{B}_{10}(W^2)$ and
$\tilde{B}_{12}(W^2)$ introduced in Eq.~(\ref{expansion-transform}).

\begin{figure}
   \begin{center}
      \setlength{\unitlength}{0.45\hsize}
      \begin{picture}(1,0.78)(0,0)
	\put(0.05,0.71){$\delta_0$~[deg]}
        \epsfxsize=0.45\hsize
        \put(0,0){\epsfbox{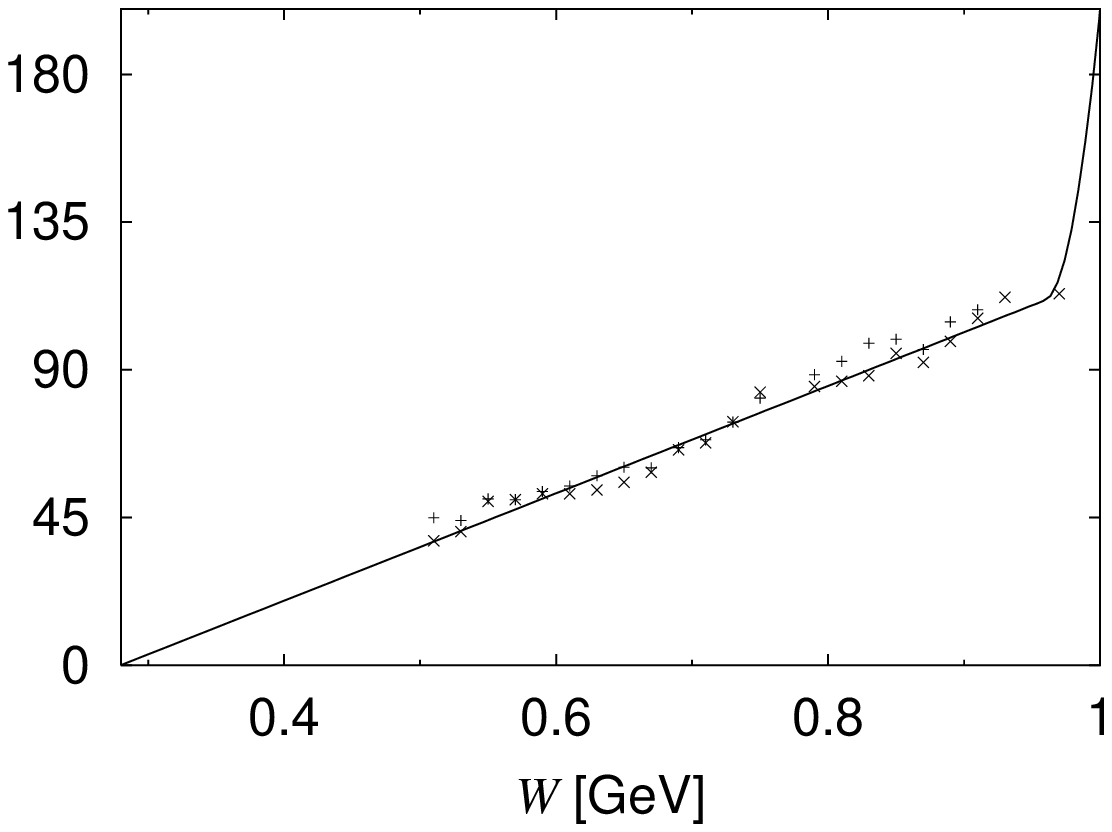}}
      \end{picture}
      \begin{picture}(1,0.78)(0,0)
	\put(0.05,0.71){$\delta_1$~[deg]}
        \epsfxsize=0.45\hsize  
	\put(0,0){\epsfbox{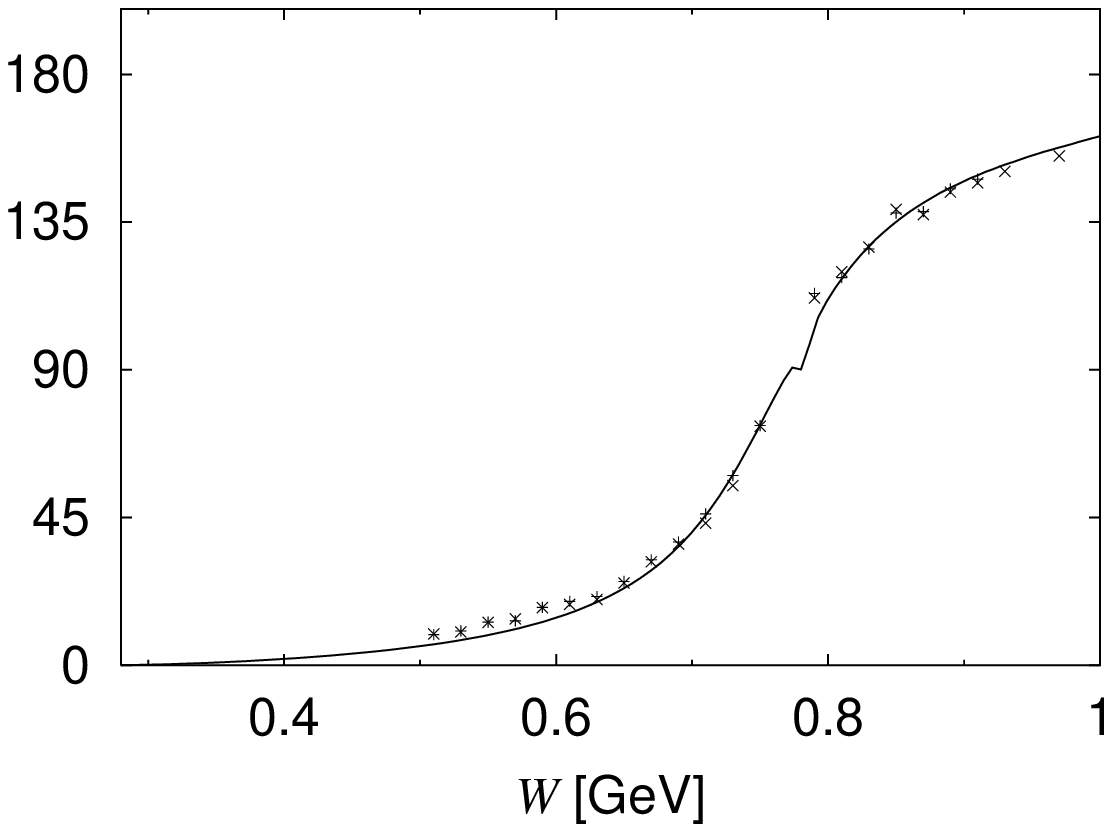}}
      \end{picture}
   \end{center}
   \begin{center}
      \setlength{\unitlength}{0.45\hsize}
      \begin{picture}(1,0.78)(0,0)
	\put(0.05,0.71){$\delta_2$~[deg]}
        \epsfxsize=0.45\hsize  
	\put(0,0){\epsfbox{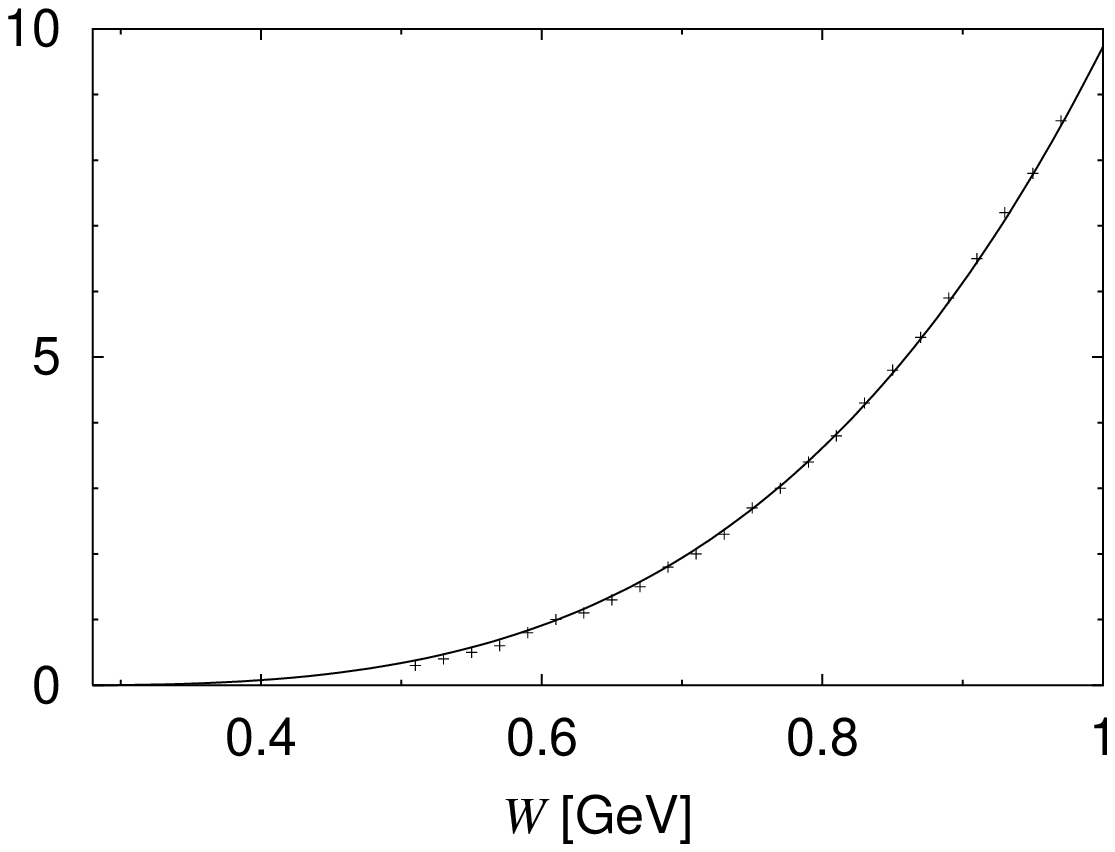}}
      \end{picture}
   \end{center}
\caption{\label{phase-shifts} The phase shifts $\delta_0$ for the
$S$-wave, $\delta_1$ for the $P$-wave and $\delta_2$ for the $D$-wave
of elastic $\pi\pi$ scattering. $\delta_0$ and $\delta_2$ refer to the
$I=0$ channel. The points are taken from \protect\cite{Estabrooks},
and the curves for $\delta_0$ and $\delta_2$ are simple
parameterizations. The curve for $\delta_1$ corresponds to the
parameterization $N=1$ of the pion form factor $F_\pi(W^2)$ in
\protect\cite{Kuehn}.}
\end{figure}

For their phases, given by Eq.~(\ref{Watson}), we use simple
parameterizations of the isosinglet $S$- and $D$-wave phase shifts
$\delta_0$ and $\delta_2$ obtained in \cite{Estabrooks}. They are
shown in Fig.~\ref{phase-shifts}, where for later use the phase shift
$\delta_1$ of the $P$-wave is also displayed. The result
(\ref{Watson}) only holds below the inelastic threshold in $\pi\pi$
scattering, therefore we restrict all our studies to invariant masses
$W$ below 1 GeV. Around that mass, corresponding to the $K\bar{K}$
threshold, the phase shift $\delta_0$ of the $S$-wave drastically
increases. While the analysis of \cite{Estabrooks} stops at
$W=0.97$~GeV and does not exhibit this abrupt change, the
investigations in Ref.~\cite{Hyams} find values of order 200$^{\circ}$
at $W=1$~GeV. Our parameterization of $\delta_0$ in that region is
meant to be indicative rather than a precise description of this
quantity. Through interference effects, the rapid variation of a phase
shift leads to a characteristic behavior in the $W$-spectrum of
appropriate observables in our process, as we shall see in
Sect.~\ref{phenomenology}.

The analyticity properties of the $\tilde{B}_{nl}$ and the phase
information from Watson's theorem (\ref{Watson}) may be used to obtain
the $W^2$-dependence of $\tilde{B}_{nl}$ via dispersion relations,
which has been exploited in~\cite{Kivel,Pol}. Note, however, that
while the complex phases are simple for the $\tilde{B}_{nl}$, it is
the $B_{nl}$ that have simple analytic properties in the $W^2$-plane,
given their definition through operator matrix elements. The
transformation from $B_{nl}$ to $\tilde{B}_{nl}$ introduces extra
poles at $W^2=0$, cf., e.g., the factors $\beta^2=(W^2-4 m_\pi^2)
/W^2$ in Eq.~(\ref{expansion-transform}). Furthermore, the evaluation
of the integrals that solve the dispersion relations requires
knowledge of the phases at energies above the value of $W$ where
$\tilde{B}_{nl}$ is evaluated. This further restricts the range of $W$
where $\tilde{B}_{nl}$ can be obtained using the $\pi\pi$ phase shifts
as input.

To keep our model simple we will make a less sophisticated ansatz. We
keep the energy dependent phases $\delta_0$ and $\delta_2$ from
Watson's theorem (\ref{Watson}). To determine $|\tilde{B}_{10}|$,
$|\tilde{B}_{12}|$, and the overall signs $\eta_{10}$, $\eta_{12}$ in
Eq.~(\ref{Watson}), we retain only the kinematical factors $\beta^2$
in the relation (\ref{expansion-transform}) and replace $B_{10}(W^2)$
and $B_{12}(W^2)$ with their values at $W=0$. Close to $W=1$~GeV one
will not expect this to be a good approximation for the $S$-wave,
given the presence of the $f_0(980)$. Below this there is however no
prominent $\pi\pi$ resonance in the $I=0$ channel, and the phase
shifts show a smooth behavior. It seems therefore reasonable to assume
that the isosinglet form factors $\tilde{B}_{10}$ and $\tilde{B}_{12}$
do not have a strong energy dependence in that region, certainly not
as strong as the electromagnetic pion form factor $F_\pi$ with its
large variations in modulus and phase due to the $\rho(770)$. We do
however not claim our simple model to be better than, say, a factor of
2.

For the input value of $B_{12}(0)$ we use the constraint
(\ref{constraint}) with $R_\pi$ evaluated from the parton
distributions in the pion. Taking the LO parameterization of GRS
\cite{GRS} we find $R_\pi$ ranging from 0.5 to 0.6 at a factorization
scale $\mu^2$ between 1~GeV$^2$ and 20~GeV$^2$. In our numerical
studies we use $R_\pi = 0.5$. Note that this is very far from the
asymptotic value (\ref{asy-ratio}), which for $n_f=2,3,4$ is
$R_\pi=0.27$, $0.36$ and $0.43$, respectively. While using the
asymptotic form of the $z$-dependence of the GDA for simplicity (and
lack of experimental information) we thus retain a clear
non-asymptotic effect in the coefficient $B_{12}(0)$. We also remark
that in the GRS LO parameterization the contribution of strange quarks
and antiquarks to $R_\pi$ is at the level of 5\% to 10\% in a wide
range of the factorization scale. This corroborates our restriction to
$u$- and $d$-quarks in the GDA, although with the caveat that the sea
quark distribution in the pion is not constrained from experimental
data \cite{GRS}.

For the coefficient $B_{10}(0)$ we make use of the relation
\begin{equation}
  B_{10}(0) = -B_{12}(0) ,
 \label{soft-pion}
\end{equation}
which has been obtained in~\cite{Pol} using chiral symmetry in the
form of a soft-pion theorem. Notice that our ansatz then has the
property that for $\beta\to 1$ the $S$- and $D$-wave components of the
GDA have equal size and opposite sign, as is easily seen from
Eq.~(\ref{expansion-transform}).

Putting everything together, we will take the following model GDAs in
our numerical studies:
\begin{equation}
\Phi_u^+ = \Phi_d^+ = 10 z(1-z) (2z-1)\, R_\pi\, 
 \left[ - \frac{3-\beta^2}{2}\, e^{i\delta_0(W^2)} 
        + \beta^2\, e^{i\delta_2(W^2)}\, P_2(\cos\theta) \right] 
  \label{model-gda}
\end{equation}
with $R_\pi = 0.5$.

With this we can easily calculate the scattering amplitude for
$\gamma^*\gamma\to \pi\pi$ to leading order in $\alpha_S$. We shall
neglect here the radiative corrections to the hard scattering, which
have been worked out to one loop in~\cite{Kivel}. Taking the
asymptotic form (\ref{asy-solution}) of the quark and gluon GDAs,
including the asymptotic value (\ref{asy-ratio}) of the ratio $R_\pi$,
they were found to reduce the leading-order amplitude for equal photon
helicities by 30\% if $\alpha_S=0.3$, with most of the correction
being due to the contribution from $\Phi_g$. Finally, we recall from
the end of Sect.~\ref{zeta-expansion} that we will neglect the
contribution of the helicity-two gluon GDA to the photon double
helicity-flip amplitude, which is also a one-loop effect.

\section{Comparison with $\gamma^*\gamma\to \pi^0$}
\label{one-pion}

Given the close analogy between the production of one and of two pions
it is natural to compare the production rates of these two processes.
Since our estimations for $\pi\pi$ production are at lowest order in
$\alpha_S$ we will compare with the corresponding expression for the
one-pion case for consistency, although experimental data and more
refined theory analyses are available there. From the leading-order
expression (\ref{tensor-one-pi}) we obtain the cross section for the
process $e\gamma\to e\pi^0$ as
\begin{equation}
\frac{d\sigma_{e\gamma\to e \pi^0}}{dQ^2} =
  \frac{\alpha^3}{s_{e\gamma}^2}\,
  \frac{1}{Q^2 (1-\epsilon)}\, 2\pi^2 f_\pi^2\,
 \label{sigma-one-pion}
\end{equation}
where we have used the asymptotic distribution amplitude $\phi^{\pi}_u
= - \phi^{\pi}_d = 3\sqrt{2} f_\pi\, z(1-z)$ with $f_\pi \approx
131$~MeV. For a lowest-order approximation, the cross section
(\ref{sigma-one-pion}) is in fair agreement with the data~\cite{CLEO}.

To compare with two-pion production, we integrate the cross section
for $e\gamma\to e\,\pi^0\pi^0$ from threshold up to
$W_{{\mathit{max}}}$. With our model GDA (\ref{model-gda}) we find
\begin{eqnarray}
\frac{d\sigma_{e\gamma\to e \pi^0\pi^0}}{dQ^2} &=&
  \frac{25\, \alpha^3}{72\, s_{e\gamma}^2}\,
  \frac{1}{Q^2 (1-\epsilon)}\,
   \int_{4m_\pi^2}^{W_{{\mathit{max}}}^2} dW^2\,
        \sqrt{1 - \frac{4m_\pi^2}{W^2}}
   \left( |\tilde{B}_{10}|^2 + \frac{1}{5} |\tilde{B}_{12}|^2 \right )
\nonumber \\
&=& \frac{125\, \alpha^3}{243\, s_{e\gamma}^2}\,
  \frac{1}{Q^2 (1-\epsilon)}\,
        R_\pi^2 m_\pi^2\,
        \sqrt{1 - \frac{4m_\pi^2}{W_{{\mathit{max}}}^2}}
        \left( \frac{W_{{\mathit{max}}}^2}{4m_\pi^2} - \frac{3}{4}
               - \frac{m_\pi^2}{W_{{\mathit{max}}}^2} \right) .
 \label{sigma-two-pions}
\end{eqnarray}
A consequence of the identical scaling behavior of the two processes
is that the ratio of the cross sections (\ref{sigma-two-pions}) and
(\ref{sigma-one-pion}) is independent of $Q^2$ in the Born
approximation.

\begin{figure}
   \begin{center}
      \setlength{\unitlength}{0.45\hsize}
      \begin{picture}(1,0.71)(0,0)
	\put(-0.22,0.615){$\displaystyle
                       \frac{d\sigma_{e\gamma\to e\pi^0\pi^0}}{
                             d\sigma_{e\gamma\to e\pi^0}}$}
        \epsfxsize=0.45\hsize  
	\put(0,0){\epsfbox{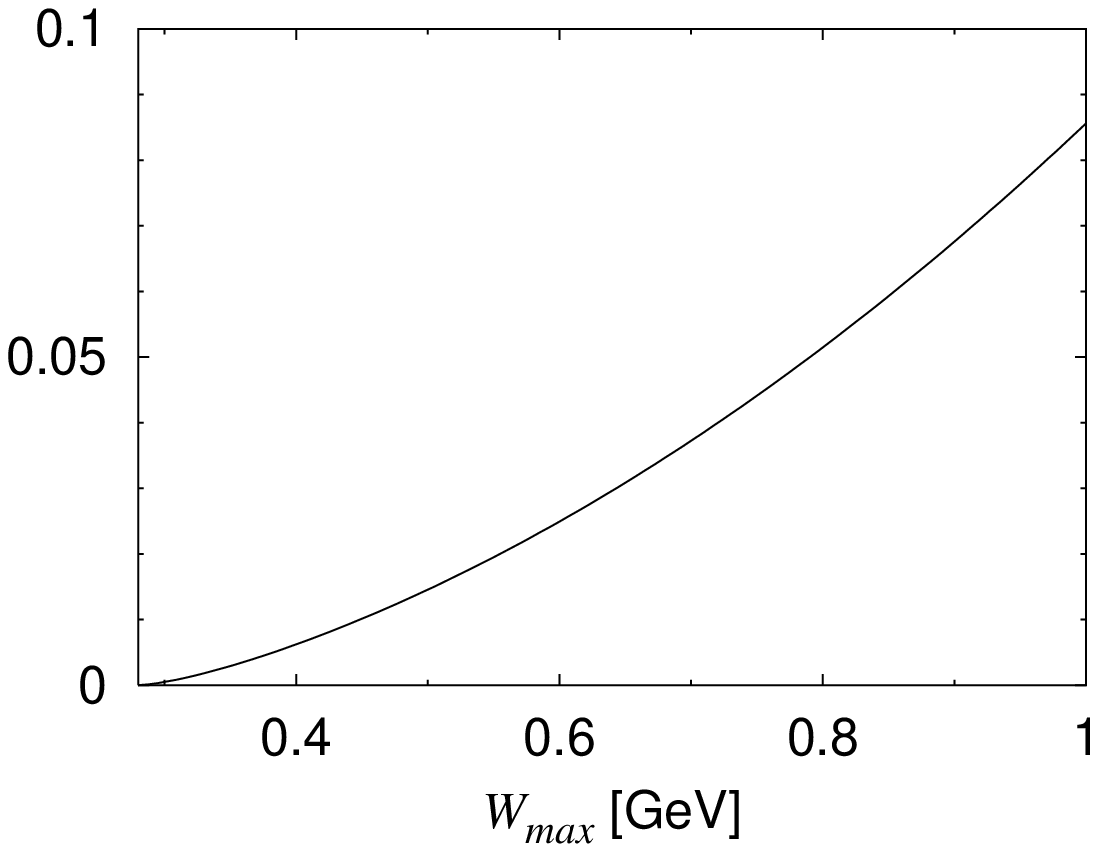}}
      \end{picture}
   \end{center}
\caption{\label{ratio-fig} The ratio of the cross sections
(\protect\ref{sigma-two-pions}) and (\protect\ref{sigma-one-pion}) for
the production of $\pi^0\pi^0$ and of $\pi^0$ in the limit of large
$Q^2$. The cross section for $e\gamma\to e\,\pi^0\pi^0$ is integrated
over $W$ from threshold up to $W_{{\mathit{max}}}$.}
\end{figure}

Fig.~\ref{ratio-fig} shows the ratio of the cross sections
(\ref{sigma-two-pions}) and (\ref{sigma-one-pion}) as a function of
the upper integration limit $W_{{\mathit{max}}}$. We see that, even
when integrating up to $W=1$~GeV, the single-pion production comes out
as clearly dominant. We remark that the measured production
rates~\cite{CLEO} for a single $\eta$ or $\eta'$ are comparable to
that of a $\pi^0$. With our isospin relation (\ref{isospin})
the cross section for $\gamma^*\gamma\to \pi^+\pi^-$ is twice that of
$\gamma^*\gamma\to \pi^0\pi^0$, the relative factor $1/2$ for
$\pi^0\pi^0$ being due to the phase space of identical particles. Due
to phase space one does not expect the production of more than two
pions to be important for $W$ below 1~GeV, except for the decays
$\eta\to3\pi$ and $\eta'\to5\pi$. The picture thus emerges that with
our estimation for $\gamma^*\gamma\to \pi\pi$ the production of
hadrons in $\gamma^*\gamma$ collisions up to 1~GeV is dominated by the
pseudoscalar channel, in other words by the parity-odd sector as
opposed to the parity-even one. This is reminiscent of the special
role played by the axial current in low-energy QCD.

At this point we wish to comment on the end-point regions of the
integrals over $z$ in the factorized expressions (\ref{tensor-two-pi})
and (\ref{tensor-one-pi}) for two-pion and one-pion production. For
$z\to 0$ and $z\to 1$ the hard-scattering kernels are divergent,
corresponding to the quark exchanged between the $\gamma$ and
$\gamma^*$ going on-shell. These poles are canceled by the end-point
zeroes of the two-pion and one-pion distribution amplitudes, so that
the end-point regions give a finite contribution to the scattering
amplitude in both cases. Quantitatively, the quark virtualities in the
hard-scattering diagrams are $z Q^2$ and $(1-z) Q^2$, and it is clear
that for a given finite $Q^2$ there is a region in $z$ where our
leading-order expressions should receive important corrections. At
small virtualities the strong coupling becomes large, increasing the
size of $\alpha_S$ corrections, and when $z Q^2$ or $(1-z) Q^2$
becomes comparable to the square of typical transverse quark momenta
in a pion, then power corrections due to the effect of the transverse
momentum of the produced $q\bar{q}$- pair will be important. We recall
in this context that various theoretical attempts to evaluate such
corrections lead to fair agreement with the data for the
$\gamma$--$\pi$ transition form factor~\cite{CLEO,Feld} down to
rather low values of $Q^2$.

For pion pair production both the hard-scattering kernel and the
distribution amplitude are zero at $z=1/2$ due to the constraints from
charge conjugation invariance, so that compared to the single-pion
case the integral in $z$ is more sensitive to the end-point
regions. We thus expect that for intermediate values of $Q^2$
corrections to the lowest-order results will be more important in
$\gamma^*\gamma\to \pi\pi$ than they are in $\gamma^*\gamma\to
\pi^0$. The experimental comparison of the $Q^2$-dependence of these
two processes will therefore be interesting and may help us to better
understand the origin of these corrections, which are a subject of
considerable importance in the physics of exclusive processes.

Taking the asymptotic $z$-dependence of the distribution amplitudes as
an example, we can explicitly see how important the end-point
contributions are in the leading-order expressions
(\ref{tensor-two-pi}) and (\ref{tensor-one-pi}). For single pion
production the integrand in Eq.~(\ref{tensor-one-pi}) is a constant
then, so that 50\% of the $z$-integral comes from the regions where
$z$ or $1-z$ is smaller than 0.25. For two-pion production the
integrand is proportional to $(2z-1)^2$, and 50\% of the integrand
comes from the regions with $z$ or $1-z$ smaller than $(1-2^{-1/3})/2
\approx 0.1$. Given these numbers, one can expect that corrections to
our leading-order calculation will not be negligible for $Q^2$ around
4~GeV$^2$, which is the lowest value considered in our numerical
estimates in Sect.~\ref{cross-sections}.

\section{Relations with the photon structure function}
\label{structure-function}

The exclusive process we consider here contributes of course to the
inclusive reaction $\gamma^\ast\gamma\to X$. As we mentioned in the
previous section, the inclusive process is built up from a limited
number of exclusive channels in the mass region of $W$ below
1~GeV. Let us examine the connection between our discussion of one-
and two-pion production with the familiar description of inclusive
$\gamma^* \gamma$ scattering in the kinematical limit we are taking
here.

The unpolarized cross section for inclusive deep inelastic scattering
on a photon, $e \gamma \to e X$, can be parameterized by two photon
structure functions $F_T$ and $F_L$ as
\begin{equation}
\frac{d\sigma_{e\gamma\to e X}}{dQ^2\, dW^2} =
  \frac{2\pi\alpha^2}{s_{e\gamma}^2}\,
  \frac{1}{x Q^2 (1-\epsilon)}\,
  \left( 2x F_T(x,Q^2) + \epsilon F_L(x,Q^2) \right) ,
\end{equation}
where $F_T$ and $F_L$ respectively give the contribution from
transverse and longitudinal polarization of the exchanged
$\gamma^*$. The transverse structure function $F_T$ is often traded for
$F_2 = 2x F_T + F_L$.

At the level of partons inclusive hadron production is described by
$\gamma^*\gamma\to q\bar{q}$ to leading order in $\alpha_S$, which
gives the well-known expressions~\cite{Bud}
\begin{eqnarray}
  \label{structure-general}
F_T^{q\bar{q}} &=& \frac{3\alpha}{2\pi} \, \sum_q e_q^4\,
\left\{ \ln\frac{1+\beta_q}{1-\beta_q} \left[
    x^2 + (1-x)^2 + 4x(1-x) \frac{m_q^2}{Q^2}
    - 8 x^2 \frac{m_q^4}{Q^4} \right] - \beta_q
    \left[ (1-2x)^2 + 4x(1-x) \frac{m_q^2}{Q^2} \right] \right\} ,
\nonumber \\
F_L^{q\bar{q}} &=& \frac{12\alpha}{\pi} \, \sum_q e_q^4\,
    x^2 (1-x) \left[ \beta_q -
        \frac{2 m_q^2}{W^2}  \ln\frac{1+\beta_q}{1-\beta_q} \right] ,
\end{eqnarray}
where $\beta_q = (1 - 4m_q^2/W^2)^{1/2}$. Note that $m_q$ is to be
understood here as a cutoff parameter, which regulates the collinear
divergence in the box diagram with massless quarks.

The limit of large $Q^2$ at fixed small $W^2$ we are taking here
implies $x\to 1$ and is different from the Bjorken limit, where $W^2$
is scaled up with $Q^2$ so that $x$ remains constant. Neglecting terms
of order $1-x\sim W^2/Q^2$ and $m_q^2/Q^2$ the expressions
(\ref{structure-general}) become
\begin{equation}
  \label{structure-limit}
F^{q\bar{q}}_T = \frac{3\alpha}{2\pi}\sum_q e_q^4 \left\{
\ln\frac{1+\beta_q}{1-\beta_q} - \beta_q \right\} , \hspace{4em}
F^{q\bar{q}}_L = O\left(\frac{W^2}{Q^2}\right) .
\end{equation}
We observe that in our limit the leading-order expression for $F_T$
becomes independent of $Q^2$, i.e., it has the same scaling behavior
as the exclusive channels $\gamma^*\gamma\to \pi$ and
$\gamma^*\gamma\to \pi\pi$. This is to be contrasted with the Bjorken
limit, where $\ln[(1+\beta_q)/(1-\beta_q)] \sim \ln[Q^2 / m_q^2] +
\ln[(1-x)/x]$ gives rise to the well-known logarithmic scaling
violation of $F_T$ at zeroth order in $\alpha_S$.

Just as in the case of $\gamma^*\gamma\to \pi\pi$, the contribution
$F_L$ from longitudinal photons is power suppressed in our limit.  Let
us add that in the Bjorken limit the hadronic part of $F_T$, often
parameterized using vector meson dominance, is only suppressed by a
factor $\ln Q^2$ with respect to the pointlike part
(\ref{structure-general}), but does not survive our limiting procedure
here: since hadronic structure functions typically decrease like a
power of $1-x$ for $x\to 1$, it becomes a correction in $W^2/Q^2$.

The contribution of our process to the structure functions is, with
our ansatz (\ref{model-gda}) for $\Phi^+_q$,
\begin{equation}
F_T^{\pi^+\pi^- + \pi^0\pi^0} = \frac{25\,\alpha}{96\,\pi}\, \beta
  \left( |\tilde{B}_{10}|^2 + \frac{1}{5} |\tilde{B}_{12}|^2 \right )
= \frac{625\,\alpha}{3456\,\pi}\, R_\pi^2\, \beta
  \left( 1 - \frac{2}{3} \beta^2 + \frac{1}{5} \beta^4 \right) .
  \label{structure-two-pi}
\end{equation}
As a function of $W$ this quickly rises from the threshold at
$2m_\pi$, levels off for $W$ around 400 to 500~MeV, and then remains
flat with a value $F_T^{\pi^+\pi^- + \pi^0\pi^0} /\alpha \approx
0.0077$. Let us compare this with the result (\ref{structure-limit})
of the $q\bar{q}$ calculation for $u$- and $d$-quarks (including
strange quarks would only lead to a minute change due to the charge
factor $e_q^4$). With the quark masses $m_u=m_d= 290$~MeV from the
parameterization of the photon structure function by Gordon and
Storrow~\cite{GorSto} we get a value of $F_T^{q\bar{q}} /\alpha
\approx 0.15$ at $W=1$~GeV, much larger than the one we obtain for
pion pairs.

It is worth remembering that $\gamma^*\gamma\to q\bar{q}$ also is the
hard-scattering subprocess in our factorized expression for
$\gamma^*\gamma\to \pi\pi$. As we discussed at the end of the previous
section, the collinear divergence of this process shows up as
singularities at the end-points of the $z$-integration in
Eq.~(\ref{tensor-two-pi}) and is canceled by the end-point zeroes of
the GDA, i.e., by the hadronization process. In the calculation of
open $q\bar{q}$ production no such cancellation takes place and the
divergence of the diagram has to be regulated. This reflects the fact
that even in the limit $Q^2\to\infty$ inclusive hadron production from
$\gamma^*\gamma$ cannot be calculated in perturbation theory alone
(unlike for instance inclusive hadron production from a single
timelike photon) and that the separation of $F_T$ into a perturbative
pointlike and a non-perturbative hadronic part is not
unambiguous. While more sophisticated procedures have been elaborated
in the literature, we consider it sufficient for our purpose to use
the quark mass regulator in Eq.~(\ref{structure-limit}). One might
also take massless quarks and a lower cutoff $\kappa_\perp$ on the
transverse quark momentum, obtaining the same result
(\ref{structure-limit}) with $m_q$ replaced by $\kappa_\perp$ in the
expression of $\beta_q$. While keeping us away from the region where
perturbation theory breaks down, such phenomenological regulators
become of course inadequate as one approaches the ``threshold'' where
$\beta_q=0$. For our numbers this is at $W=580$~MeV. One should bear
this in mind when using the expression (\ref{structure-limit}) for
invariant masses $W$ around 1~GeV.

On the other hand we saw in Sect.~\ref{one-pion} that with our
estimate of two-pion production the hadronic mass spectrum below 1~GeV
is dominated by the single-meson states $\pi^0$, $\eta$, $\eta'$. It
is clear that in such a region the parton-level result can only hold
in the sense of parton-hadron duality, averaged over a sufficiently
large interval in $W$. We therefore integrate the cross section for
$e\gamma\to e X$ over $W$ from threshold up to 1~GeV. The parton-level
result, obtained from Eq.~(\ref{structure-limit}) with
$m_u=m_d=290$~MeV, amounts to 2.42 times the cross section
(\ref{sigma-one-pion}) for one-pion production. This factor should be
compared with the factor $1 + 0.26 + 0.97 + 2.64$ for the individual
contributions of the exclusive channels $\pi + \pi\pi + \eta +
\eta'$. Here we used Eq.~(\ref{structure-two-pi}) for two-pion
production, whereas for $\eta$ and $\eta'$ we replaced $f_\pi=131$~MeV
in Eq.~(\ref{sigma-one-pion}) with the respective values 129~MeV and
213~MeV taken from the analysis of~\cite{Feld}. Given the caveats of
parton-hadron duality (below 1~GeV there are very few resonances in
the two-photon channel, and $W = 1$~GeV is just above the $\eta'$
threshold) and those of the parton-level calculation itself (discussed
above), we find the agreement remarkably fair.

\section{Phenomenology}
\label{phenomenology}

We will now discuss the phenomenology of our process in $e\gamma$ and
in $e^+e^-$ collisions. The production of neutral and charged pion
pairs is rather different in this respect, since $\pi^0\pi^0$ is only
produced by the $\gamma^*\gamma$ subprocess we have discussed so far,
whereas for $\pi^+\pi^-$ production this process interferes with
bremsstrahlung, i.e., the production of the pion pair from a timelike
photon radiated off the beam lepton. We start with the simpler case of
neutral pions, and then discuss charged pairs. In the following we
will restrict ourselves to unpolarized photon and lepton beams. A
brief discussion of beam polarization will be given in Appendix
\ref{beam-polar}.

\subsection{Helicity amplitudes}
\label{helicity}

The building blocks of our investigation are the helicity amplitudes
for $\gamma^* \gamma \to \pi \pi$, which describe the dynamics of this
process in a model independent way. They are obtained from the
hadronic tensor $T^{\mu\nu}$ by multiplying the reduced amplitudes
\begin{equation}
A_{ij}(Q^2,W^2,\theta) = 
  \epsilon^\mu_{i} \, T_{\mu\nu} \epsilon'{}^\nu_{j} , 
\hspace{3em} i = +,0,-, \hspace{3em} j = +,-
\end{equation}
with the squared elementary charge $e^2$. In the $\gamma^*\gamma$
c.m.\ our photon polarization vectors read
\begin{equation}
\epsilon_{0}   = \frac{1}{Q}\, (|{\mathbf q}\,|, 0, 0, q^0) , \hspace{3em}
\epsilon_{\pm} = \frac{1}{\sqrt{2}}\, (0, \mp 1, -i, 0)
\end{equation}
for the virtual and
\begin{equation}
\epsilon'{}_{\pm} = \frac{1}{\sqrt{2}}\, (0, \mp 1, +i, 0)
\end{equation}
for the real photon, where we have used the coordinate system
described in Sect.~\ref{kinematics}. By parity invariance, there are
only three independent helicity amplitudes, which we choose to be
$A_{++}$, $A_{-+}$ and $A_{0+}$.

Each of these three amplitudes plays a distinctive dynamical role in
the kinematical region $Q^2 \gg W^2, \Lambda^2$. It is $A_{++}$ that
receives the leading twist contribution we have discussed in detail,
and which in the scaling regime gives access to the generalized quark
distribution amplitudes $\Phi^{\pi\pi}_q$,
\begin{equation}
  \label{leading-amp}
A_{++} = \sum_q \frac{e_q^2}{2}\int_0^1\! dz\,{2z-1\over z(1-z)}\,
\Phi^{\pi\pi}_q(z,\zeta,W^2)
\end{equation}
to zeroth order in $\alpha_S$. The amplitude $A_{-+}$ has a
leading-twist part at order $\alpha_S$, due to the helicity-two gluon
GDA. We briefly discussed this at the end of
Sect.~\ref{zeta-expansion}; for more detail we refer
to~\cite{Kivel}. Finally, the contribution $A_{0+}$ from a
longitudinal $\gamma^*$ is nonleading twist. The predicted power
behavior in $Q^2$ at fixed $W^2$ and $\zeta$ is therefore that
$A_{++}$ becomes independent of $Q^2$, whereas $A_{0+}$ decreases at
least like $1/Q$. The amplitude $A_{-+}$ should become
$Q^2$-independent. If the helicity-two gluon GDA is however not
sufficiently large, $A_{-+}$ may be dominated by higher-twist
contributions at accessible values of $Q^2$ and should decrease like a
power of $1/Q$ in the corresponding $Q^2$-range. Of course all these
predictions are to be understood as up to corrections in $\log
Q^2$. At sufficiently large $Q^2$, the longitudinal amplitude $A_{0+}$
is thus predicted to be small compared with $A_{++}$. One can also
expect that $A_{-+}$ will be smaller than $A_{++}$, since its
leading-twist part is suppressed by $\alpha_S$.

To discuss the different partial waves in which the pion pair can be
produced, we expand each of the amplitudes $A_{++}$, $A_{0+}$,
$A_{-+}$ as
\begin{equation}
 \label{part-expansion}
A_{ij} = 
  \sum_{\scriptstyle l=j-i \atop \scriptstyle {\rm even}}^{\infty}
  A_{ijl}(Q^2,W^2) \, P_l^{j-i}(\cos\theta) ,
\hspace{3em} i = +,0,-, \hspace{3em} j = + ,
\end{equation}
where $P^m_l$ denotes the associated Legendre polynomial corresponding
to the value of $J_z$ of the $\pi\pi$ system in its c.m.

\subsection{The $\gamma^* \gamma$ subprocess and $\pi^0 \pi^0$ production}
 \label{neutral-pi}

The differential $e\gamma$ cross section for neutral pion pair
production reads
%
%
\begin{eqnarray}
\left. \frac{d\sigma_{e\gamma\to e\, \pi\pi}}{
             dQ^2\, dW^2\, d(\cos\theta)\, d\varphi}
\right|_{G} =
   \frac{\alpha^3}{16\pi}\, \frac{\beta}{s_{e\gamma}^2}\,
   \frac{1}{Q^2 (1-\epsilon)}
& \Big( & |A_{++}|^2 + |A_{-+}|^2 + 2\epsilon\, |A_{0+}|^2
\nonumber \\
\phantom{\frac{1}{s_{e}}} 
&-& \cos\varphi\; \Re \left\{ A^*_{++} A^{\phantom{*}}_{0+}
                            - A^*_{-+} A^{\phantom{*}}_{0+} \right\}
   2\sqrt{\epsilon(1+\epsilon)}
\nonumber \\
\phantom{\frac{1}{s_{e}}} 
&-& \cos2\varphi\; \Re \left\{ A^*_{++} A^{\phantom{*}}_{-+} \right\}
    2\epsilon \, \Big) ,
 \label{gamma-gamma}
\end{eqnarray}
where the subscript $G$ indicates that the pions are produced in a
$\gamma^*\gamma$ subprocess. For $\pi^0\pi^0$ production the phase
space in Eq.~(\ref{gamma-gamma}) is understood as restricted to
$\cos\theta \in (0,1)$, $\varphi \in (0,2\pi)$ because there are two
identical particles in the final state. We notice the close similarity
of the expression (\ref{gamma-gamma}) with the cross section of the
crossed channel process of virtual Compton scattering, and much of
what we discuss in the following has its counterpart there
\cite{DGPR}.

To obtain the $e^+e^-$ cross section we use the equivalent photon
approximation~\cite{Bud}, 
\begin{equation}
  \frac{d\sigma_{ee\to ee\, \pi\pi}}{
        dQ^2\, dW^2\, d(\cos\theta)\, d\varphi\, dx_2} =
  \frac{\alpha}{\pi}\, \frac{1}{x_2} \left(
    \frac{1+(1-x_2)^2}{2} \ln\left[
          \frac{Q'^2_{{\mathit{max}}}(x_2)}{Q'^2_{{\mathit{min}}}(x_2)}
    \right] - (1-x_2) \right) \,
  \frac{d\sigma_{e\gamma\to e\, \pi\pi}}{
        dQ^2\, dW^2\, d(\cos\theta)\, d\varphi} ,
  \label{EPA}
\end{equation}
where $Q'^2_{{\mathit{min}}}$ and $Q'^2_{{\mathit{max}}}$ are the
minimal and maximal virtuality of the photon $q'$, respectively.  We
have a lower kinematical limit $Q'^2_{{\mathit{min}}} = x_2^2\, m_e^2
/(1-x_2)$ determined by the electron mass $m_e$, whereas
$Q'^2_{{\mathit{max}}}$ depends on experimental cuts and will be
discussed in more detail in Sect.~\ref{lab-kin}. We remark that for a
given $ee$ collider energy the variables $x_2$ and $y$ are not
independent at fixed $Q^2$ and $W^2$, since
\begin{equation}
y x_2 = \frac{Q^2+W^2}{s_{ee}} ,
 \label{y-x}
\end{equation}
and that in Eq.~(\ref{EPA}) one can easily trade $dx_2$ for $dy$.

Since the helicity amplitudes $A_{ij}$ are independent of $\varphi$
they can be partially disentangled from the $\varphi$-dependence of
the cross section, which is completely explicit in
Eq.~(\ref{gamma-gamma}). In particular, the relative size and the
$Q^2$-behavior of the $\varphi$-independent term and of the terms with
$\cos\varphi$ and $\cos2\varphi$ allow detailed tests of the scaling
predictions. This provides indicators on how close one is to the
asymptotic regime at finite values of $Q^2$. The $\varphi$-independent
term in the large parentheses of Eq.~(\ref{gamma-gamma}) receives
contributions from leading-twist amplitudes and should thus display
scaling behavior. The coefficient of $\cos\varphi$ is the interference
of leading-twist and non-leading twist amplitudes and should be
suppressed by at least one power of $1/Q$. Finally, the $\cos2\varphi$
term should scale or be power suppressed depending on the size of the
helicity-two gluon GDA.

Apart from standard fitting techniques a way to separate terms with
different angular dependence is the use of weighted cross
sections. Weighting each event with a function $w(\varphi,\theta)$ we
define
\begin{equation}
 \label{weight-def}
S_{e\gamma}(w) = \int dQ^2 dW^2 d\Omega\,
\frac{d\sigma_{e\gamma}}{dQ^2 dW^2 d\Omega}\, w(\varphi,\theta) ,
\end{equation}
where $d\Omega = d(\cos\theta)\, d\varphi$. Notice that since it is
not normalized, $S_{e\gamma}(w)$ is not just the average value of the
function $w(\varphi,\theta)$, and includes information about the size
of the cross section itself. Interpreting $S_{e\gamma}(w)$ as a
statistical variable one can calculate its standard deviation and
finds for its relative statistical error (cf., e.g., \cite{opt-obs})
\begin{equation}
\delta(w) = \frac{1}{\sqrt{N}}\;
\frac{ \displaystyle  \sqrt{\int dQ^2 dW^2 d\Omega\,
   \frac{d\sigma_{e\gamma}}{dQ^2 dW^2 d\Omega}\, w^2(\varphi,\theta)}
                   \; \sqrt{\int dQ^2 dW^2 d\Omega\, 
   \frac{d\sigma_{e\gamma}}{dQ^2 dW^2 d\Omega}} }{
\displaystyle \left| \int dQ^2 dW^2 d\Omega\,
\frac{d\sigma_{e\gamma}}{dQ^2 dW^2 d\Omega}\, w(\varphi,\theta) 
\right| } ,
  \label{stat-error}
\end{equation}
where
\begin{equation}
N = {\mathcal{L}} \int dQ^2 dW^2 d\Omega\,
\frac{d\sigma_{e\gamma}}{dQ^2 dW^2 d\Omega}
\end{equation}
is the expected number of events for a given integrated luminosity
$\mathcal{L}$. Eq.~(\ref{stat-error}) generalizes the well-known
result that the relative statistical error of the cross section, i.e.,
of $S_{e\gamma}(1)$, is $1 /\sqrt{N}$. We emphasize that the method of
weighted cross sections is very flexible, and that the choice of
weights $w(\varphi,\theta)$ can for instance be adapted to
experimental conditions such as limited angular acceptance or
cuts. One can of course take weighting functions that depend on other
variables than only $\theta$ and $\varphi$. In the following we will
also use weighted differential cross sections, where only some of the
kinematical variables have been integrated out while others are held
fixed. In a data analysis, one may thus use the weighting technique
for some variables and fitting for others.

The weighting technique is convenient to project out different terms
in the cross section. As an immediate example we note that the terms
constant in $\varphi$, with $\cos\varphi$, and with $\cos2\varphi$ in
the $e\gamma$ cross section are obtained from
\begin{equation}
 \label{phi-moments}
\frac{d S_{e\gamma}(\cos m\varphi)}{dQ^2\,dW^2\,d(\cos\theta)}
\end{equation}
with $m=0$, 1 and 2, respectively. If the moments with $m=1$ and 2 are
measured to be small compared with the moment $m=0$, this can be
because any two of the amplitudes $A_{++}$, $A_{0+}$, $A_{-+}$ are
much smaller than the third, or it may be due to their relative
phases. From the theoretical considerations in Sect.~\ref{helicity}
the most natural hypothesis in this case is however that $A_{0+}$ and
$A_{-+}$ are small compared with $A_{++}$.

While the $\varphi$-dependence of the cross section
(\ref{gamma-gamma}) gives access to the various helicity combinations
of the real and virtual photon, its dependence on $\theta$ contains
information on the angular momentum states in which the pion pair is
produced. A priori there can be arbitrarily high partial waves, but to
analyze the $\theta$-distribution in practice one will assume that at
a given $W$ only a finite number of them is important, if only for
reasons of phase space. It is easy to see from
Eqs.~(\ref{part-expansion}) and~(\ref{gamma-gamma}) that for a
superposition of partial waves $l=0$, 2, \ldots $L$ the moment of
$\cos m\varphi$ in (\ref{phi-moments}) is a linear combination of
polynomials $P_{2l}^m(\cos\theta)$ with highest degree $2L$. Weighting
the cross section with $\cos (m\varphi)\, P_{2L+2}^m(\cos\theta)$ and
integrating over $\varphi$ and $\theta$ then gives a zero
result. Using these weights thus provides one way to estimate from
experimental data how many partial waves are relevant. Let us recall
the physical relevance of this information: in the scaling regime the
highest partial wave relevant in $A_{++}$ provides a constraint on how
far the two-pion distribution amplitude is from its asymptotic form,
as we discussed in Sect.~\ref{partial-waves}.

Let us assume that only partial waves with $l\le L$ effectively
contribute in the cross section~(\ref{gamma-gamma}). The
$\theta$-dependence of the moments in (\ref{phi-moments}) is then
determined by $L+1$ coefficients for $m=0$, $L$ coefficients for $m=1$
and $L$ coefficients for $m=2$, corresponding to the number of
polynomials $P_{2l}^m(\cos\theta)$ with $l\le L$. On the other hand,
there are $3L/2+1$ complex amplitudes $A_{ijl}$ with $l\le L$ in the
expansion (\ref{part-expansion}), so that there are $3L+2$ real
quantities one would like to determine. A global phase is however
unobservable in the cross section (\ref{gamma-gamma}), and one may for
instance refer all phases to the phase of $A_{++0}$. The $3L+1$
coefficients one can extract from the dependence of the cross section
on $\varphi$ and $\theta$ thus allow one to reconstruct the
$|A_{ijl}|$ and their relative phases. Since the relation between the
angular coefficients and the amplitudes is quadratic, there will
however be multiple solutions in general. More information can be
obtained with polarized beams, which we briefly discuss in
Appendix~\ref{beam-polar}.

The situation is simplest if the $\theta$-dependence of the cross section
is compatible with the $\pi^0\pi^0$ system being produced only in an
$S$- and a $D$-wave, and if in addition the $\varphi$-dependence is
flat. Assuming that $A_{0+}$ and $A_{-+}$ are negligible compared to
$A_{++}$, one can then decompose
\begin{equation}
\frac{d\sigma_{e\gamma\to e\,\pi^0\pi^0}}{dQ^2\,dW^2\,d(\cos\theta)}
 = C_{00} + C_{02}\, P_2(\cos\theta)
 + C_{22}\, [P_2(\cos\theta)]^2
 \label{gamma-partial}
\end{equation}
and project out the coefficients, using that $C_{ll'} = d
S_{e\gamma}(w_{l l'}) /(dQ^2\,dW^2)$ with weights
\begin{eqnarray}
w_{00} &=& - \frac{5}{16}\, ( 1 - 42 \cos^2\theta + 49 \cos^4\theta ) , 
\nonumber \\
w_{02} &=& - \frac{35}{8}\, ( 1 - 6 \cos^2\theta + 5 \cos^4\theta ) ,
\nonumber \\
w_{22} &=&   \frac{35}{16}\, ( 3 - 30 \cos^2\theta + 35 \cos^4\theta ) .
\end{eqnarray}
>From these coefficients one can readily extract the amplitudes
$|A_{++0}|$, $|A_{++2}|$, and the cosine of their relative phase.

\begin{figure}
   \begin{center}
      \setlength{\unitlength}{0.49\hsize}
      \begin{picture}(1,0.87)(0,0)
	\put(0.5,0.8){\large (a)}
	\put(0.05,0.72){$2W C_{ll'}$~[fb GeV$^{-3}$]}
	\put(0.36,0.2255){$C_{00}$}
	\put(0.36,0.185){$C_{02}$}
	\put(0.36,0.1445){$C_{22}$}
        \epsfxsize=0.49\hsize  
      	\put(0,0){\epsfbox{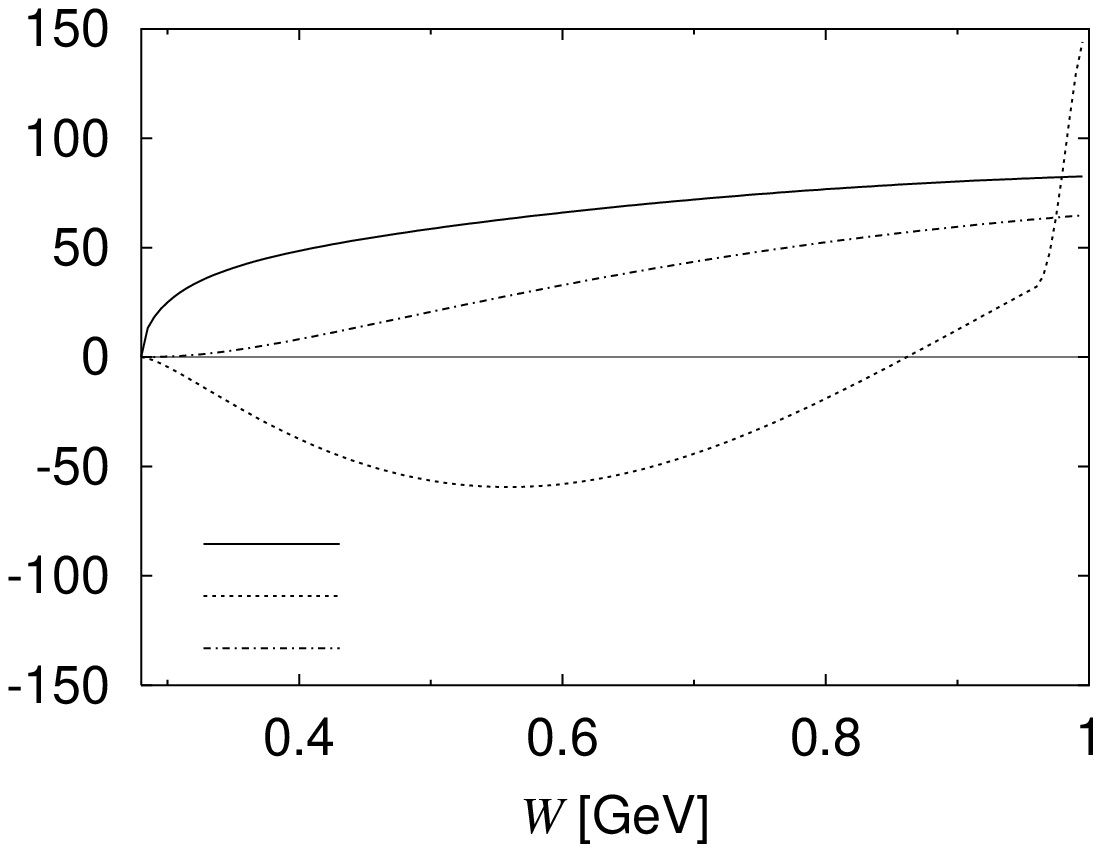}}
      \end{picture}
      \begin{picture}(1,0.87)(0,0)
	\put(0.5,0.8){\large (b)}
	\put(0.05,0.72){$2W C_{ll'}$~[fb GeV$^{-3}$]}
	\put(0.36,0.2255){$C_{00}$}
	\put(0.36,0.185){$C_{02}$}
	\put(0.36,0.1445){$C_{22}$}
        \epsfxsize=0.49\hsize  	
	\put(0,0){\epsfbox{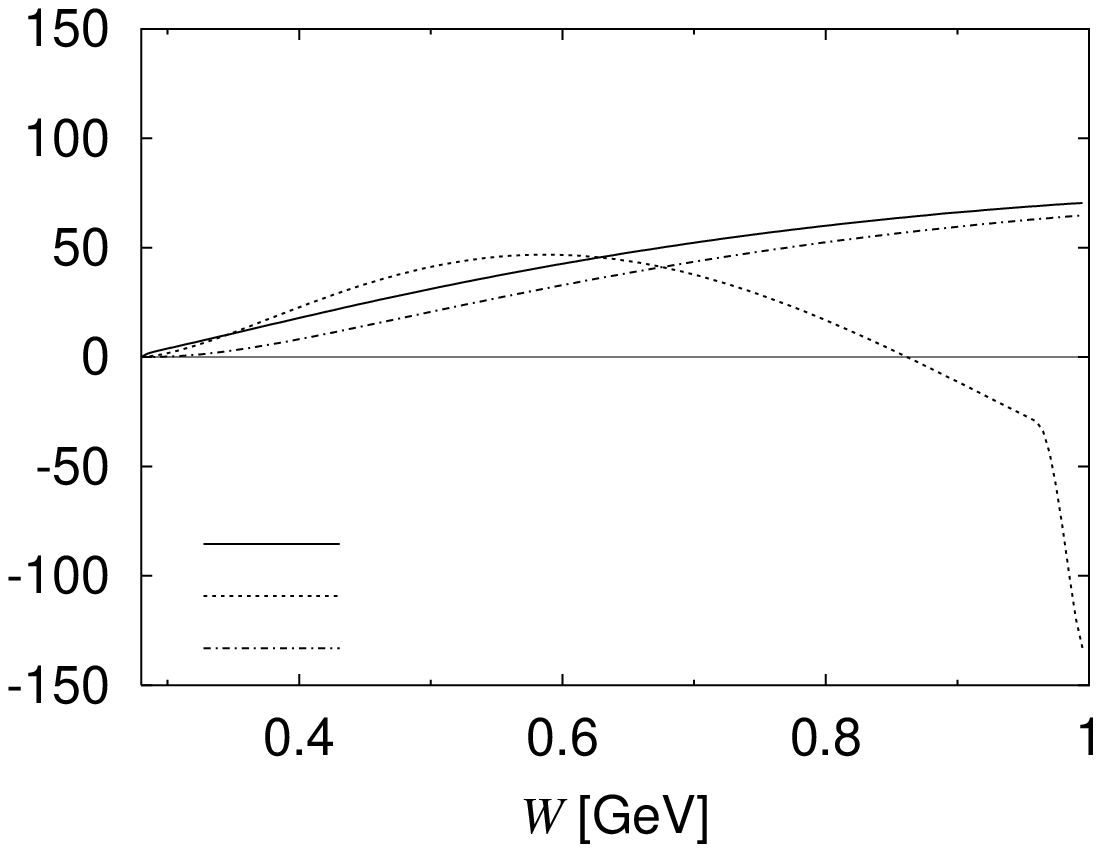}}
      \end{picture}
   \end{center}
\caption{\label{gds} (a) The coefficients $C_{l l'}$ in the
differential cross section (\protect\ref{gamma-partial}), evaluated
with our model GDA (\protect\ref{model-gda}). They are plotted against
$W$ instead of $W^2$ and therefore have been multiplied with a
Jacobian $2W$. The values of the remaining kinematical variables are
$s_{e\gamma}=50$~GeV$^2$ and $Q^2=5$~GeV$^2$. (b) The same with the
alternative ansatz for the GDA described in the text.}
\end{figure}

In Fig.~\ref{gds} (a) we show the coefficients $C_{00}$, $C_{02}$ and
$C_{22}$ obtained with our model GDA (\ref{model-gda}). The
interference term between the $S$- and $D$-waves contains a factor
$\cos(\delta_0 - \delta_2)$ and thus is sensitive to the phase
shifts. Characteristic features in the $W$-dependence of $C_{02}$ are
the point where $\delta_0 - \delta_2=90^\circ$, and the sudden change
just below $W=1$~GeV due to the behavior of the $S$-wave. To explore
the dependence of these observables on our input GDA we have made an
ad hoc modification, changing the sign in the prediction
(\ref{soft-pion}) from chiral dynamics and taking instead $B_{10}(0) =
B_{12}(0)$ with $B_{12}(0)$ fixed by the constraint (\ref{constraint})
as before. Notice that this flips the overall sign $\eta_{10}$ of the
$S$-wave in our model. The result is shown in Fig.~\ref{gds} (b) and
illustrates the sensitivity, especially of the $S$-$D$ interference,
to the detailed dynamics of the $\gamma^*\gamma$ process.

\subsection{Production of $\pi^+\pi^-$ and interference with
bremsstrahlung}

\begin{figure}
   \begin{center}
        \leavevmode
        \epsfxsize=0.77\hsize  \epsfbox{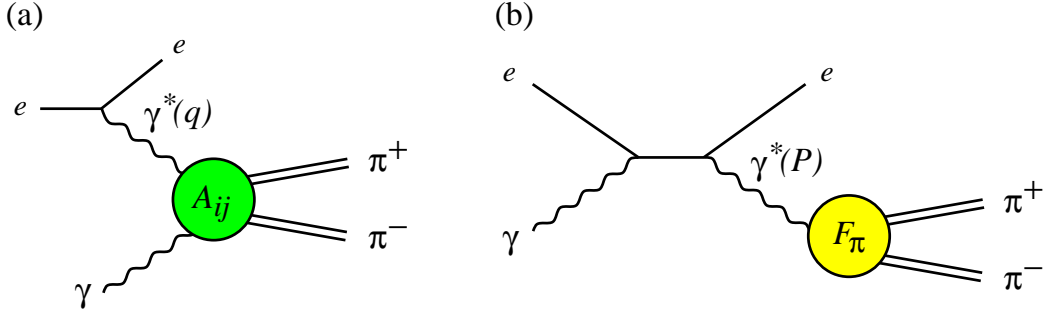}
   \end{center}
\caption{\label{brems-fig} The two subprocesses contributing to the
reaction $e\gamma\to e\, \pi^+\pi^-$: (a) $\gamma^*\gamma$ scattering
and (b) bremsstrahlung. There is a second bremsstrahlung diagram,
where the photon vertices are interchanged on the lepton line.}
\end{figure}

\begin{figure}
   \begin{center}
      \setlength{\unitlength}{0.5\hsize}
      \begin{picture}(1,0.71)(0,0)
	\put(-0.16,0.38){$|F_\pi(W^2)|^2$}
	\epsfxsize=0.5\hsize  
	\put(0,0){\epsfbox{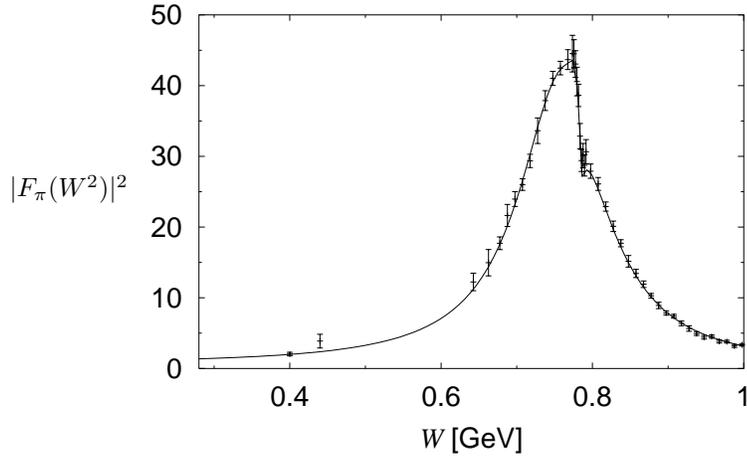}}
      \end{picture}
  \end{center}
\caption{\label{fpi-fig} The square of the electromagnetic pion form
factor in the timelike region. The data points are from
\protect\cite{Fpi-data} and the curve is the parameterization $N=1$ of
\protect\cite{Kuehn}.}
\end{figure}

For the production of $\pi^+\pi^-$ pairs in $e\gamma$ collisions, the
$\gamma^\ast \gamma$ reaction we want to study competes with
bremsstrahlung, where the pion pair originates from a virtual photon
radiated off the lepton~\cite{Bud}, cf.\ Fig.~\ref{brems-fig}. This
process produces the pion pair in the $C$-odd channel and hence does
not contribute for $\pi^0\pi^0$. Its amplitude can be fully computed
for values of $W$ where the timelike electromagnetic pion form factor
$F_\pi(W^2)$ is known. The modulus of $F_\pi$ has been well measured
in $e^+ e^- \to \pi^+\pi^-$. By Watson's theorem its phase is equal to
the $P$-wave phase shift $\delta_1$, provided that $W$ is in the range
where $\pi\pi$ scattering is elastic. This is rather well satisfied
for $W$ up to 1~GeV. In our numerical studies we use for $F_\pi$ the
parameterization $N=1$ of~\cite{Kuehn}, which is in good agreement
with the data for $|F_\pi|^2$ shown in Fig.~\ref{fpi-fig}. It also
gives a fair description of the phase of $F_\pi$ in the $W$-range
where we use it, as we see from the comparison with the phase shift
$\delta_1$ in Fig.~\ref{phase-shifts}.

The contribution of the $\gamma^*\gamma$ subprocess to the cross
section of $e\gamma\to e\, \pi^+\pi^-$ has the same form
(\ref{gamma-gamma}) as for $e\gamma\to e\, \pi^0\pi^0$. We recall that
with the isospin relation (\ref{isospin}) the leading-twist helicity
amplitude $A_{++}$ in (\ref{leading-amp}) is the same for neutral and
for charged pion pairs. The bremsstrahlung contribution
reads
%
%
\begin{eqnarray}
  \label{brems}
\left. \frac{d\sigma_{e\gamma\to e\, \pi\pi}}{
             dQ^2\, dW^2\, d(\cos\theta)\, d\varphi}
\right|_{B} =
  \frac{\alpha^3}{16\pi}\, \frac{\beta}{s_{e\gamma}^2}\,
  \frac{2\beta^2}{W^2\, \epsilon}\, |F_\pi(W^2)|^2
& \Big( & [1 - 2x(1-x)] \sin^2\theta + 4x(1-x)\, \epsilon
                                       \cos^2\theta
\nonumber \\
\phantom{\frac{1}{s_{e}}} 
&+& \cos\varphi\; \sqrt{2x(1-x)}\, (2x-1) \sqrt{\epsilon(1+\epsilon)}\,
    2 \sin\theta \cos\theta \nonumber \\
\phantom{\frac{1}{s_{e}}} 
&-& \cos2\varphi\; x(1-x)\, 2\epsilon \sin^2\theta\, \Big) .
\end{eqnarray}
Finally, the interference term of the two subprocesses can be written
as\footnote{A C program containing the expressions
(\protect\ref{gamma-gamma}), (\protect\ref{brems}),
(\protect\ref{interfere}), (\protect\ref{inter-coefficients}), as well
as the amplitude $A_{++}$ calculated with our model
GDA~(\ref{model-gda}), can be obtained from the authors.}
\begin{equation}
\left. \frac{d\sigma_{e\gamma\to e\, \pi\pi}}{
             dQ^2\, dW^2\, d(\cos\theta)\, d\varphi}
\right|_{I} = - 2 e_l\,
  \frac{\alpha^3}{16\pi}\, \frac{\beta}{s_{e\gamma}^2}\,
  \frac{\sqrt{2} \beta}{\sqrt{W^2 Q^2 \epsilon(1-\epsilon) }} \,
  \Big( C_0 + C_1\, \cos\varphi + C_2 \cos2\varphi + 
         C_3\cos3\varphi \Big)
 \label{interfere}
\end{equation}
with $e_l = 1$ for positrons and $-1$ for electrons, and coefficients
%
%
\begin{eqnarray}
C_0 &=& \Re \Big\{ F_\pi^* A_{++} \Big\} \,
        \sqrt{2x(1-x)} \sqrt{\epsilon(1+\epsilon)} \cos\theta
\nonumber \\
&&    + \Re \Big\{ F_\pi^* A_{0+} \Big\} \,
        (1-x) \sqrt{\epsilon(1+\epsilon)} \sin\theta ,
\nonumber \\
C_1 &=& \Re \Big\{ F_\pi^* A_{++} \Big\} \,
        [1 - (1-x)(1+\epsilon)] \sin\theta
\nonumber \\
&&    - \Re \Big\{ F_\pi^* A_{0+} \Big\} \,
        \sqrt{2x(1-x)}\, 2\epsilon \cos\theta
\nonumber \\
&&    + \Re \Big\{ F_\pi^* A_{-+} \Big\} \,
        (1-x) \sin\theta ,
\nonumber \\
C_2 &=& - \Re \Big\{ F_\pi^* A_{0+} \Big\} \,
        x \sqrt{\epsilon(1+\epsilon)} \sin\theta
\nonumber \\
&&    - \Re \Big\{ F_\pi^* A_{-+} \Big\} \,
        \sqrt{2x(1-x)} \sqrt{\epsilon(1+\epsilon)} \cos\theta ,
\nonumber \\
C_3 &=& - \Re \Big\{ F_\pi^* A_{-+} \Big\} \,
        x \epsilon \sin\theta .
  \label{inter-coefficients}
\end{eqnarray}
Remember that in our kinematical limit $1-x \sim W^2 /Q^2$ is
small. The structure of the bremsstrahlung contribution (\ref{brems})
then becomes rather simple, since at $Q^2 \gg W^2$ the terms in large
parentheses reduce to $\sin^2\theta$. With the scaling predictions for
the $\gamma^*\gamma$ amplitudes discussed in Sect.~\ref{helicity} we
also obtain the $Q^2$-behavior for each of the coefficients $C_n$ in
the interference term (\ref{interfere}).

The relative dependence on $Q^2$, $W^2$ and on $\epsilon$ of the three
contributions to the cross section is controlled by the prefactors
\begin{equation}
\frac{1}{Q^2 (1-\epsilon)} , \hspace{2em}
\frac{2 \beta^2}{W^2\, \epsilon} , \hspace{2em}
\frac{\sqrt{2}\beta}{\sqrt{W^2 Q^2 \epsilon(1-\epsilon) }}
 \label{factors}
\end{equation}
for $\gamma^*\gamma$, bremsstrahlung and their interference,
respectively, and by the pion form factor $F_\pi(W^2)$, which appears
linearly in the interference and squared in the pure bremsstrahlung
term. The factors $Q^2$ and $W^2$ in (\ref{factors}) can be traced
back to the propagator of the virtual photon in each subprocess, and
the extra factor of $\beta$ in the bremsstrahlung amplitude reflects
the fact that the pion pair is produced in the $P$-wave there.

\begin{figure}
   \begin{center}
      \setlength{\unitlength}{0.49\hsize}
      \begin{picture}(1,0.93)(0,0)
	\put(0.5,0.85){\large (a)}
	\put(0.03,0.75){$\displaystyle
             \frac{d\sigma_{ee\to ee\,\pi^+\pi^-}}{
                   dQ^2\, dW^2\, d\varphi\, dy}$~[fb GeV$^{-4}$]}
        \epsfxsize=0.49\hsize  
	\put(0,0){\epsfbox{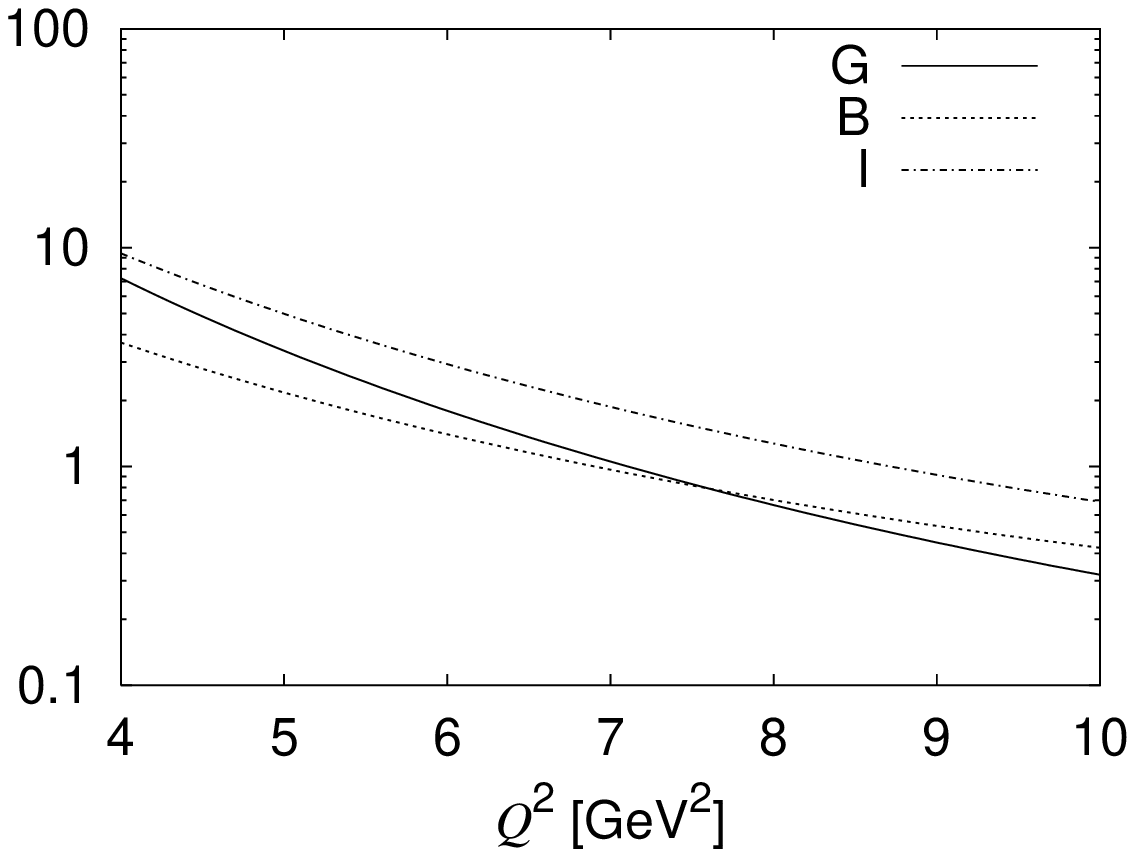}}
      \end{picture}
      \begin{picture}(1,0.82)(0,0)
	\put(0.5,0.85){\large (b)}
	\put(0.03,0.75){$\displaystyle
             \frac{d\sigma_{ee\to ee\,\pi^+\pi^-}}{
                   dQ^2\, dW^2\, d\varphi\, dy}$~[fb GeV$^{-4}$]}
        \epsfxsize=0.49\hsize  
      \put(0,0){\epsfbox{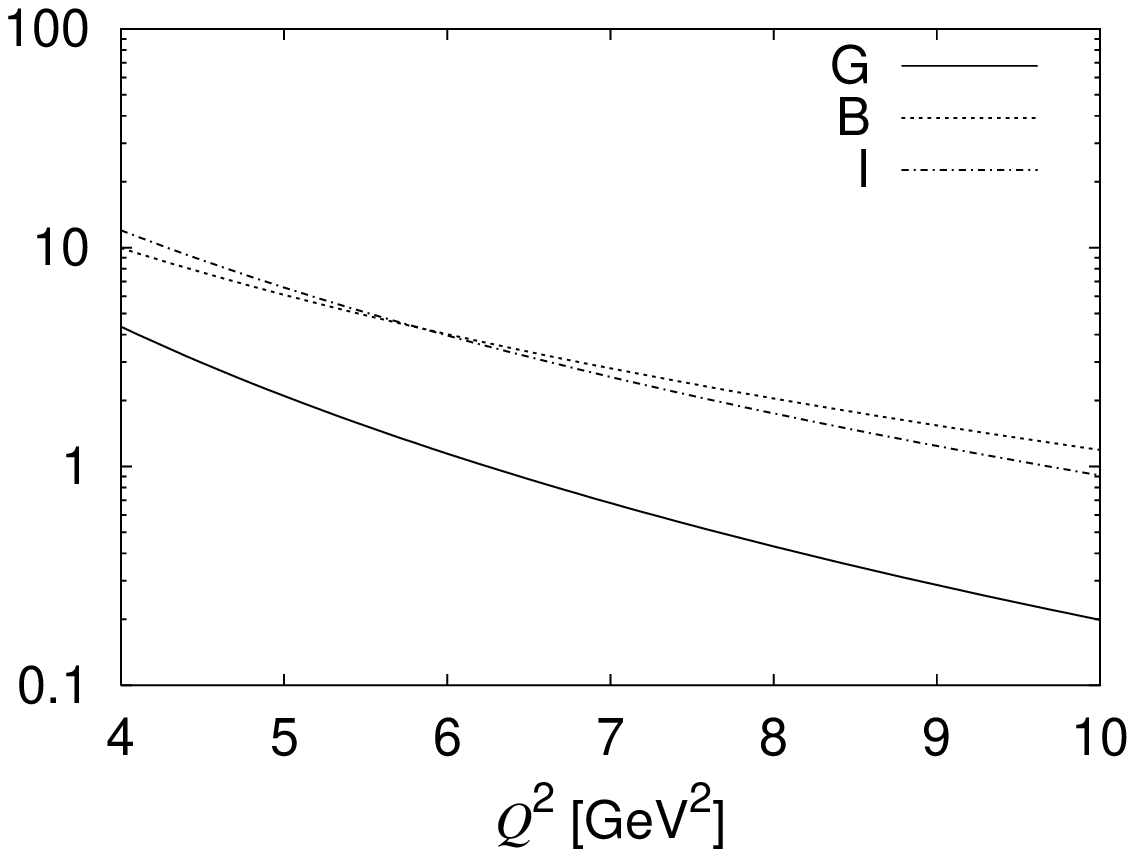}}
      \end{picture}
   \end{center}
\caption{\label{Q-dep} (a) The contributions to the differential $ee$
cross section from $\gamma^*\gamma$, bremsstrahlung and their
interference. Kinematical variables are $W=400$~MeV, $\varphi=0$,
$y=0.1$, $E_1=3.1$~GeV, $E_2=9$~GeV. For the real photon flux in
(\protect\ref{EPA}) we take $\alpha_{2L}^{{\mathit{max}}}=300$~mrad
and $l'^{{\mathit{max}}}_{\perp L}=100$~MeV as explained in
Sect.~\protect\ref{lab-kin}. The sign of the interference term
corresponds to an $e^+\gamma$ subprocess. (b) The same as (a), but
with $y=0.2$.}
\end{figure}

\begin{figure}
   \begin{center}
      \setlength{\unitlength}{0.49\hsize}
      \begin{picture}(1,0.84)(0,0)
	\put(0.03,0.75){$\displaystyle
             \frac{d\sigma_{ee\to ee\,\pi^+\pi^-}}{
                   dQ^2\, dW^2\, d\varphi\, dy}$~[fb GeV$^{-4}$]}
        \epsfxsize=0.49\hsize  
	\put(0,0){\epsfbox{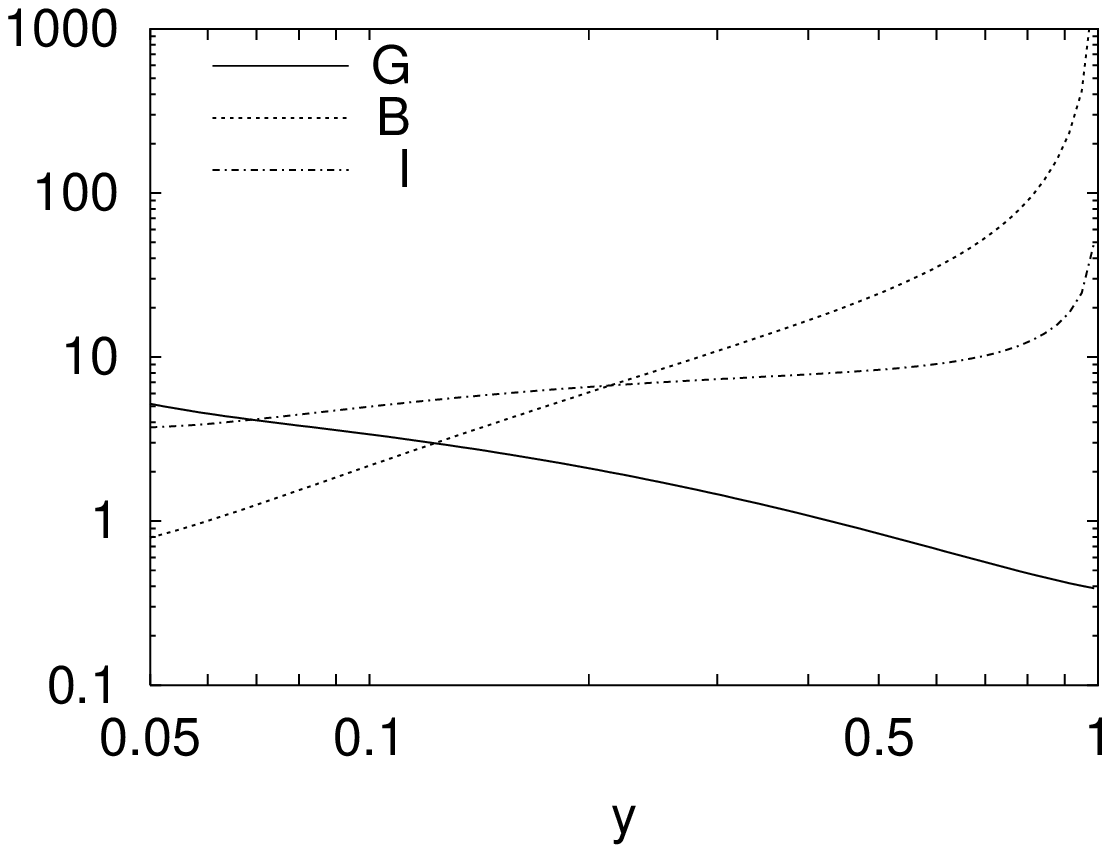}}
      \end{picture}
   \end{center}
\caption{\label{y-dep} The same as Fig.~\protect\ref{Q-dep}, but as a
function of $y$ at $Q^2=5$~GeV$^2$.}
\end{figure}

{}From the factors (\ref{factors}) it follows that the
$\gamma^*\gamma$ contribution decreases faster with $Q^2$ than
bremsstrahlung. On the other hand the $\gamma^*\gamma$ process is
enhanced at large $\epsilon$, whereas bremsstrahlung profits from
small $\epsilon$. To study the amplitudes $A_{ij}$ either in the
$\gamma^*\gamma$ contribution or in the interference term, one will
therefore go to larger values of $\epsilon$, corresponding to small or
intermediate values of $y$ (notice that $\epsilon=0.8$ corresponds to
$y=0.5$). The behavior in $Q^2$ and $y$ of the different contributions
to the $ee$ cross section can be seen in Figs.~\ref{Q-dep} and
\ref{y-dep}, respectively, which have again been obtained with our
model GDA (\ref{model-gda}). Notice that apart from the factors
(\ref{factors}) just discussed, there is a global dependence on $y$
and $Q^2$ through the factor $1/s_{e\gamma}^2$ in the $e\gamma$ cross
section and through the variable $x_2$ in the real photon flux, cf.\
Eqs.~(\ref{y-def}) and (\ref{y-x}).

\begin{figure}
   \begin{center}
      \setlength{\unitlength}{0.5\hsize}
      \begin{picture}(1,0.84)(0,0)
	\put(0.03,0.75){$\displaystyle 2W\,
             \frac{d\sigma_{e^+\gamma\to e^+\,\pi^+\pi^-}}{
                   dQ^2\, dW^2\, d\varphi}$~[fb GeV$^{-3}$]}
        \epsfxsize=0.5\hsize  
	\put(0,0){\epsfbox{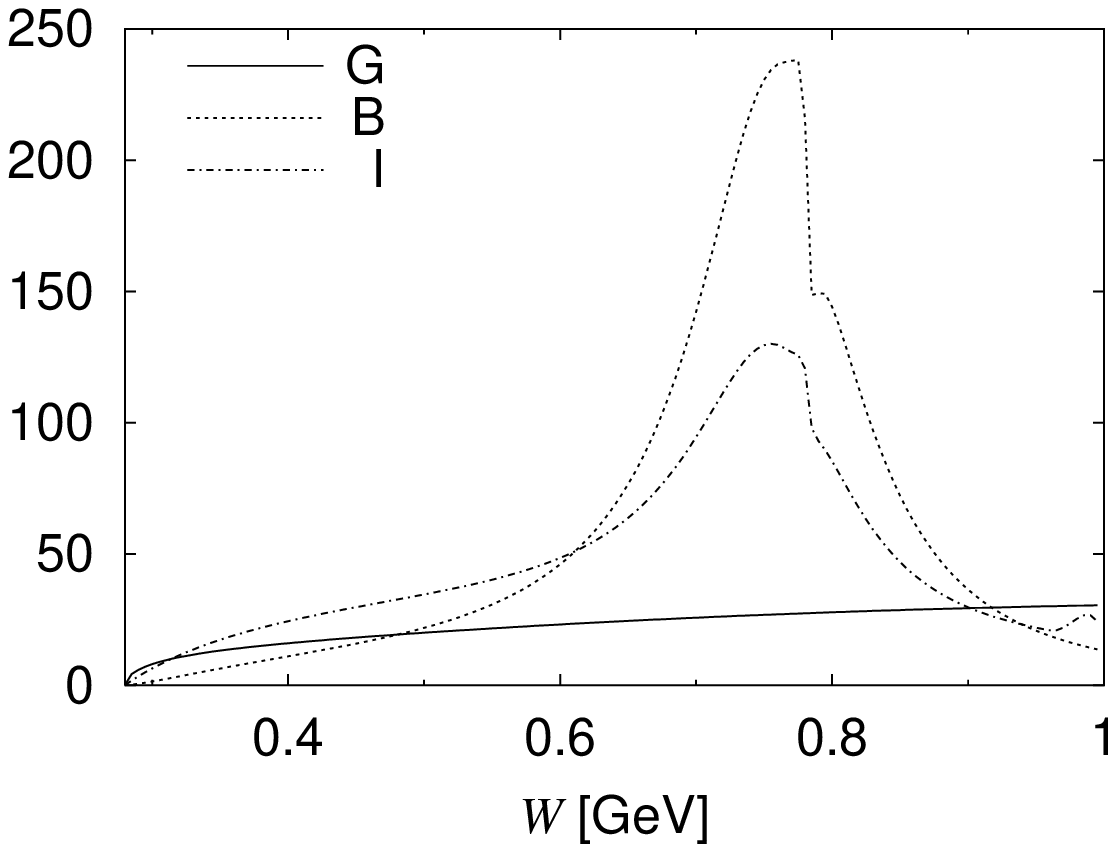}}
      \end{picture}
   \end{center}
\caption{\label{W-dep} The $W$-dependence of the different
contributions to the differential $e^+\gamma$ cross section at
$Q^2=5$~GeV$^2$, $\varphi=0$ and $s_{e\gamma}=50$~GeV$^2$. The
corresponding values of $y$ range from $0.1$ to $0.12$.}
\end{figure}

A very strong effect on the relative weight of the different
contributions comes from the pion form factor $F_\pi(W^2)$. As one can
anticipate from Fig.~\ref{fpi-fig} it leads to a considerable
enhancement of the bremsstrahlung term in a broad $W$ interval around
the $\rho$ mass, thereby also enhancing the interference. The
$W^2$-dependence of the different terms, obtained with of our ansatz
(\ref{model-gda}) for the GDA, are shown in Fig.~\ref{W-dep}. As we
discussed in Sect.~\ref{simple-model} this ansatz most likely
oversimplifies the $W^2$-dependence of the coefficients $B_{10}$ and
$B_{12}$ in $\Phi^+_q$, but the corresponding error in estimating the
$W^2$-behavior of $A_{++}$ should not change the qualitative picture
of Fig.~\ref{W-dep}.

\begin{figure}
   \begin{center}
      \setlength{\unitlength}{0.49\hsize}
      \begin{picture}(1,0.93)(0,0)
	\put(0.5,0.85){\large (a)}
	\put(0.03,0.75){$\displaystyle
             \frac{d\sigma_{e^+\gamma\to e^+\,\pi^+\pi^-}}{
                   dQ^2\, dW^2\, d\varphi}$~[fb GeV$^{-4}$]}
        \epsfxsize=0.49\hsize  
	\put(0,0){\epsfbox{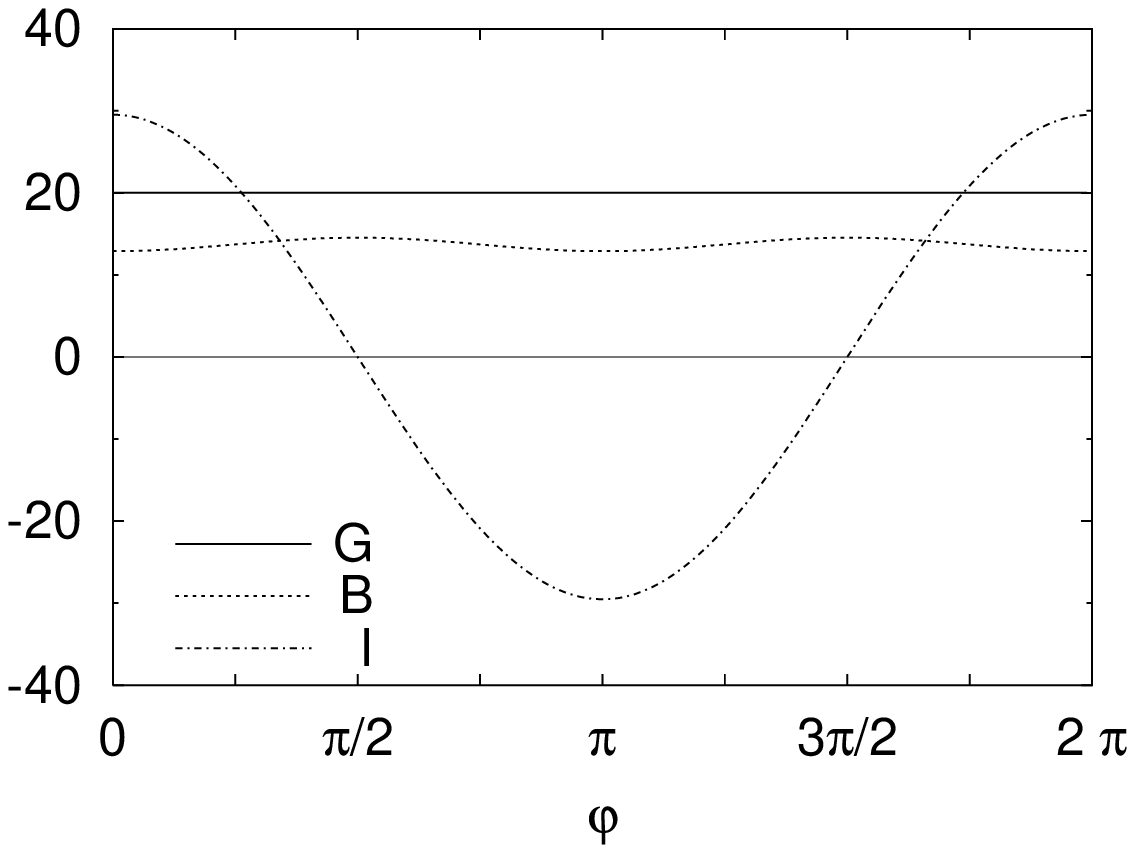}}
      \end{picture}
      \begin{picture}(1,0.93)(0,0)
	\put(0.5,0.85){\large (b)}
 	\put(0.03,0.75){$\displaystyle
             \frac{d\sigma_{e^+\gamma\to e^+\,\pi^+\pi^-}}{
                   dQ^2\, dW^2\, d\varphi}$~[fb GeV$^{-4}$]}
     	\epsfxsize=0.49\hsize  
	\put(0,0){\epsfbox{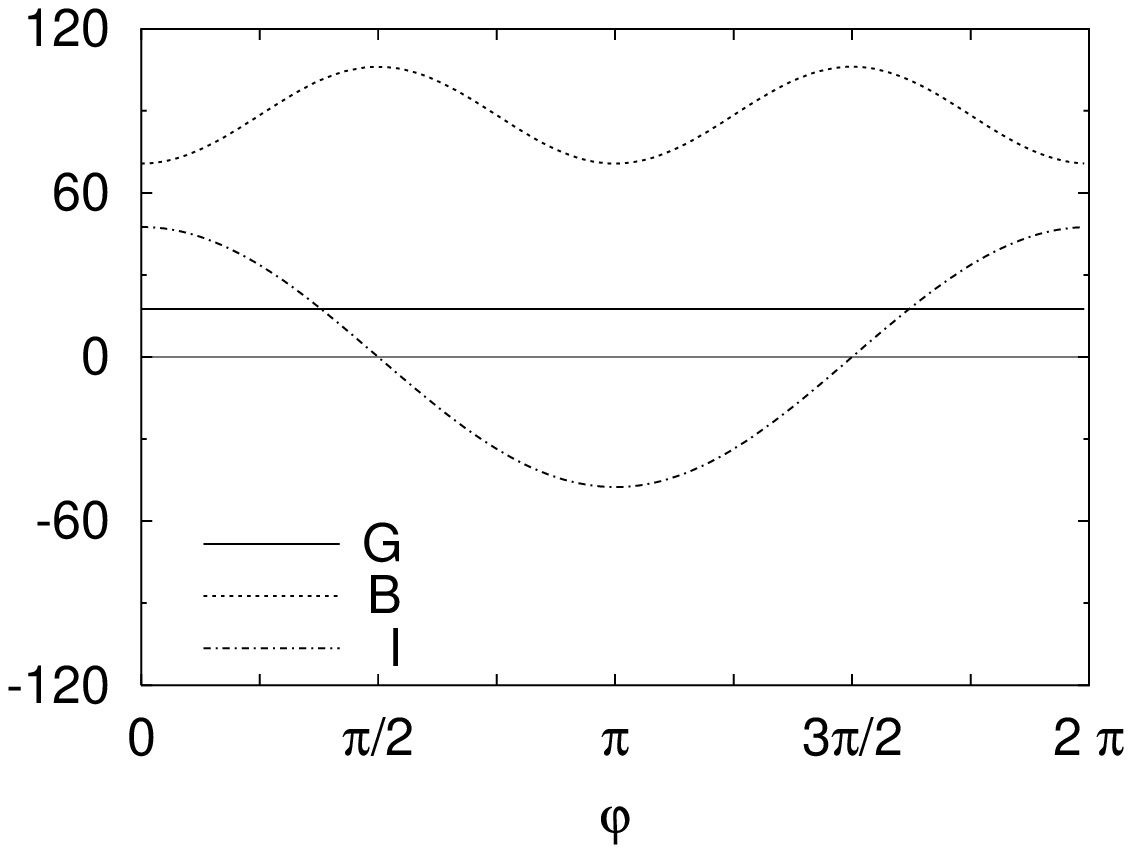}}
      \end{picture}
   \end{center}
\caption{\label{phi-dep} (a) The $\varphi$-dependence of the different
contributions to the differential $e^+\gamma$ cross section at
$Q^2=5$~GeV$^2$, $W=400$~MeV and $y=0.1$. (b) The same for
$W=800$~MeV.}
\end{figure}

\begin{figure}
   \begin{center}
      \setlength{\unitlength}{0.49\hsize}
      \begin{picture}(1,0.93)(0,0)
	\put(0.5,0.85){\large (a)}
	\put(0.03,0.75){$\displaystyle
             \frac{d\sigma_{e^+\gamma\to e^+\,\pi^+\pi^-}}{
             dQ^2\, dW^2\, d\cos(\theta)\, d\varphi}$~[fb GeV$^{-4}$]}
        \epsfxsize=0.49\hsize  
	\put(0,0){\epsfbox{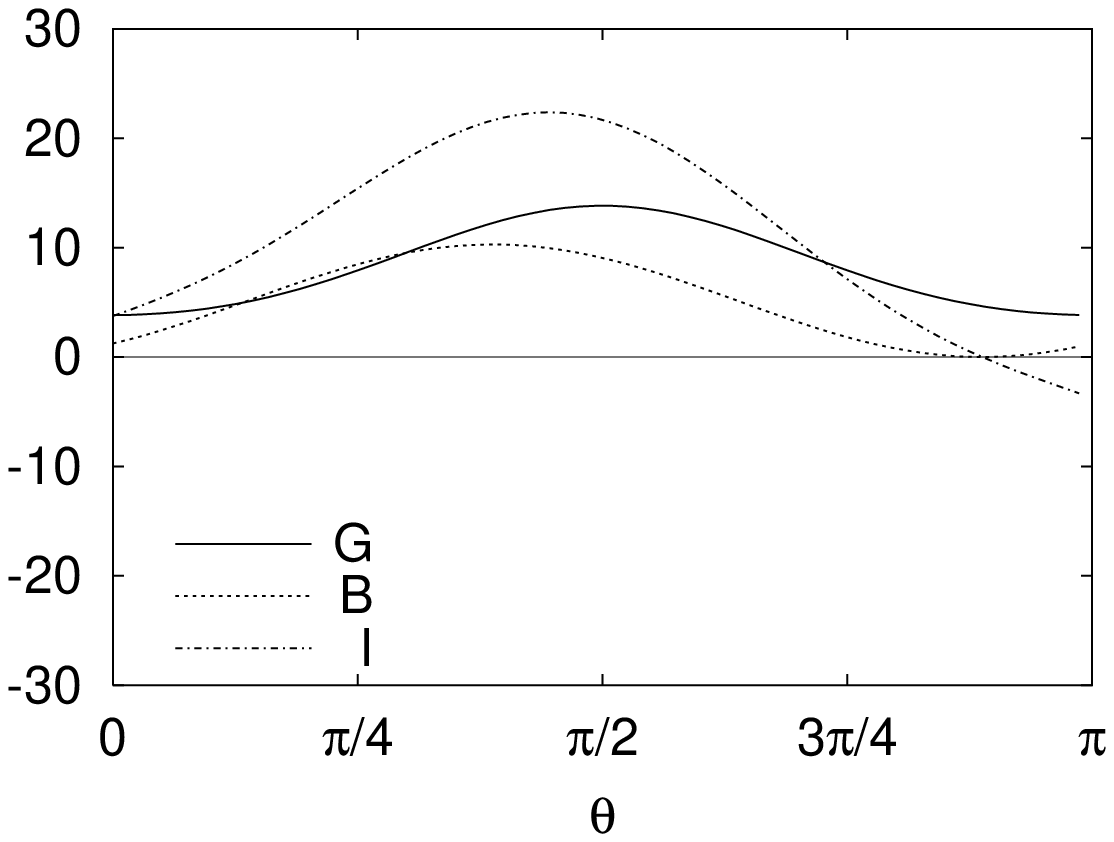}}
      \end{picture}
      \begin{picture}(1,0.93)(0,0)
	\put(0.5,0.85){\large (b)}
 	\put(0.03,0.75){$\displaystyle
             \frac{d\sigma_{e^+\gamma\to e^+\,\pi^+\pi^-}}{
             dQ^2\, dW^2\, d\cos(\theta)\, d\varphi}$~[fb GeV$^{-4}$]}
     	\epsfxsize=0.49\hsize  
	\put(0,0){\epsfbox{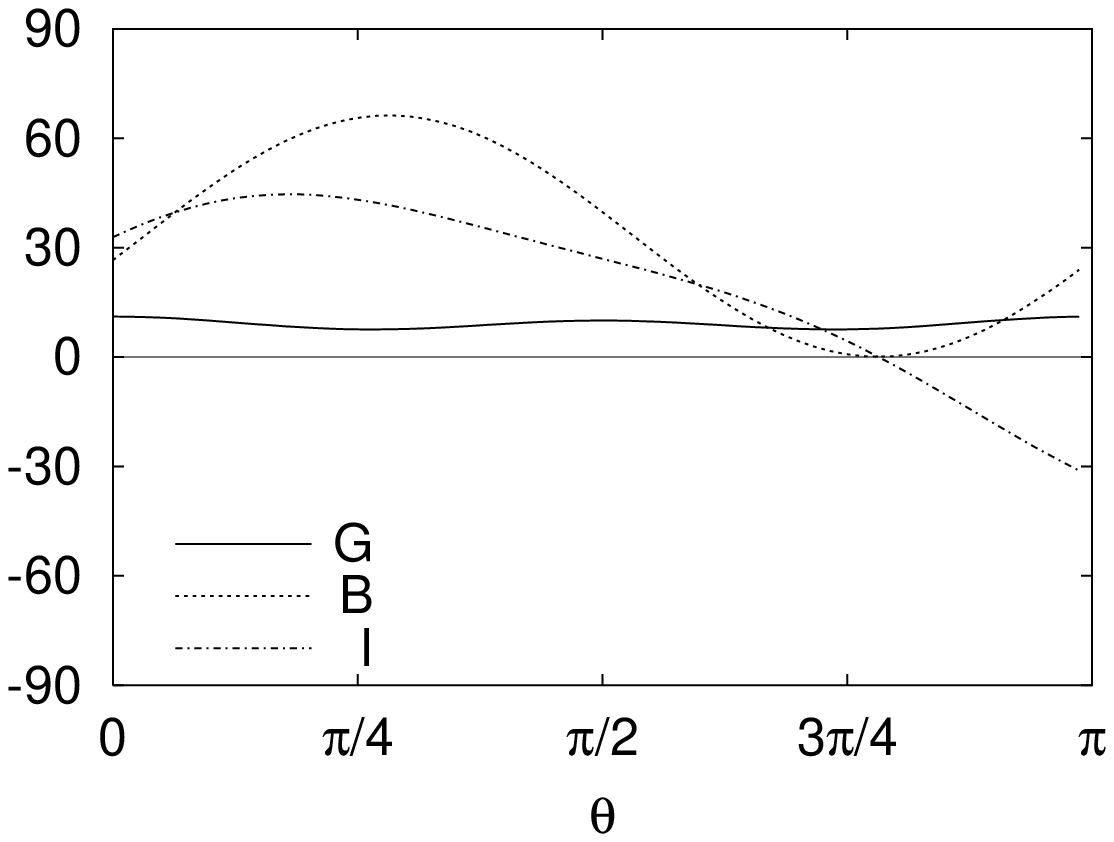}}
      \end{picture}
   \end{center}
\caption{\label{theta-dep} (a) The $\theta$-dependence of the
different contributions to the differential $e^+\gamma$ cross section
at $Q^2=5$~GeV$^2$, $W=400$~MeV, $\varphi=0$ and $y=0.1$. (b) The same
for $W=800$~MeV.}
\end{figure}

In the limit of large $Q^2$ the different contributions to the cross
section have distinctive dependences on $\varphi$. The
$\gamma^*\gamma$ contribution is predicted to be constant in $\varphi$
with a $\cos2\varphi$ modulation due to the product $A_{++}^*
A^{\phantom{*}}_{-+}$. The bremsstrahlung term should be flat, and the
interference between them should be dominated by $\cos\varphi$ and
$\cos3\varphi$, going with $A_{++}$ and $A_{-+}$, respectively. We
show examples of the $\varphi$-behavior in Fig.~\ref{phi-dep},
remembering that in our model $A_{-+}$ is zero because we have
neglected the contribution of the helicity-two gluon GDA. We notice
that the $\cos2\varphi$ term in bremsstrahlung, which is kinematically
suppressed by $1-x \sim W^2 /Q^2$, is clearly visible at the larger
energy $W=800$~MeV. The $\theta$-dependence, shown in
Fig.~\ref{theta-dep}, is also quite different for the three components
of the cross section. For the $\gamma^*\gamma$ term and the
interference it depends in detail on the coefficients of the different
partial waves contributing to the amplitudes $A_{ij}$.

\subsection{Studying the $\gamma^*\gamma$ subprocess through the
interference term}
\label{interference-term}

The interference between the $\gamma^*\gamma$ and bremsstrahlung
subprocesses provides an opportunity to study the $\gamma^*\gamma$
contribution at \emph{amplitude} level. On one hand this means that
one can completely separate the contributions $A_{++}$, $A_{-+}$ and
$A_{0+}$ from different photon polarizations. On the other hand it
gives access to the phases of these amplitudes relative to the phase
of the pion form factor $F_\pi$, which is equal to the $\pi\pi$ phase
shift $\delta_1$ in the range of $W$ we are considering. In
kinematical regions where the bremsstrahlung amplitude is large,
especially for $W$ around the $\rho$ mass peak, the interference can
also be used to ``amplify'' the $\gamma^*\gamma$ signal.

For this to be useful it is essential that one can cleanly separate
the interference term (\ref{interfere}) from the pure $\gamma^*\gamma$
and bremsstrahlung contributions in the cross section. This is
possible since the $\gamma^*\gamma$ collision produces the pion pair
in the $C$-even channel, whereas in bremsstrahlung $\pi\pi$ occurs in
the $C$-odd projection. The interference term can therefore be
separated by reversing the charge of the lepton in the $e\gamma$
collision, a possibility that is automatically provided at $e^+e^-$
colliders. Alternatively, any observable that is odd under exchange of
the $\pi^+$ and $\pi^-$ momenta is only sensitive to the interference
term, which in turn drops out in any observable even under this
exchange. In terms of the variables we are using, this exchange
corresponds to the substitution $(\theta,\varphi)\to
(\pi-\theta,\pi+\varphi)$. This means that we have direct access to
the interference through the angular distribution of the pion pair in
its rest frame. We emphasize that on the experimental level this does
not require a perfect angular measurement, but only that the detection
and reconstruction does not introduce a bias between positive and
negative pions.

{}From the $\varphi$-dependence of the cross section one can extract
the four coefficients $C_n$ in Eq.~(\ref{interfere}), which determine
the three quantities $\Re \{ F_\pi^* A_{++} \}$, $\Re \{ F_\pi^*
A_{0+} \}$ and $\Re \{ F_\pi^* A_{-+} \}$. In fact, they
over-determine them, and one can for instance use only $C_1$, $C_2$,
$C_3$, and keep the information from $C_0$ for a cross check. We
remark in passing that this is owed to the fact that pions have zero
spin, otherwise there would be more helicity amplitudes for the
$\gamma^*\gamma$ reaction than independent observables one can extract
from the $\varphi$-dependence. Using the $\varphi$-moments
(\ref{phi-moments}) with $m=1$, 2, 3 and inverting the relation
between $C_1$, $C_2$, $C_3$ and the helicity amplitudes we obtain
\begin{eqnarray}
 \label{project-amp}
\frac{K}{1- (1-x)(1+\epsilon)}\,
\frac{d S_{e\gamma}(w_{+})}{dQ^2\,dW^2\,d(\cos\theta)} 
+ \Big\{ \theta \leftrightarrow \pi-\theta \Big\}
  &=& 2\, \Re \Big\{ F_\pi^* A_{++} \Big\}\, 
      \sin^3\theta ,
 \nonumber \\
\frac{K}{x\sqrt{\epsilon(1+\epsilon)}}\,
\frac{d S_{e\gamma}(w_{0})}{dQ^2\,dW^2\,d(\cos\theta)}
+ \Big\{ \theta \leftrightarrow \pi-\theta \Big\}
  &=& 2\, \Re \Big\{ F_\pi^* A_{0+} \Big\}\,
      \sin^2\theta\cos\theta ,
 \nonumber \\
\frac{K}{x\epsilon}\,
\frac{d S_{e\gamma}(w_{-})}{dQ^2\,dW^2\,d(\cos\theta)}
+ \Big\{ \theta \leftrightarrow \pi-\theta \Big\}
  &=& 2\, \Re \Big\{ F_\pi^* A_{-+} \Big\}\,
      \sin\theta
\end{eqnarray}
with a global factor 
\begin{equation}
K(Q^2,W^2,\epsilon) = - e_l\, \left(
  \frac{\alpha^3}{8}\, \frac{(\beta x y)^2}{Q^4}\,
  \frac{\sqrt{2}}{\sqrt{W^2 Q^2 \epsilon(1-\epsilon) }} \right)^{-1}
\end{equation}
and weights
\begin{eqnarray}
 \label{half-weights}
w_{+} &=& \sin^2\theta \cos\varphi -
   \sqrt{\frac{2(1-x)}{x}}\, \sqrt{\frac{\epsilon}{1+\epsilon}}\,
   2\cos\theta\sin\theta \cos2\varphi
 + \frac{1-x}{x \epsilon}\,
   (\sin^2\theta + 4\epsilon \cos^2\theta) \cos3\varphi ,
 \nonumber \\
w_{0} &=& {} - \sin\theta\cos\theta \cos2\varphi +
  \sqrt{\frac{2(1-x)}{x}}\, \sqrt{\frac{1+\epsilon}{\epsilon}}\,
  \cos^2\theta \cos3\varphi ,
 \nonumber \\
w_{-} &=& {} - \cos3\varphi . \phantom{\sqrt{\frac{1}{2}}}
\end{eqnarray}
By taking weights that are odd under the exchange of the $\pi^+$ and
$\pi^-$ momenta and summing over configurations with $\theta$ and
$\pi-\theta$ we have canceled the contributions from the pure
$\gamma^*\gamma$ and bremsstrahlung terms in the cross section. We
remark that our method can easily be adapted to the case where one
does not have full acceptance in $\varphi$, since the moments of
$\cos\varphi$, $\cos2\varphi$ and $\cos3\varphi$ are always linear
combinations of $\Re \{ F_\pi^* A_{ij} \}$.

The functions $w_i$ have been chosen such that they are finite,
because the use of unbounded weighting functions is problematic. As a
consequence, the terms $\Re \{ F_\pi^* A_{ij} \}$ on the r.h.s.\ of
Eq.~(\ref{project-amp}) are still multiplied with functions of
$\theta$. One can avoid the rather strong suppression of angles
$\theta$ near 0 or $\pi$ in $\Re \{ F_\pi^* A_{++} \} \sin^3\theta$
if the measurement of the moments (\ref{project-amp}) indicates that
$A_{-+}$ and $A_{0+}$ are small compared with $A_{++}$. In this case
one may replace the weight $w_{+}$ with $\cos\varphi$, whose moment is
dominated by $A_{++} \sin\theta$ with corrections of order
$\sqrt{1-x}\, A_{0+}$ and $(1-x) A_{-+}$. Alternatively, the moment of
\begin{equation}
w'{}_{\!\!+} = \sin\theta \cos\varphi - \sqrt{\frac{2(1-x)}{x}}\,
   \sqrt{\frac{\epsilon}{1+\epsilon}}\, 2\cos\theta \cos2\varphi ,
\end{equation}
projects on $A_{++} \sin^2\theta$ with corrections only of order
$(1-x) A_{-+}$. In a similar way the moment of $\cos\theta
\cos2\varphi$ approximately projects on $A_{0+} \sin\theta \cos\theta$
if $A_{-+}$ is sufficiently small.

The $\theta$-dependence of the moments (\ref{project-amp}) contains
information on the partial wave decomposition of the pion pair. One
way to extract the partial waves is of course to fit the
$\theta$-dependence of the weighted differential cross sections
(\ref{project-amp}). Alternatively, one can use weighted cross
sections integrated over both $\varphi$ and $\theta$. The weight
$\cos3\varphi\, P^2_l(\cos\theta) /\sin\theta$ readily projects out
the $l$th partial wave in $A_{-+}$ as we easily see from
Eq.~(\ref{project-amp}). Note that, since $P^2_l(\cos\theta) \propto
\sin^2\theta$, this weighting function is a trigonometric
polynomial. Similarly, $\cos2\varphi\, P^1_l(\cos\theta)/ \sin\theta$
can be used to obtain the $l$th partial wave in $A_{0+}$ if the
contribution from $A_{-+}$ is small enough.

For $A_{++}$ the situation is more complicated, because the functions
$w_{+} P_l(\cos\theta)/ \sin^3\theta$, $w'{}_{\!\!+} P_l(\cos\theta)/
\sin^2\theta$ and $\cos\varphi\, P_l(\cos\theta)/ \sin\theta$ are all
unbounded. The same problem occurs for the function $w_{0}\,
P^1_l(\cos\theta)/ (\sin^2\theta\, \cos\theta)$. In practice one may
proceed as we discussed in Sect.~\ref{neutral-pi} and restrict the
analysis to a finite number of partial waves, which has to be
determined from the data. Decomposing the coefficient $C_n$ in
(\ref{inter-coefficients}) on polynomials $P^n_{l+1}(\cos\theta)$ one
can see that if only partial waves with $l\le L$ are relevant in the
amplitudes $A_{ij}$, then weighting the cross section with $\cos
n\varphi\, P^n_{L+3}(\cos\theta)$ and integrating over $\varphi$ and
$\theta$ must give zero. For a restricted number of partial waves one
can then find weights to project out the corresponding amplitudes. In
the case where $A_{-+}$ and $A_{0+}$ are negligible and only the
partial waves $l=0$ and $l=2$ are important in $A_{++}$, we have for
instance
\begin{equation}
\frac{K}{1- (1-x)(1+\epsilon)}\,
\frac{d S_{e\gamma}(w_{+l})}{dQ^2\,dW^2}
  = \Re \Big\{ F_\pi^* A_{++l} \Big\} , \hspace{4em} l = 0, 2
\end{equation}
with 
\begin{equation}
 \label{full-weights}
w_{+0} = \frac{4}{3\pi}\, \cos\varphi\, (1 + 2 \cos^2\theta) , \hspace{4em}
w_{+2} = - \frac{16}{3\pi}\, \cos\varphi\, (1 - 4 \cos^2\theta) .
\end{equation}

\begin{figure}
   \begin{center}
      \setlength{\unitlength}{0.49\hsize}
      \begin{picture}(1,0.93)(0,0)
	\put(0.5,0.85){\large (a)}
	\put(0.03,0.75){$\displaystyle 2W
             \frac{dS_{e\gamma}(w)}{dQ^2\, dW^2}$~[fb GeV$^{-3}$]}
	\put(0.36,0.232){$\cos\varphi$}
	\put(0.36,0.192){$w_{+0}$}
	\put(0.36,0.151){$w_{+2}$}
        \epsfxsize=0.49\hsize
	\put(0,0){\epsfbox{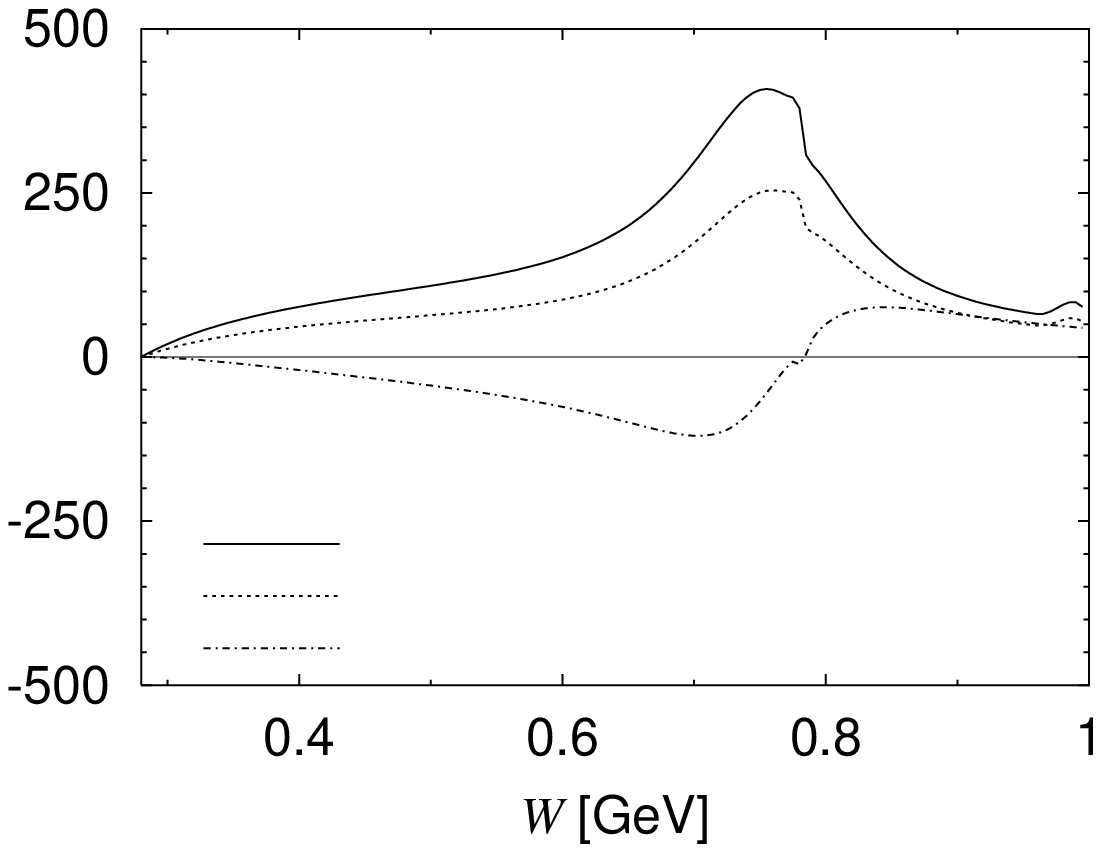}}
      \end{picture}
      \begin{picture}(1,0.93)(0,0)
	\put(0.5,0.85){\large (b)}
	\put(0.03,0.75){$\displaystyle 2W
             \frac{dS_{e\gamma}(w)}{dQ^2\, dW^2}$~[fb GeV$^{-3}$]}
	\put(0.36,0.232){$\cos\varphi$}
	\put(0.36,0.192){$w_{+0}$}
	\put(0.36,0.151){$w_{+2}$}
        \epsfxsize=0.49\hsize
	\put(0,0){\epsfbox{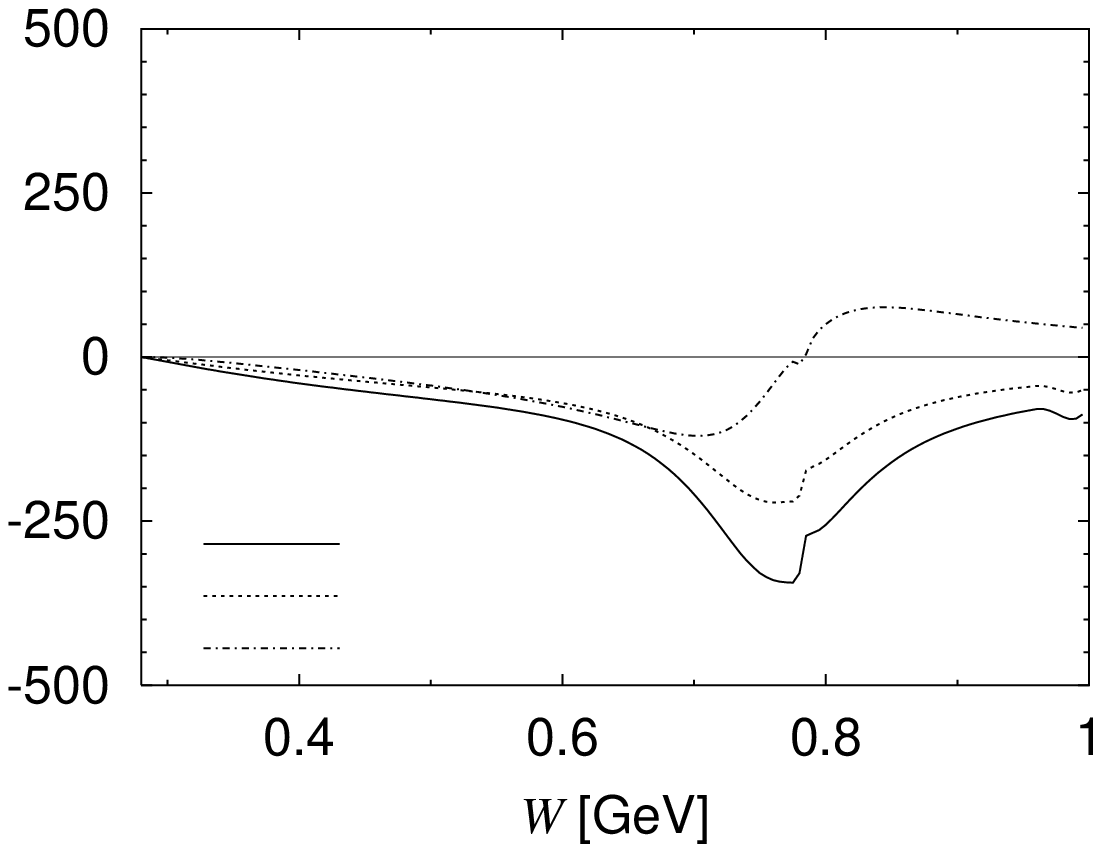}}
      \end{picture}
   \end{center}
\caption{\label{ids} (a) Differential cross sections weighted with
$\cos\varphi$, $w_{+0}$ and $w_{+2}$. The curves are calculated for an
$e^+\gamma$ collision at $s_{e\gamma}=50$~GeV$^2$ and $Q^2=5$~GeV$^2$
with the model GDA in (\protect\ref{model-gda}). (b) The same for the
alternative GDA described at the end of
Sect.~\protect\ref{neutral-pi}.}
\end{figure}

In Fig.~\ref{ids} we show the moments of $\cos\varphi$, $w_{+0}$ and
$w_{+2}$ as a function of $W$ for our model GDA (\ref{model-gda}) and
also for the alternative ansatz described at the end of
Sect.~\protect\ref{neutral-pi}. We clearly see the sensitivity of our
observables to the detailed phase structure of the $\gamma^*\gamma$
amplitude.

\subsection{Comparison with lepton pair production}

In this section we compare our process $e\gamma\to e\, \pi^+\pi^-$
with the production of a muon pair, $e\gamma\to e\, \mu^+\mu^-$, in
the same kinematics. This is interesting in itself because
$\mu^+\mu^-$ production is the QED analogue of the reaction we are
studying, but also because it constitutes an experimental background
to the extent that a muon pair can be misidentified as a pair of
charged pions.

The helicities of the muons can couple to 0 or $\pm 1$ along the
direction of the $\mu^+$ momentum in the $\gamma^*\gamma$ c.m. From
angular momentum conservation in the subprocesses $\gamma^*\gamma\to
\mu^+\mu^-$ and $\gamma^*\to \mu^+\mu^-$ (the latter occurring in
bremsstrahlung) it is clear that the dependence on $\theta$ and
$\varphi$ must be different in the cross sections for pion and for
muon pair production. We therefore restrict ourselves here to the
cross sections integrated over these angles. For the bremsstrahlung
contribution we have
\begin{eqnarray}
 \label{mumu-brems-def}
\left. \frac{d\sigma_{e\gamma\to eX}}{dQ^2\, dW^2}
\right|_{B} =
  \frac{\alpha^3}{3 s_{e\gamma}^2}\, 
  \frac{1 - 2x(1-x)(1-\epsilon)}{\epsilon}\, f^X_B(W^2) ,
\end{eqnarray}
where
\begin{equation}
 \label{mumu-brems}
f^{\pi^+\pi^-}_B = \frac{\beta^3 |F_\pi(W^2)|^2}{W^2} , \hspace{3em}
f^{\mu^+\mu^-}_B = \frac{2 \beta_\mu (3 - \beta_\mu^2)}{W^2}
\end{equation}
with the muon velocity $\beta_{\mu} = (1 - 4m_{\mu}^2/W^2)^{1/2}$ in
the $\gamma^*\gamma$ c.m. For the $\gamma^*\gamma$ process we can
easily adapt the result (\ref{structure-limit}) for open
$q\bar{q}$-production to the $\mu^+\mu^-$ case and find
\begin{equation}
 \label{mumu-gamma-def}
\left. \frac{d\sigma_{e\gamma\to eX}}{dQ^2\, dW^2}
\right|_{G} =
  \frac{\alpha^3}{4 s_{e\gamma}^2}\, \frac{1}{Q^2 (1-\epsilon)}\,
  f^X_G(W^2) ,
\end{equation}
where
\begin{equation}
 \label{mumu-gamma}
f^{\pi^+\pi^-}_G = \left( \frac{25 R_\pi}{18} \right)^2\, \beta
  \left( 1 - \frac{2}{3} \beta^2 + \frac{1}{5} \beta^4 \right) ,
\hspace{3em}
f^{\mu^+\mu^-}_G = 8 \left(
  \ln\frac{1+\beta_{\mu}}{1-\beta_{\mu}} - \beta_{\mu} \right) ,
\end{equation}
up to corrections of order $W^2 /Q^2$. Notice that both for
bremsstrahlung and for $\gamma^*\gamma$, the $Q^2$-dependence is the
same in the pion and the muon case.

\begin{figure}
   \begin{center}
      \setlength{\unitlength}{0.49\hsize}
      \begin{picture}(1,0.85)(0,0)
	\put(0.5,0.78){\large (a)}
	\put(0.05,0.72){$2W f_B^X$~[GeV$^{-1}$]}
	\put(0.58,0.59){$f_{B}^{\pi\pi}$}
	\put(0.26,0.29){$f_{B}^{\mu\mu}$}
        \epsfxsize=0.49\hsize
	\put(0,0){\epsfbox{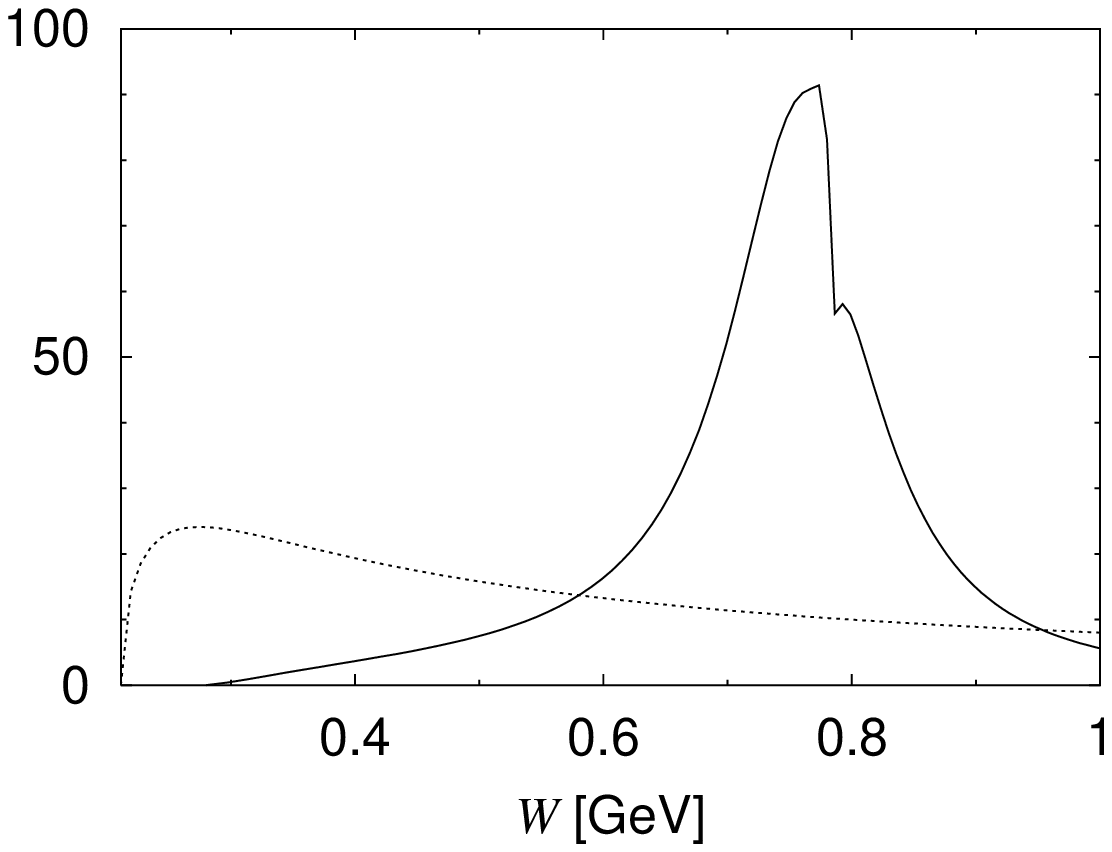}}      
      \end{picture}
      \begin{picture}(1,0.85)(0,0)
	\put(0.5,0.78){\large (b)}
	\put(0.05,0.72){$2W f_G^X$~[GeV]}
	\put(0.2,0.37){$100\times f_{G}^{\pi\pi}$}
	\put(0.49,0.25){$f_{G}^{\mu\mu}$}
        \epsfxsize=0.49\hsize  	
	\put(0,0){\epsfbox{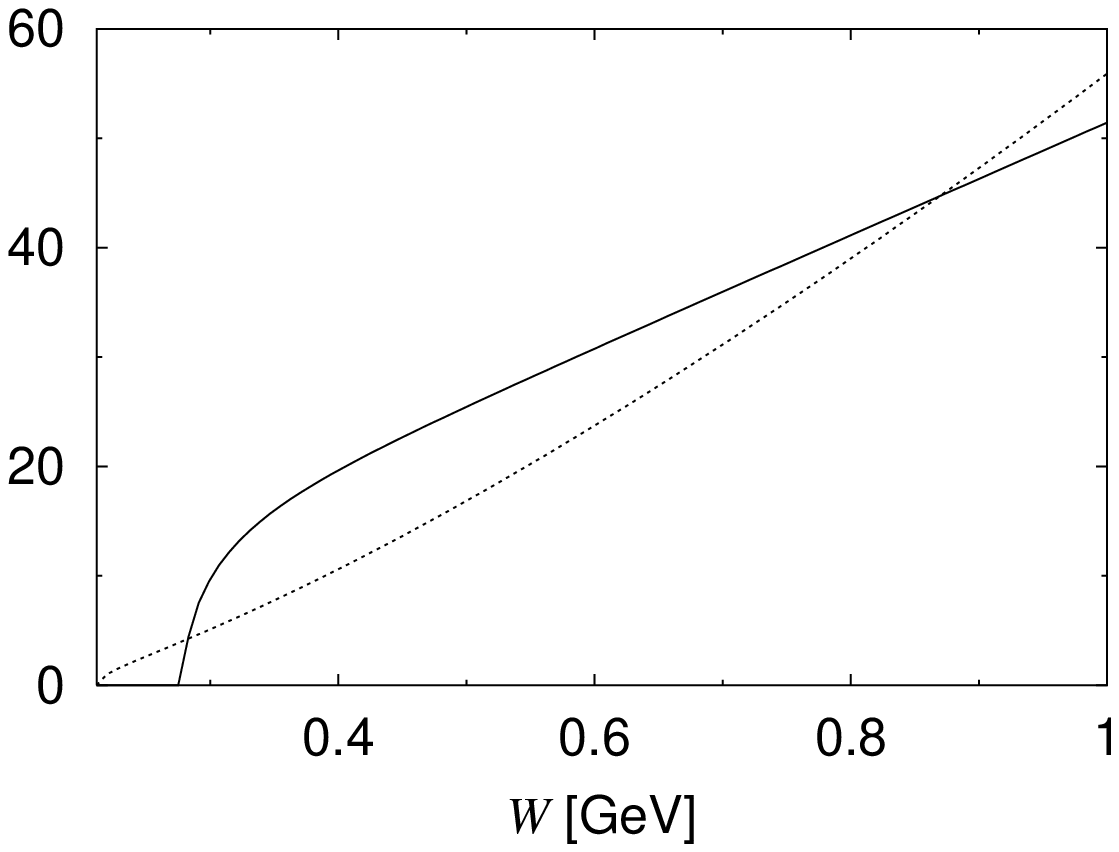}}
      \end{picture}
   \end{center}
\caption{\label{mumu-fig} (a) The functions $f^X_B$ occurring in the
bremsstrahlung contribution (\protect\ref{mumu-brems-def}) to
$\pi^+\pi^-$ and $\mu^+\mu^-$ production. They are plotted against $W$
instead of $W^2$ and therefore have been multiplied with a Jacobian
$2W$. (b) The same for the functions $f^X_G$ in the $\gamma^*\gamma$
contribution (\protect\ref{mumu-gamma-def}). Note that the curve for
pions, obtained with our model GDA (\protect\ref{model-gda}), has been
multiplied by a factor 100.}
\end{figure}

The functions $f^X_B$ and $f^X_G$ are compared in
Fig.~\ref{mumu-fig}. We see that for the bremsstrahlung contribution
pion production is enhanced by the strong resonance effect around the
$\rho$ mass, as manifested in $F_\pi(W^2)$. In the $\gamma^*\gamma$
subprocess, on the other hand, we find that with our estimate of the
GDA, pion production is suppressed compared to muons by a factor 50 to
100. This is mostly due to the numerical constants in the expressions
(\ref{mumu-gamma}). In part it also comes from the logarithm
$\log(1-\beta_\mu)$ in $f^{\mu^+\mu^-}_G$, which is generated by the
collinear regions around $\theta=0$ and $\pi$ as discussed in
Sect.~\ref{structure-function}. Notice that for this reason the
$\mu^+\mu^-$ cross section will be relatively sensitive to cuts that
affect $\theta$. The same will apply to the interference between
bremsstrahlung and $\gamma^*\gamma$, which drops of course out after
angular integration. From the results on $f^X_B$ and $f^X_G$ we expect
that the ratio of muon to pion pair production will be appreciable in
the interference term.

Another experimental background, again due to particle
misidentification, is $e^\pm \gamma\to e^\pm\, e^+e^-$. Compared with
$\mu^+\mu^-$ production there are further Feynman diagrams, which can
be obtained from the muon case by interchanging the lines with momenta
$k'$ and either $p$ or $p'$, now corresponding to identical
particles. We shall not analyze these diagrams here, but will at least
assess the contributions from those diagrams that are also present in
muon production. Replacing $\beta_\mu$ with $\beta_e$ we obtain
velocities extremely close to~1. Nothing dramatic happens in the
bremsstrahlung part (\ref{mumu-brems}), but the logarithm in the
$\gamma^*\gamma$ subprocess (\ref{mumu-gamma}) is now much larger than
for muons. This large logarithm is however generated by transverse
momenta $p_\perp$ of order $m_e$ in the $\gamma^*\gamma$ c.m., which
correspond to extremely small angles $\theta$ of order $m_e/W$. For
any cut that effectively leads to a minimum angle
$\theta_{\mathit{min}}$ much larger than that, one has to replace
$\beta_\mu$ with $\cos\theta_{\mathit{min}}$ in
Eq.~(\ref{mumu-gamma}), which can significantly reduce the size of the
logarithm.

We finally note that the differential cross sections for $e^+e^- \to
e^+e^-\, e^+e^-$ and $e^+e^- \to e^+e^-\, \mu^+\mu^-$ have been fully
calculated to first order in QED and are available in the form of
Monte Carlo generators~\cite{BDK}.

\section{Cross section estimates}
\label{cross-sections}

\subsection{Laboratory kinematics and experimental cuts}
\label{lab-kin}

Before giving our estimates for the cross section of our process at
various $e^+e^-$ colliders, we give a brief discussion of the
kinematics in the laboratory frame and the effects of some
experimental cuts. Starting with the kinematics of the scattered
lepton $k'$, we remark that there is a simple transformation between
the variables $(Q^2, y)$ and $(E_1{}^{\!\!'}, \alpha_{1L})$, where
$E_1{}^{\!\!'}$ and $\alpha_{1L}$ respectively are the energy and
scattering angle of $k'$ in the laboratory frame. Imposing minimum
values on both quantities we have
\begin{equation}
 y = 1 + \frac{Q^2}{4 E_1^2} - \frac{E'_1}{E_1}
 \le 1 + \frac{Q^2}{4 E_1^2} - \frac{E'^{\mathit min}_1}{E_1}
 \label{y-max}
\end{equation}
and
\begin{equation}
 y = 1 - \frac{Q^2}{4 E_1^2} \frac{1+\cos\alpha_{1L}}{1-\cos\alpha_{1L}}
 \ge 1 - \frac{Q^2}{4 E_1^2} \frac{1+
     \cos\alpha_{1L}^{\mathit min}}{1-\cos\alpha_{1L}^{\mathit min}} .
 \label{y-min}
\end{equation}
The condition (\ref{y-max}) cuts on large values of $y$ and is
generally not very serious, because most information on the
$\gamma^*\gamma$ process is obtained from low or intermediate $y$ as
we discussed after Eq.~(\ref{factors}). The lower cut (\ref{y-min}),
on the other hand, severely restricts the interesting $y$-range in
some experimental setups if $Q^2$ is not large enough. We will
encounter an example of this in Sect.~\ref{b-factories}.

\begin{figure}
   \begin{center}
      \setlength{\unitlength}{0.49\hsize}
      \begin{picture}(1,0.85)(0,0)
	\put(0.5,0.78){\large (a)}
	\put(0.04,0.72){$p_L^0$~[GeV]}
        \epsfxsize=0.49\hsize
        \put(0,0){\epsfbox{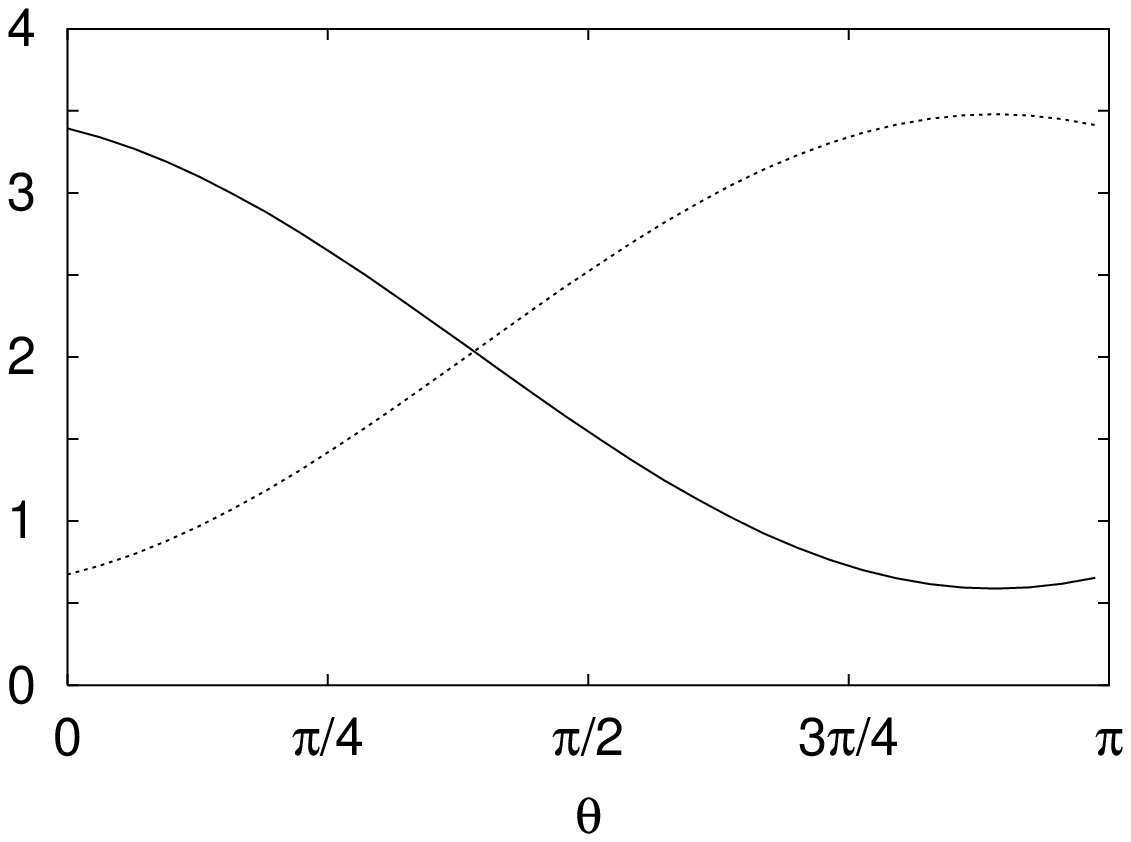}}
      \end{picture}
      \begin{picture}(1,0.85)(0,0)
	\put(0.5,0.78){\large (b)}
	\put(0.05,0.72){$p_{\perp L}^{\protect\phantom{'}}$~[GeV]}
        \epsfxsize=0.49\hsize
        \put(0,0){\epsfbox{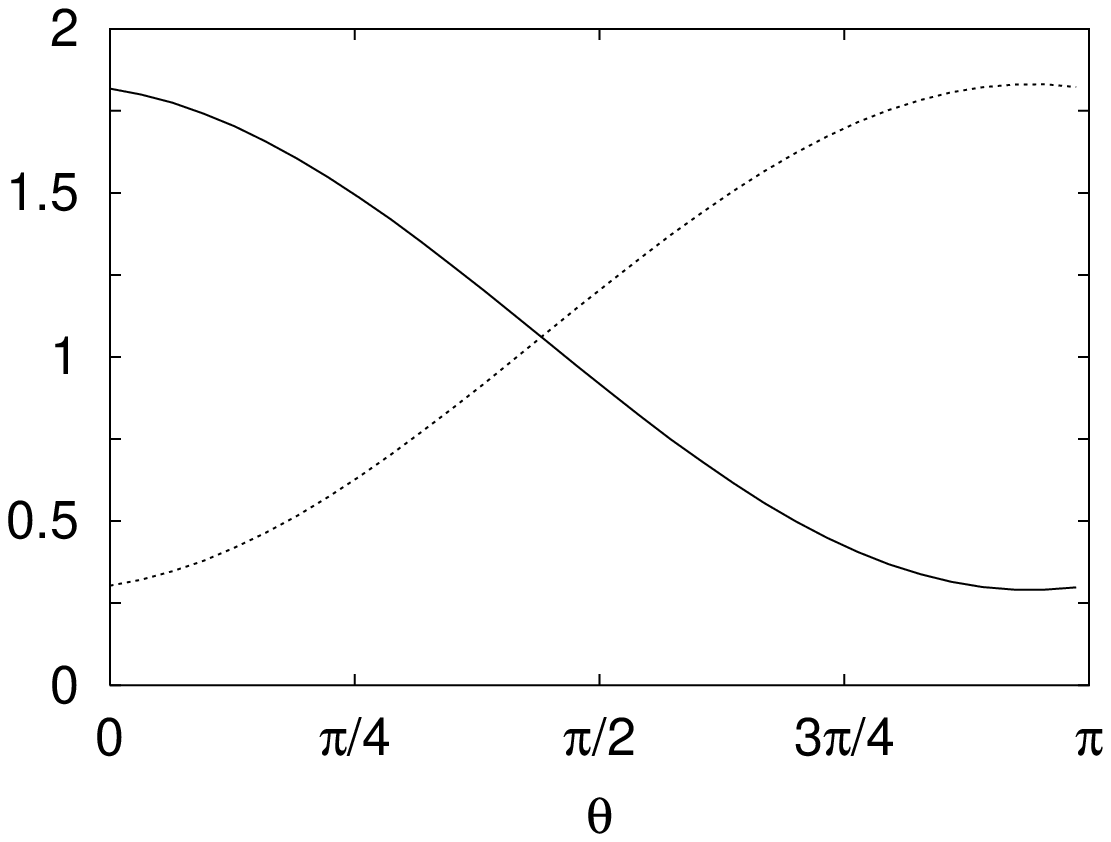}}
      \end{picture}
   \end{center}
\caption{\label{lab-theta} (a) The pion energies $p^0_L$ (solid) and
$p'^0_L$ (dotted) in the laboratory as a function of the angle
$\theta$ in the two-pion c.m. The values of the remaining kinematical
variables are $E_1=3.1$~GeV, $E_2=9$~GeV, $Q^2=5$~GeV$^2$,
$W=400$~MeV, $y=0.1$, $\varphi=0$. (b) The same for the transverse
pion momenta $p_{\perp L}^{\protect\phantom{'}}$ and $p'_{\perp L}$ in
the laboratory.}
\end{figure}

The transformation of the pion momenta into the laboratory system
leads to rather lengthy expressions, which we will not give
here. Notice that the lepton $k'$ has a large transverse momentum
$k'_{\perp L} = Q \sqrt{1-y}$ in the laboratory, which must be
compensated by the two pions. Even though the $\pi\pi$ system has a
rather low invariant mass, the pions thus carry large transverse
momentum which helps to detect them. An exception are configurations
with the the c.m.\ angle $\theta$ close to 0 or $\pi$, which in the
laboratory correspond to an asymmetric sharing of momentum between the
two pions. This is illustrated in Fig.~\ref{lab-theta}.

It is instructive to consider the point where there the $\pi\pi$
system has zero longitudinal momentum $P_{L}^3$ in the
laboratory. With the approximation $W^2 \ll Q^2$ we find
\begin{equation}
P_{L}^3 = y E_1 - \frac{1-y}{y}\, \frac{Q^2}{4 E_1} ,
\end{equation}
so that $P_{L}^3=0$ when $y$ equals
\begin{equation}
 \label{y-zero}
y_0 = \frac{Q}{2 E_1} \left( \sqrt{1 + \frac{Q^2}{16 E_1^2}} 
                           - \frac{Q}{4 E_1} \right) .
\end{equation}
For $Q \ll E_1$ this simplifies to $y_0 = Q /(2 E_1)$. If $y$ is very
different from $y_0$ the $\pi\pi$ system is strongly boosted along the
beam axis, and if this boost is too large then one or both pions will
go out of the detector acceptance.

We finally have to discuss the kinematics of the scattered lepton $l'$
in the laboratory. In terms of its scattering angle $\alpha_{2L}$ we
have, up to electron mass corrections,
\begin{equation}
 \label{transverse-mom}
l'_{\perp L} = (1-x_2)\, E_2\, \sin\alpha_{2L}
\end{equation}
for the transverse component of $l'$, and
\begin{equation}
 \label{photon-vir}
Q'^2 = -q'^2 = 
  (1-x_2)\, E_2^2\, \left( 2 \sin\frac{\alpha_{2L}}{2} \right)^2
\end{equation}
for the photon virtuality. For small $\alpha_{2L}$ we obtain the
simple relation
\begin{equation}
 \label{approx-vir} 
Q'^2 = \frac{l'^2_{\perp L}}{1-x_2} .
\end{equation}
It turns out that an antitagging condition on the lepton $l'$, i.e.,
$\alpha_{2L}^{\phantom{m}} \le \alpha_{2L}^{\mathit max}$ with
$\alpha_{2L}^{\mathit max}$ determined by the acceptance of a lepton
in the detector, is not enough to keep $Q'^2$ small. With the
parameters $E_2$ and $\alpha_{2L}^{\mathit max}$ in Tables
\ref{b-factories-uncut} and \ref{lep} we find that, except in the
region of $x_2$ very close to~1, the maximum values of $Q'^2$ and
$l'^2_{\perp L}$ are a few GeV$^2$. Under such circumstances it is
clearly inappropriate to approximate $q'^2$ as zero and the momenta
$l$, $l'$, $q'$ as collinear, which we have done throughout this
work. Both the kinematical transformation from the $e\gamma$ frame to
the laboratory and the calculation of the cross section have to be
modified then. One must not only
recalculate the two-photon and bremsstrahlung processes of
Fig.~\ref{brems-fig} but also include further diagrams contributing to
the reaction $e^+e^-\to e^+e^-\, \pi\pi$. Although this
is possible in principle, we wish to retain here the simpler
expressions for the cross section with one real photon. We therefore
require that $Q'^2$ be small compared with the other kinematical
invariants in our problem.

A way to achieve this, suggested by Eq.~(\ref{approx-vir}), is to
impose an upper cut on $l'_{\perp L}$, i.e., in practical terms on the
sum $| {\mathbf k}'_{\perp L} + {\mathbf p}^{\phantom{'}}_{\perp L} +
{\mathbf p}'_{\perp L}|$ of the reconstructed transverse momenta,
possibly supplemented by a lower cut on $1-x_2$. In our numerical
studies we determine the maximum virtuality $Q'^2_{{\mathit{max}}}$ in
the photon flux of Eq.~(\ref{EPA}) through Eqs.~(\ref{transverse-mom})
and (\ref{photon-vir}) by requiring both $\alpha_{2L}^{\phantom{m}}
\le \alpha_{2L}^{\mathit max}$ and $l'_{\perp L} \le
l'^{{\mathit{max}}}_{\perp L} = 100$~MeV. This leads to considerably
smaller virtualities than the antitagging condition alone, although
for $x_2$ very close to~1 the resulting $Q'^2_{{\mathit{max}}}$ is
still not very much smaller than $W^2$. In practice one may therefore
consider an additional cut on $x_2$, but we have refrained from this
in our estimates. Notice that the $Q'^2$-spectrum of the photon flux
is logarithmic so that a substantial part of the cross section comes
from $Q'^2$ much smaller than $Q'^2_{{\mathit{max}}}$.

\subsection{$B$-factories}
\label{b-factories}

We have now all elements to give cross section estimates for existing
$e^+e^-$ facilities. We start with the $B$-factories, BABAR, BELLE and
CLEO, running at a c.m.\ energy $\sqrt{s_{ee}}$ around 10~GeV. Using
our model GDA (\ref{model-gda}) we calculate the integrated cross
section $\sigma$ and the individual contributions $\sigma_G$ and
$\sigma_B$ from the $\gamma^*\gamma$ and bremsstrahlung
subprocesses. To project out their interference term we take simple
examples of weighted $e^+e^-$ cross sections,
$S_{ee}(\mathrm{sgn}(\cos\varphi))$ and $S_{ee}(\cos\varphi)$, defined
in complete analogy with the weighted $e\gamma$ cross sections
(\ref{weight-def}). We remark that $S_{ee}(\mathrm{sgn}(\cos\varphi))$
is simply the left-right asymmetry of the pions in their c.m. We
integrate over $y$ from its lower kinematical limit
\begin{equation}
 \label{y-limit}
y \ge \frac{Q^2 + W^2}{4 E_1 E_2}
\end{equation}
up to $y=0.5$. Choosing a larger value increases the cross section,
but the gain is mainly due to bremsstrahlung. Up to which values of
$y$ one can extract useful information on the $\gamma^*\gamma$ process
depends of course on the detailed kinematics and must be studied in
each particular case. The same is true for the upper limit of the
$Q^2$-integration. For its lower limit we take 4~GeV$^2$ as a minimum
value where one might expect a lowest-order calculation to be
reliable, cf.\ our discussion in Sect.~\ref{one-pion}. To determine
the value of $Q'^2_{{\mathit{max}}}$ in the equivalent photon flux we
impose the cuts discussed at the end of Sect.~\ref{lab-kin}. Our
results for $e^+e^- \to e^+e^-\, \pi^+\pi^-$ are given in
Table~\ref{b-factories-uncut}, where apart from the quantities just
discussed we also give the coefficients in the relative statistical
errors $\delta(w) \sim {\mathrm{const}} / \sqrt{N}$ of the weighted
cross sections $S_{ee}(w)$. We see that the results for the different
kinematical situations are practically identical. This indicates that
it is the cut $l'_{\perp L} \le 100$~MeV which determines the real
photon flux in most of the relevant parameter space, and not the cut
on $\alpha_{2L}$, which is different in each of the five cases. We
also find that $S_{ee}(\cos\varphi)$ has a slightly smaller relative
statistical error than $S_{ee}(\mathrm{sgn}(\cos\varphi))$ and thus
greater sensitivity to the interference term.

\begin{table}
%
%
\begin{center}
\begin{tabular}{lrrrrr}
 & BABAR & BABAR & BELLE & BELLE & CLEO \\
 & $e^-$ tagged & $e^+$ tagged & $e^-$ tagged & $e^+$ tagged & \\
$E_1$ [GeV] & 9 & 3.1 & 8 & 3.5 & 5.3 \\
$E_2$ [GeV] & 3.1 & 9 & 3.5 & 8 & 5.3 \\
$\alpha_{2L}^{\mathit max}$ [mrad] \rule[-1ex]{0ex}{1ex} 
                                   & 684 & 300 & 154 & 112 & 227 \\
\hline
$\sigma$     [fb] & 452 & 452 & 452 & 452 & 453 \\
$\sigma_{G}$ [fb] &  15 &  15 &  14 &  15 &  15 \\
$\sigma_{B}$ [fb] & 437 & 438 & 437 & 438 & 438 \\
$S_{ee}(\mathrm{sgn}(\cos\varphi))$ [fb] & $-51$ & 52 & $-51$ & 51 & 51 \\
$S_{ee}(\cos\varphi)$               [fb] & $-40$ & 40 & $-40$ & 40 & 41 \\
$\sqrt{N}\, \delta(\mathrm{sgn}(\cos\varphi))$ &
  8.8 & 8.8 & 8.8 & 8.8 & 8.8 \\
$\sqrt{N}\, \delta(\cos\varphi)$ &
  7.7 & 7.7 & 7.7 & 7.7 & 7.7 \\
\end{tabular}
\caption{\label{b-factories-uncut} Cross sections for $e^+e^- \to
e^+e^-\, \pi^+\pi^-$, integrated over the range $W = 300$~MeV to 1000
MeV, $Q^2 = 4$~GeV$^2$ to $20$~GeV$^2$, and $y$ from its lower
kinematical limit (\protect\ref{y-limit}) up to~0.5. The cut
parameters $\alpha_{2L}^{\mathit max}$ and $l'^{{\mathit{max}}}_{\perp
L}=100$~MeV determine the real photon flux as described in
Sect.~\protect\ref{lab-kin}. In the column for CLEO, the sign of the
weighted cross sections $S_{ee}(\mathrm{sgn}(\cos\varphi))$ and
$S_{ee}(\cos\varphi)$ corresponds to a tagged $e^+$.}
\vspace{2\baselineskip}
\begin{tabular}{lrrrrr}
 & BABAR & BABAR & BELLE & BELLE & CLEO \\
 & $e^-$ tagged & $e^+$ tagged & $e^-$ tagged & $e^+$ tagged & \\
$\alpha_{1L}^{\mathit min}$ [mrad]       & 300 & 684 & 112 & 154 & 227 \\
$(\pi-\alpha_{1L}^{\mathit max})$ [mrad] & 684 & 300 & 154 & 112 & 227 \\
$\theta_{L}^{\mathit min}$ [mrad]        & 300 & 684 & 297 & 524 & 314 \\
$(\pi-\theta_{L}^{\mathit max})$ [mrad]  \rule[-1ex]{0ex}{1ex} 
                                         & 684 & 300 & 524 & 297 & 314 \\
\hline
$\sigma$     [fb] & 329 & 433 & 433 & 443 & 446 \\
$\sigma_{G}$ [fb] &   6 &  12 &  13 &  13 &  14 \\
$\sigma_{B}$ [fb] & 323 & 422 & 420 & 430 & 433 \\
$S_{ee}(\mathrm{sgn}(\cos\varphi))$ [fb] & $-31$ & 48 & $-50$ & 51 & 52 \\
$S_{ee}(\cos\varphi)$               [fb] & $-24$ & 38 & $-39$ & 40 & 41 \\
$\sqrt{N}\, \delta(\mathrm{sgn}(\cos\varphi))$ & 
 10.5 & 8.9 & 8.7 & 8.7 & 8.6 \\
$\sqrt{N}\, \delta(\cos\varphi)$ &
  9.0 & 7.8 & 7.6 & 7.6 & 7.5 \\
\end{tabular}
\caption{\label{b-factories-cut} As Table
\protect\ref{b-factories-uncut} but with cuts imposed on the detection
angles as specified, and in addition a minimum transverse momentum for
the tagged lepton and for both pions of 100~MeV in the laboratory.
$E_1$, $E_2$ and $\alpha_{2L}^{\mathit max}$ for each column are the
same as in Table~\protect\ref{b-factories-uncut}. }
\end{center}
\end{table}

To estimate the effects of experimental acceptance for the detected
particles we impose
\begin{itemize}
\item a cut $\alpha_{1L}^{\mathit min} \le \alpha_{1L}^{\phantom{m}}
\le \alpha_{1L}^{\mathit max}$ on the scattering angle $\alpha_{1L}$
of the tagged lepton $k'$,
\item a cut $\theta_{L}^{\mathit min} \le (\theta_{L}^{\phantom{m}},
\theta'_{L}) \le \theta_{L}^{\mathit max}$ on the polar angles
$\theta^{\phantom{m}}_{L}$ and $\theta'_{L}$ of the pion momenta $p$
and $p'$, measured with respect to the direction of the initial beam
lepton $k$,
\item a minimum transverse momentum of 100~MeV for the tagged lepton
and for each of the pions.
\end{itemize}
All quantities refer of course to the laboratory frame. The results
are shown in Table~\ref{b-factories-cut}.

Comparing with Table~\ref{b-factories-uncut} we see that the effects
of these cuts are generally quite moderate. The strongest effect is
observed for BABAR kinematics in the case where the $e^-$ is
tagged. This can be traced back to the constraint
$\alpha_{1L}^{\mathit min} \le \alpha_{1L}^{\phantom{m}}$. The minimum
value of $y$ implied by Eq.~(\ref{y-min}) for $Q^2 = 4$~GeV$^2$ is
0.46 in this case, which effectively cuts away all phase space where
the $\gamma^*\gamma$ process is relevant. The situation improves
rapidly as $Q^2$ goes up, and for $Q^2=6$~GeV$^2$ our cut implies
$y\ge 0.19$. For the other experimental configurations the same cut is
much less restrictive: for BABAR kinematics with a tagged $e^+$ our
cut on $\alpha_{1L}$ implies $y\ge 0.18$ at $Q^2=4$~GeV$^2$, whereas
in the cases of BELLE and CLEO there is not restriction on $y$ from
the inequality (\ref{y-min}) at all, not even at $Q^2=4$~GeV$^2$.

We find that in the kinematics of $B$-factories the interference term
is clearly larger than the contribution from $\gamma^*\gamma$ alone.
With several 10~fb$^{-1}$ integrated luminosity our estimated cross
sections give event rates of order 10,000. As we see from the tables,
the relative statistical error on the interference term, extracted
through the moments $S_{ee}(\mathrm{sgn}(\cos\varphi))$ or
$S_{ee}(\cos\varphi)$ is about 8 to 10 times larger than for
integrated cross sections (where it is $1/ \sqrt{N}$), so that the
interference could be measured with statistical errors in the 10\%
range.

For the production of neutral pion pairs we easily obtain the cross
section without cuts by multiplying $\sigma_G$ in
Table~\ref{b-factories-uncut} with a factor 1/2, due to the restricted
phase space of identical particles. We refrain from a discussion of
the experimental reconstruction of the four-photon state coming from
two pion decays, but for an order-of-magnitude indication of event
rates one may take half of the cross sections $\sigma_G$ in
Table~\ref{b-factories-cut}. We then estimate hundreds of events with
several 10~fb$^{-1}$, corresponding again to a statistical error
around 10\%. Thus studies of both charged and neutral pair production
seem promising to us.

\subsection{LEP}
  \label{LEP}

Let us now investigate the situation at high-energy colliders, taking
as examples LEP1 at $E_1 = E_2 = 45$~GeV and LEP2 at $E_1 = E_2 =
95$~GeV.

In the columns labeled ``no cuts'' in Table \ref{lep} we list our
predicted cross sections, with cuts only on $l'_{\perp L}$ and
$\alpha_{2L}$ so that the real photon flux is defined. For the
kinematics we have chosen, the cross sections come out about a factor
2 to 3 larger than at the $B$-factories. Luminosities at LEP are
however much smaller, so that unfortunately we estimate rather low
achievable event rates, and it is not clear to what extent studies of
our process in this kinematical regime will be feasible.

\begin{table}
\begin{center}
\begin{tabular}{lrrrr}
 & \multicolumn{2}{c}{LEP1} & \multicolumn{2}{c}{LEP2} \\
 & no cuts & with cuts & no cuts & with cuts \\
$E_1=E_2$                  [GeV] & 45 & 45 & 95 & 95 \\
$Q^2_{{\mathit{max}}}$ [GeV$^2$] \rule[-1ex]{0ex}{1ex} 
                                 & 20 & 20 & 40 & 40 \\
\hline
$\sigma$     [fb] & 1023 & 167 & 1333 & 50 \\
$\sigma_{G}$ [fb] &   86 &  53 &  124 & 17 \\
$\sigma_{B}$ [fb] &  937 & 114 & 1209 & 33 \\
$S_{ee}(\mathrm{sgn}(\cos\varphi))$ [fb] & 128 & 41 & 159 & 13 \\
$S_{ee}(\cos\varphi)$               [fb] & 101 & 32 & 125 & 10 \\
$\sqrt{N}\, \delta(\mathrm{sgn}(\cos\varphi))$ &
  8.0 & 4.0 & 8.4 & 3.7 \\
$\sqrt{N}\, \delta(\cos\varphi)$ &
  7.0 & 3.5 & 7.4 & 3.3 \\
\end{tabular}
\caption{\label{lep} Cross sections for $e^+e^- \to e^+e^-\,
\pi^+\pi^-$, integrated over the range $W = 300$~MeV to 1000 MeV, $Q^2
= 4$~GeV$^2$ to $Q^2_{{\mathit{max}}}$, and $y$ from its lower
kinematical limit (\protect\ref{y-limit}) up to~0.5. The columns
marked ``no cuts'' correspond to imposing only the cuts that determine
the real photon flux as explained in Sect.~\protect\ref{lab-kin}, with
parameters $\alpha_{2L}^{\mathit max}=30$~mrad and
$l'^{{\mathit{max}}}_{\perp L}=100$~MeV. The columns ``with cuts''
refer to the additional cuts described in the text. The sign of the
weighted cross sections $S_{ee}(\mathrm{sgn}(\cos\varphi))$ and
$S_{ee}(\cos\varphi)$ is for a tagged $e^+$.}
\end{center}
\end{table}

To see the effect of cuts on the detected particles we require
\begin{itemize}
\item $\alpha_{1L}^{\mathit min} \le \alpha_{1L} \le
\pi-\alpha_{1L}^{\mathit min}$ with $\alpha_{1L}^{\mathit
min}=30$~mrad and $E_1{}^{\!\!'} \ge 0.7\, E_1$ for the tagged lepton,
\item $\theta_{L}^{\mathit min} \le (\theta_{L}^{\phantom{m}},
\theta'_{L}) \le \pi-\theta_{L}^{\mathit min}$ with
$\theta_{L}^{\mathit min}=262$~mrad, corresponding to pseudorapidities
$|\eta| \le 2$, and a minimum transverse momentum of 100~MeV for each
of the pions.
\end{itemize}
The results are given in the columns ``with cuts'' of Table
\ref{lep}. The most serious restriction here is the cut on the pion
angles $\theta_{L}^{\phantom{m}}$ and $\theta'_{L}$. This can be
understood from our considerations after Eq.~(\ref{y-zero}). The value
of $y$ where the $\pi\pi$ system has zero longitudinal momentum in the
laboratory is $Q /(2 E_1)$ and thus of order 0.01 to 0.05 here. Over
most of the $y$-range the pions are therefore so strongly boosted in
the lab that they appear under extremely small angles and cannot be
detected.  We observe in fact in Table~\ref{lep} that the effect of
cuts is stronger at LEP2 with its higher beam energy, and that it is
more pronounced for bremsstrahlung than for the $\gamma^*\gamma$
process, the latter being less affected by a loss of events at larger
$y$.

At LEP1 the cut on $\alpha_{1L}$ puts no restriction on $y$, but for
LEP2 we find that for $Q^2 = 4$~GeV$^2$ it implies $y > 0.5$, so that
one must go to larger $Q^2$. For $Q^2$ of about 8~GeV$^2$ there is no
restriction on $y$ from the constraint (\ref{y-min}) any more.

We finally note that at the very large values of $Q^2$ accessible at
high-energy colliders one can afford invariant masses $W$ well above
1~GeV, while still fulfilling the basic condition $W^2 \ll Q^2$ of our
study. We have not explored this mass region, since our model for the
pion GDA is not applicable there. It is however clear that there will
be a strong enhancement of the GDAs at $W$ around the masses of
$C$-even resonances, such as the $f_2$(1270).

\section{Summary and outlook}
\label{summary}

In this paper we have analyzed in detail the process
$\gamma^*\gamma\to \pi\pi$ in the domain where the virtuality $Q$ of
the $\gamma^*$ is much larger than the invariant mass $W$ of the
two-pion system. It factorizes into a parton-level subprocess, which
is under perturbative control, and non-perturbative matrix elements
called generalized distribution amplitudes. This makes the reaction a
laboratory to study the non-perturbative dynamics of a two-pion system
forming from a well-defined partonic state, namely from a
quark-antiquark or a two-gluon pair produced at small distance. The
perturbative stage of the overall process is completely analogous to
the one in single-meson production, well studied in the case of a
$\pi^0$, $\eta$ and $\eta'$. It results in a scaling behavior of the
amplitude as $Q^2$ increases at fixed $W^2$, selects characteristic
helicity combinations of the two photons, and predicts that the two
pions are produced with total isospin zero. The dynamical content of
the non-perturbative matrix elements, on the other hand, is more
complex than for a single particle. Even the lowest Fock state of
$|\pi\rangle \otimes |\pi\rangle$, that is, $q\bar{q} \otimes
q\bar{q}$, contains more partons that the initial $q\bar{q}$ or $gg$
system from which the two pions are formed. In this sense a GDA
describes the transition between different parton configurations in
the non-perturbative regime. The two-pion distribution amplitude
contains the full strong interactions between the two pions, leading
to dynamical phases which, by Watson's theorem, are identical to the
phase shifts in elastic $\pi\pi$ scattering as long as $W$ is below
the inelastic threshold. We use this relation as an input for our
model GDA, and therefore restrict our study to the $W$-region up to
1~GeV.

The evolution equation giving the factorization scale dependence of
the GDAs is more complex than for a single pion due to the mixing of
$q\bar{q}$ or $gg$ amplitudes, and we have given the relevant
splitting functions and anomalous dimensions for the quantum numbers
of relevance here. A simultaneous expansion of $\Phi(z,\zeta,W^2)$ in
the parton momentum fraction $z$ and partial waves of the pion system
leads to local matrix elements between the vacuum and a two-pion
state. By analytic continuation they are related to the moments of the
parton distribution functions of the pion. We have used the quark
momentum fraction $R_\pi$ in the pion, determined from a global fit of
these distributions, as an input for our model of
$\Phi(z,\zeta,W^2)$. The corresponding value of $R_\pi$ is well below
its asymptotic value under perturbative evolution, which may be an
indication that the lowest non-asymptotic terms in the crossed-channel
quantity $\Phi(z,\zeta,W^2)$ are not small at factorization scales in
the GeV range. We emphasize that the question of how close one is to
the asymptotic result of evolution is particularly interesting,
because in the case of light pseudoscalars the single-meson
distribution amplitudes may be surprisingly close to their asymptotic
form even at low scales~\cite{CLEO,Feld}.

{}From a theory point of view it is also interesting to consider
$\Phi_q(z,\zeta,W^2)$, defined by the matrix element in
Eq.~(\ref{gda-def}), for values of $W$ much larger than the scale of
non-perturbative interactions. While the dynamics in
$\Phi_q(z,\zeta,W^2)$ is entirely soft for small $W$, part of it
becomes hard when $W$ increases. In the limit $W \gg 1$~GeV and to
leading order in $\alpha_S$ one can explicitly write
$\Phi_q(z,\zeta,W^2)$ in terms of a perturbative subprocess and the
$q\bar{q}$ distribution amplitudes for each separate
pion~\cite{WU}. The resulting $\Phi_q(z,\zeta,W^2)$ is very far from
the asymptotic form in $z$. It receives substantial contributions from
high partial waves of the $\pi\pi$ system, has a power-law falloff
like $1/W^2$, and its imaginary part is small compared to its real
part.

We have constructed a model for the GDA at $W$ below 1~GeV, using
simple structure as a guide, and $R_\pi$ and the $\pi\pi$ phase shifts
as phenomenological inputs. Comparing the rates for the production of
$\pi\pi$ and of a single pseudoscalar meson, we found that the hadron
spectrum in $\gamma^*\gamma$ collisions below 1~GeV is strongly
dominated by the single resonances $\pi^0$, $\eta$, and $\eta'$.

We have further compared our process with open $q\bar{q}$ production,
which at higher invariant masses $W$ is commonly used to describe the
part of the total hadronic $\gamma^*\gamma$ cross section due to the
pointlike part of the real photon. Interestingly, we find that in our
particular kinematical limit, the corresponding scattering amplitude
has the same scaling behavior and helicity structure as the one for
the exclusive processes $\gamma^*\gamma\to \pi$ and $\gamma^*\gamma\to
\pi\pi$. The main difference is that in the $\pi$ and $\pi\pi$ cases
the collinear divergence of the lowest-order hard scattering diagrams
is regulated by the hadronization process. This is encapsulated in the
distribution amplitudes, which vanish at the end points $z=0$ and
1. In the open $q\bar{q}$ calculation, on the other hand, the
divergence has to be regulated explicitly. We also note that the
sensitivity to the soft end-point region may be larger for pion-pair
production than for a single pion, because for two pions the hard
scattering and the distribution amplitudes vanish at $z=1/2$ for
symmetry reasons. Thus one may expect the onset of the scaling
behavior to occur at different $Q^2$ in the two cases, an issue that
will be interesting to study in experiment.

An investigation of the structure of the cross section shows that in
$e\gamma$ and $e^+e^-$ collisions information on the $\gamma^*\gamma$
process can be obtained either through the square of the
$\gamma^*\gamma$ amplitude, or from its interference with the
bremsstrahlung process if the pions are charged. This interference can
readily be projected out by appropriate $C$-odd observables, and it
offers the opportunity to separate the different $\gamma^*\gamma$
helicity amplitudes. It further provides direct access to their
dynamical phases, although a full phase reconstruction requires
polarized beams (cf.\ Appendix~\ref{beam-polar}).

The angular distribution of the pion pair in its c.m.\ contains
detailed information about the dynamics of the $\gamma^*\gamma$
process. The dependence on the azimuth $\varphi$ separates the
different helicity combinations of the real and virtual photon, each
of which plays a distinct role in the scaling limit. In particular it
permits one to study leading-twist and non-leading twist amplitudes at
the same time, which should provide additional insight into how far
one is from the asymptotic regime. The $\theta$-dependence, on the
other hand, gives access to the partial waves in which the two pions
are produced. It is sensitive to the phases, which reflect the
dynamics of the $\pi\pi$ system and its resonances. Even though one
will probably not be able to perform a full extraction of the $\pi\pi$
phase shifts in this way, our process provides constraints on these
quantities that are independent of the analyses of elastic $\pi\pi$
scattering. The presence of higher partial waves would in itself be
very interesting, since it gives indirect information on the deviation
of $\Phi(z,\zeta,W^2)$ from its asymptotic form in $z$.

We have restricted ourselves to the production of pion pairs in this
work, but it is clear that many of our results are also valid for
other exclusive systems. The most obvious generalization is to charged
or neutral $K\bar{K}$ pairs, whose comparison with $\pi\pi$ would
allow one to study aspects of flavor $SU(3)$ breaking in the context
of the quark-hadron transition. At even higher values of $W^2$ there
is the production of $p\bar{p}$, where extra spin degrees of freedom
come in, as in the well-studied case of the parton distributions of the
nucleon.

Another very similar process is the production of $\mu^+\mu^-$ pairs,
i.e., the QED analogue of our reaction. Comparing the rates of
$e\gamma\to e\, \mu^+\mu^-$ with our estimate for $e\gamma\to e\,
\pi^+\pi^-$ we find that the bremsstrahlung mechanism prefers pions if
$W$ is in the vicinity of the $\rho$ mass, reflecting the strong
resonance effect in the $\pi\pi$ system. For the production from
$\gamma^*\gamma$, on the other hand, the cross section is considerably
larger in the case of muon pairs. We remark that this could not be
anticipated from a dimensional analysis. The amplitudes for
$\gamma^*\gamma\to \mu\mu$ and for $\gamma^*\gamma\to \pi\pi$ have the
same $Q^2$-dependence in our kinematical limit, and the two-pion
distribution amplitude, which describes that pions are not pointlike
but have internal structure, is a dimensionless quantity.

Using our model GDA to calculate the cross section for $e^+e^-\to
e^+e^-\, \pi\pi$, we find encouraging rates for the kinematics and
luminosity of $B$-factories. Thus there should be enough statistics
for detailed studies at these facilities. Our estimates of the effect
of cuts also indicate that in the kinematical region interesting in
our context, the pions and the tagged lepton are well within the
experimental acceptance. For high-energy colliders such as LEP, our
predictions are less optimistic, at least in the range of $W$ below
1~GeV which we have studied here, due both to the lesser luminosity
and the strong longitudinal boost of the pion system.

In conclusion, we find that the process $\gamma^*\gamma\to \pi\pi$ can
offer valuable insight into the interactions between quarks, gluons
and hadrons, and that it should well be measurable at existing
$e^+e^-$ facilities.

\acknowledgments

It is a pleasure to thank P. Aurenche, S.J.\ Brodsky, T. Feldmann,
M. Fontannaz, P. Hoyer, L. Mankiewicz, O.~Nachtmann, M. Polyakov and
O.V.\ Teryaev for discussions, and H. Marsiske, C. Munger, V. Savinov,
S.~S\"oldner-Rembold, S. Uehara and M. Wang for their interest and
valuable information about experimental aspects.

M.D.\ thanks CPhT and LPNHE of \'Ecole Polytechnique for kind
invitations.

\appendix

\section{Pion isospin states}
\label{pion-states}

We specify in this appendix our sign convention for the definition of
pion states. The relative sign for $\pi^+$ and $\pi^-$ is relevant
because it determines the relative sign of the GDAs for charged and
neutral pion pairs.

In terms of eigenstates $|\pi^i\rangle$ of the isospin operators $I^i$
($i=1,2,3$) we define
\begin{equation}
|\pi^+\rangle = \frac{1}{\sqrt{2}} \left(
                    |\pi^1\rangle + i |\pi^2\rangle \right) ,
\hspace{3em}
|\pi^-\rangle = \frac{1}{\sqrt{2}} \left(
                    |\pi^1\rangle - i |\pi^2\rangle \right) ,
\hspace{3em}
|\pi^0\rangle = |\pi^3\rangle . \phantom{\frac{1}{2}} 
 \label{state-def}
\end{equation}
Notice that the sign for $|\pi^+\rangle$ is opposite to the usual
convention for eigenstates of $SU(2)$. This has to be remembered when
writing down two-pion states with definite isospin using the
Clebsch-Gordan coefficients.

The convention (\ref{state-def}) is in line with the customs of field
theory, see for instance Sect.~12.5 of~\cite{BJD}. If, starting from
the real scalar fields associated with $|\pi^1\rangle$ and
$|\pi^2\rangle$, one constructs the complex scalar field $\varphi$
which creates $|\pi^-\rangle$ out of the vacuum, then $|\pi^+\rangle$
is created by the conjugated field $\varphi^*$. If one used the
opposite sign in defining $|\pi^+\rangle$, which is more natural in
the context of isospin, then there would be an extra minus sign
between the fields creating $|\pi^-\rangle$ and
$|\pi^+\rangle$. Through the LSZ reduction formula this sign would
show up in crossing relations. With our definition (\ref{state-def})
this does not happen, and we have for instance that the spacelike pion
form factor
\begin{equation}
 \langle \pi^+(p) |\, J^\mu_{\mathrm{em}}(0) \,| \pi^+(p') \rangle =
  (p+p')^\mu\, F_\pi(t)
\end{equation}
with $t=(p-p')^2$ becomes
\begin{equation}
 \langle \pi^+(p) \pi^-(p') |\, J^\mu_{\mathrm{em}}(0) \,| 0\rangle =
  (p-p')^\mu\, F_\pi(s)
\end{equation}
with $s=(p+p')^2$ in the timelike region. We remark in passing that if
one uses the isospin relation (\ref{more-isospin}) and neglects the
contributions from strange and heavy quarks, one has the sum rule
\begin{equation}
\int dz\, \Phi^-_u(z,\zeta,W^2) = (2\zeta-1)\, F_\pi(W^2) .
\end{equation}

The choice (\ref{state-def}) also leads to a convenient relation for
the action of the charge conjugation operator $C$, namely
\begin{equation}
C |\pi^+\rangle = |\pi^-\rangle , \hspace{3em}
C |\pi^0\rangle = |\pi^0\rangle .
\end{equation}
The impossibility to find a sign convention that is natural for both
charge conjugation and the isospin algebra is discussed at length in
Chapt.~5, \S7 of~\cite{MartSp} (where the other sign in defining
$|\pi^+\rangle$ was chosen). We also remark that the definition
(\ref{state-def}) implies
\begin{equation}
\langle \pi^+ |\, \bar{u}_\alpha(x)\, d_\beta(0) \,| 0\rangle =
\langle \pi^- |\, \bar{d}_\alpha(x)\, u_\beta(0) \,| 0\rangle ,
\end{equation}
and therefore a relative plus sign between the distribution amplitudes
for $\pi^+$ and $\pi^-$.

Our definition is the same as the one chosen by Polyakov et al., cf.\
\cite{Poly-W}, and it was also adopted in~\cite{WU}. We finally
mention that the definition leading to Eq.~(15) of~\cite{DGPT} has the
opposite sign for $|\pi^+\rangle$.

\section{Beam polarization}
\label{beam-polar}

As we have shown in Sects.\ \ref{neutral-pi} and
\ref{interference-term}, the unpolarized $e\gamma$ cross section
contains detailed information on the $\gamma^*\gamma$ helicity
amplitudes $A_{ij}$. From Eqs.~(\ref{gamma-gamma}) and
(\ref{inter-coefficients}) it is however clear that this information
is not sufficient to fully reconstruct the three independent complex
amplitudes $A_{++}$, $A_{0+}$ and $A_{+-}$. For completeness we give
in this appendix the expressions of the cross section with
longitudinally polarized lepton and photon beams, and discuss what
additional information can be obtained from single and double
polarization asymmetries.

Starting with the $\gamma^*\gamma$ contribution, we have
%
%
\begin{eqnarray}
\left. \frac{d\sigma_{e\gamma\to e\, \pi\pi}}{
          dQ^2\, dW^2\, d(\cos\theta)\, d\varphi} \right|_{G} = 
{\mathrm{eq.~(\protect\ref{gamma-gamma})}} + \frac{\alpha^3}{16\pi}
\frac{\beta}{s_{e\gamma}^2}\, \frac{1}{Q^2 (1-\epsilon)}
 & \Big( &  P_l\, \sin\varphi \; \Im \left\{ A^*_{++}
     A^{\phantom{*}}_{0+} - A^*_{-+} A^{\phantom{*}}_{0+} \right\}
     2\sqrt{\epsilon(1-\epsilon)}  \nonumber \\
\phantom{\frac{1}{s_{e}}}
 &+& P_\gamma\, \sin\varphi \; \Im \left\{ A^*_{++}
     A^{\phantom{*}}_{0+} + A^*_{-+} A^{\phantom{*}}_{0+} \right\}
     2\sqrt{\epsilon(1+\epsilon)} \nonumber \\
\phantom{\frac{1}{s_{e}}}
 &+& P_\gamma\, \sin2\varphi \; \Im \left\{ A^*_{++}
     A^{\phantom{*}}_{-+}  \right\} 2\epsilon \nonumber \\
\phantom{\frac{1}{s_{e}}}
 &+& P_l P_\gamma\,
     \left\{ |A_{++}|^2 - |A_{-+}|^2 \right\}
     \sqrt{1-\epsilon^2} \nonumber \\
\phantom{\frac{1}{s_{e}}}
 &-& P_l P_\gamma\, 
     \cos\varphi\; \Re \left\{A^*_{++} A^{\phantom{*}}_{0+}
                            + A^*_{-+} A^{\phantom{*}}_{0+} \right\}
     2\sqrt{\epsilon(1-\epsilon)} \, \Big) ,
\end{eqnarray}
where $P_l$ and $P_\gamma$ respectively denote the longitudinal
polarization of the lepton and photon beam, ranging from $-1$ to
1. Together with Eq.~(\ref{gamma-gamma}) we see that if both lepton
and photon are polarized, one has enough independent terms to
reconstruct the real and imaginary parts of the interferences
$A^*_{++} A^{\phantom{*}}_{-+}$, $A^*_{++} A^{\phantom{*}}_{0+}$ and
$A^*_{-+} A^{\phantom{*}}_{0+}$. Furthermore, the squared terms
$|A_{++}|^2$ and $|A_{-+}|^2$ come with a different relative sign in
the unpolarized cross section and the double polarization asymmetry.

The bremsstrahlung contribution to the cross section reads
\begin{eqnarray}
  \label{brems-pol}
\left. \frac{d\sigma_{e\gamma\to e\, \pi\pi}}{
             dQ^2\, dW^2\, d(\cos\theta)\, d\varphi}
\right|_{B} = {\mathrm{eq.~(\protect\ref{brems})}} +
  \frac{\alpha^3}{16\pi}\, \frac{\beta}{s_{e\gamma}^2}\,
  \frac{2\beta^2}{W^2\, \epsilon}\, |F_\pi(W^2)|^2 \,
  P_l P_\gamma\,
& \Big( & (2x-1) \sqrt{1-\epsilon^2}\, \sin^2\theta
\nonumber \\
&+& \cos\varphi\; \sqrt{2x(1-x)} \sqrt{\epsilon(1-\epsilon)}\,
   2 \sin\theta \cos\theta\, \Big) .
\end{eqnarray}
Notice that it only contributes to the unpolarized cross section and
the double polarization asymmetry, but not to single polarization
asymmetries. Finally, the interference term can be written as
\begin{eqnarray}
\left. \frac{d\sigma_{e\gamma\to e\, \pi\pi}}{
             dQ^2\, dW^2\, d(\cos\theta)\, d\varphi}
\right|_{I} = {\mathrm{eq.~(\protect\ref{interfere})}} -
  2 e_l\, \frac{\alpha^3}{16\pi}\, \frac{\beta}{s_{e\gamma}^2}\,
  \frac{\sqrt{2} \beta}{\sqrt{W^2 Q^2 \epsilon(1-\epsilon) }}
& \Big[ &
   P_l\, \Big( C^l_1\, \sin\varphi + C^l_2 \sin2\varphi \Big)
\nonumber \\
\phantom{\frac{1}{s_{e}}}
&+&  P_\gamma\, \Big( C^\gamma_1\, \sin\varphi 
     + C^\gamma_2\, \sin2\varphi + C^\gamma_3\, \sin3\varphi \Big) 
\nonumber \\
\phantom{\frac{1}{s_{e}}}
&+&  P_l\, P_\gamma \Big( C^{l\gamma}_0 
     + C^{l\gamma}_1\, \cos\varphi + C^{l\gamma}_2\, \cos2\varphi 
     \Big) \, \Big]
\end{eqnarray}
with coefficients
\begin{eqnarray}
C^l_1 &=& - \Im \Big\{ F_\pi^* A_{++} \Big\} \,
        x \sqrt{1-\epsilon^2} \sin\theta
\nonumber \\
&&    + \Im \Big\{ F_\pi^* A_{-+} \Big\} \,
        (1-x) \sqrt{1-\epsilon^2} \sin\theta ,
\nonumber \\
C^l_2 &=& \Im \Big\{ F_\pi^* A_{0+} \Big\} \,
        x \sqrt{\epsilon(1-\epsilon)} \sin\theta
\nonumber \\
&&    - \Im \Big\{ F_\pi^* A_{-+} \Big\} \,
        \sqrt{2x(1-x)} \sqrt{\epsilon(1-\epsilon)} \cos\theta 
\end{eqnarray}
for lepton polarization,
\begin{eqnarray}
C^\gamma_1 &=& - \Im \Big\{ F_\pi^* A_{++} \Big\} \,
        [ 1 - (1-x) (1-\epsilon) ] \sin\theta
\nonumber \\
&&    + \Im \Big\{ F_\pi^* A_{0+} \Big\} \,
        \sqrt{2x(1-x)}\, 2\epsilon \cos\theta
\nonumber \\
&&    - \Im \Big\{ F_\pi^* A_{-+} \Big\} \,
        (1-x) \sin\theta
\nonumber \\
C^\gamma_2 &=& \Im \Big\{ F_\pi^* A_{0+} \Big\} \,
        x \sqrt{\epsilon(1+\epsilon)} \sin\theta
\nonumber \\
&&    + \Im \Big\{ F_\pi^* A_{-+} \Big\} \,
        \sqrt{2x(1-x)} \sqrt{\epsilon(1+\epsilon)} \cos\theta ,
\nonumber \\ 
C^\gamma_3 &=& \Im \Big\{ F_\pi^* A_{-+} \Big\} \,
        x \epsilon \sin\theta
\end{eqnarray}
for photon polarization, and
\begin{eqnarray}
C^{l\gamma}_0 &=& \Re \Big\{ F_\pi^* A_{++} \Big\} \,
        \sqrt{2x(1-x)} \sqrt{\epsilon(1-\epsilon)}\, \cos\theta 
\nonumber \\
&&    - \Re \Big\{ F_\pi^* A_{0+} \Big\} \,
        (1-x) \sqrt{\epsilon(1-\epsilon)} \sin\theta ,
\nonumber \\
C^{l\gamma}_1 &=& \Re \Big\{ F_\pi^* A_{++} \Big\} \,
        x \sqrt{1-\epsilon^2} \sin\theta
\nonumber \\
&&    - \Re \Big\{ F_\pi^* A_{-+} \Big\} \,
        (1-x) \sqrt{1-\epsilon^2} \sin\theta ,
\nonumber \\
C^{l\gamma}_2 &=& - \Re \Big\{ F_\pi^* A_{0+} \Big\} \,
        x \sqrt{\epsilon(1-\epsilon)} \sin\theta
\nonumber \\
&&    + \Re \Big\{ F_\pi^* A_{-+} \Big\} \,
        \sqrt{2x(1-x)} \sqrt{\epsilon(1-\epsilon)} \cos\theta
\end{eqnarray}
if both lepton and photon are polarized. We see that with polarized
photons one can extract $\Im \{ F_\pi^* A_{++} \}$, $\Im \{ F_\pi^*
A_{0+} \}$ and $\Im \{ F_\pi^* A_{-+} \}$, which together with the
unpolarized interference term makes it possible to reconstruct all
three complex $\gamma^*\gamma$ amplitudes for values of $W$ where the
pion form factor $F_\pi$ is known. One cannot achieve the same with a
polarized lepton beam alone, since there are only two terms in the
$\varphi$-dependence. In this case one can still use the suppression
by $1-x$ of the second term in $C^l_1$ in order to approximately
extract $\Im \{ F_\pi^* A_{++} \}$. Finally, the double polarization
asymmetry gives access to the same quantities one can already obtain
in the unpolarized case.


\begin{thebibliography}{99}

\bibitem{Terazawa} H.~Terazawa, Rev.\ Mod.\ Phys.\ {\bf 45}, 615
(1973).

\bibitem{Bud} V.M. Budnev {\it et al.}, Phys.\ Rept.\ {\bf 15C}, 181
(1975).

\bibitem{photon-conf} S.J.~Brodsky, hep-ph/9708345, talk presented at
PHOTON~97, Egmond aan Zee, Netherlands, May 1997; \\
M.R.~Pennington, Nucl.\ Phys.\ {\bf B} \ (Proc.\ Suppl.) {\bf 82}, 291
(2000), hep-ph/9907353.

\bibitem{DGPT} M. Diehl, T. Gousset, B. Pire and O.V. Teryaev, Phys.\
Rev.\ Lett.\ {\bf 81}, 1782 (1998), hep-ph/9805380; \\ 
M. Diehl, T. Gousset and B. Pire, Nucl.\ Phys.\ {\bf B} \ (Proc.\
Suppl.) {\bf 82}, 322 (2000), hep-ph/9907453.

\bibitem{MRG} D. M\"uller {\it et al.}, Fortschr.\ Phys.\ {\bf 42},
101 (1994), hep-ph/9812448.

\bibitem{Freund} A. Freund, Phys.\ Rev.\ {\bf D61}, 074010 (2000),
hep-ph/9903489.

\bibitem{LepageBrodsky} G.P. Lepage and S.J. Brodsky, Phys. Rev. {\bf
D22}, 2157 (1980).

\bibitem{TFF} S. Ong, Phys.\ Rev.\ {\bf D52}, 3111 (1995); \\ 
R. Jakob, P. Kroll and M. Raulfs, J. Phys. {\bf G22}, 45 (1996),
hep-ph/9410304; \\
P. Kroll and M. Raulfs, Phys.\ Lett.\ {\bf B387}, 848 (1996),
hep-ph/9605264; \\
A.V. Radyushkin and R.T. Ruskov, Nucl.\ Phys.\ {\bf B481}, 625 (1996),
hep-ph/9603408; \\
I.V. Musatov and A.V. Radyushkin, Phys.\ Rev.\ {\bf D56}, 2713 (1997),
hep-ph/9702443.

\bibitem{CLEO} CLEO Collab., J. Gronberg {\it et al.}, Phys.\ Rev.\
{\bf D57}, 33 (1998), hep-ex/9707031.

\bibitem{DVCS} X. Ji, J.\ Phys.\ {\bf G24}, 1181 (1998),
hep-ph/9807358, and references therein.

\bibitem{rho-pi-pi} M. Diehl, T. Gousset and B. Pire, to appear in the
Procs.\ of the Workshop on Exclusive and Semiexclusive Processes at
High Momentum Transfer, Jefferson Lab, Newport News, VA, USA, May
1999, hep-ph/9909445; \\
B. Lehmann-Dronke {\it et al.}, Phys.\ Lett.\  {\bf B475}, 147 (2000),
hep-ph/9910310.

\bibitem{Wat} K. Watanabe, Prog.\ Theor.\ Phys.\ {\bf 67}, 1834 (1982).

\bibitem{Leipzig} J. Bl\"umlein, B. Geyer and D. Robaschik, Nucl.\
Phys.\ {\bf B560}, 283 (1999), hep-ph/9903520, and references therein.

\bibitem{DGPR} M. Diehl, T. Gousset, B. Pire and J.P. Ralston, Phys.\
Lett.\ {\bf B411}, 193 (1997), hep-ph/9706344.

\bibitem{CD} J.C. Collins and M. Diehl, Phys.\ Rev.\ {\bf D61}, 114015
(2000), hep-ph/9907498.

\bibitem{Kivel} N. Kivel, L. Mankiewicz and M.V. Polyakov, Phys.\
Lett.\ {\bf B467}, 263 (1999), hep-ph/9908334.

\bibitem{BG-rev} V.N. Baier and A.G. Grozin, Sov.\ J.\ Part.\ Nucl.\
{\bf 16}, 1 (1985), and references therein.

\bibitem{Poly-W} M.V. Polyakov and C. Weiss, Phys.\ Rev.\ {\bf D60},
114017 (1999), hep-ph/9902451.

\bibitem{Pol} M.V. Polyakov, Nucl.\ Phys.\ {\bf B555}, 231 (1999),
hep-ph/9809483.

\bibitem{private} M.V. Polyakov, private communication.

\bibitem {BB} I.I. Balitsky and V.M. Braun, Nucl.\ Phys.\ {\bf B311},
541 (1988/89).

\bibitem{ERBL} G.P. Lepage and S.J. Brodsky, Phys.\ Lett.\ {\bf B87},
359 (1979); \\
A.V. Efremov and A.V. Radyushkin, Phys.\ Lett.\ {\bf B94}, 245 (1980).

\bibitem{BG-odd} V.N. Baier and A.G. Grozin, Nucl.\ Phys.\ {\bf B192},
476 (1981).

\bibitem{BG-even} V.N. Baier and A.G. Grozin, Sov.\ J.\ Nucl.\ Phys.\
{\bf 35}, 596 (1982).

\bibitem{Chase} M.K. Chase, Nucl.\ Phys.\ {\bf B174}, 109 (1980).

\bibitem{CZ} V.L. Chernyak and A.R. Zhitnitsky, Phys. Rept. {\bf 112},
173 (1984).

\bibitem{Alt} G.~Altarelli, Phys.\ Rept.\ {\bf 81}, 1 (1982).

\bibitem{Estabrooks} P.~Estabrooks and A.D.~Martin, Nucl.\ Phys.\
{\bf B79}, 301 (1974).

\bibitem{Hyams} B.~Hyams {\it et al.}, Nucl.\ Phys.\ {\bf B64},
134 (1973); \\
D.V.~Bugg, B.S.~Zou and A.V.~Sarantsev, Nucl.\ Phys.\ {\bf B471}, 59
(1996); \\
R.~Kaminski, L.~Lesniak and K.~Rybicki, hep-ph/9912354.

\bibitem{Kuehn} J.H. K\"uhn and A. Santamaria, Z.\ Phys.\ {\bf C48},
445 (1990).

\bibitem{GRS} M. Gl\"uck, E. Reya and I. Schienbein, Eur.\ Phys.\ J.\
{\bf C10}, 313 (1999), hep-ph/9903288.

\bibitem{Feld} T.~Feldmann, Int.\ J.\ Mod.\ Phys.\ {\bf A15}, 159
(2000), hep-ph/9907491.

\bibitem{GorSto} L.E.~Gordon and J.K.~Storrow, Z.\ Phys.\ {\bf C56},
307 (1992).

\bibitem{opt-obs} M. Diehl and O. Nachtmann, Z.\ Phys.\ {\bf C62}, 397
(1994).

\bibitem{Fpi-data} L.M. Barkov {\it et al.}, Nucl.\ Phys.\ {\bf
B256}, 365 (1985).

\bibitem{BDK} F.A.~Berends, P.H.~Daverveldt and R. Kleiss, Nucl.\
Phys.\ {\bf B253}, 441 (1985); \\ 
Comput.\ Phys.\ Commun.\ {\bf 40}, 285 (1986); Comput.\ Phys.\
Commun.\ {\bf 40}, 309 (1986).

\bibitem{WU} M. Diehl, T. Feldmann, P. Kroll and C. Vogt, Phys.\ Rev.\
{\bf D61}, 074029 (2000), hep-ph/9912364.

\bibitem{BJD} J.D. Bjorken and S.D. Drell, \emph{Relativistic Quantum
Fields}, McGraw-Hill, 1965.

\bibitem{MartSp} A.D. Martin and T.D. Spearman, \emph{Elementary
Particle Theory}, North Holland, 1970.


\end{thebibliography}
\end{document}